\documentstyle[aps,pre,epsf,twocolumn,graphicx]{revtex}

\def\be{\begin{equation}}
\def\ee{\end{equation}}
\def\ba{\begin{eqnarray}}
\def\ea{\end{eqnarray}}

\def\la{\langle}
\def\ra{\rangle}

\def\pr{{\rm Prob}}

\def\bx{{\bf x}}

\def\fh{F_{c_{{\rm hyst}}}^{\uparrow}}
\def\fa{F_{c_{{\rm aval}}}^{\uparrow}}
\def\fd{F_c^{\downarrow}}

\def\fpm{f_p^{max}}

\begin{document}
\title{Depinning with dynamic stress overshoots: A hybrid of critical and pseudohysteretic behavior}
\date{\today}
\author{J.M. Schwarz$^1$ and Daniel S. Fisher$^2$}
\address{$^1$ Department of Physics, 201 Physics Building, Syracuse University, 
Syracuse, New York  
13244-1130}
\address{$^2$ Lyman Laboratory of Physics, Harvard University, Cambridge, 
Massachusetts 02138} 
\maketitle

\begin{abstract}
A model of an elastic manifold driven through a random medium by an applied force $F$ is introduced and studied. The focus is on the effects of inertia and 
elastic waves, in particular {\it stress overshoots} in which motion of one segment of the manifold causes a temporary stress on its neighboring segments in addition to the static stress. 
Such stress 
overshoots  
decrease the critical force for depinning and make the
depinning transition hysteretic with static and pinned configurations coexisting with the steadily moving phase for a range of $F$. We find that the steady state velocity of the moving phase is, nevertheless, history independent and the critical behavior as the force is decreased is in the same universality class as in the absence of stress overshoots --- the dissipative limit in which hysteresis cannot occur and theoretical analysis has been possible.  
To reach this conclusion, finite-size scaling analyses have been performed  and a variety of quantities studied, including velocities, roughnesses, distributions of critical forces, and universal amplitude ratios.  \

If the force is increased slowly from zero, the behavior is complicated with a spectrum of avalanche sizes occurring that seems to be  quite different from the dissipative limit. Related behavior is seen as the force is increased back up again to restart the motion of samples that have been stopped from the moving phase. The restarting process itself involves both fractal and bubble-like nucleation.   Hysteresis loops in small and intermediate size samples  can be understood in terms of a depletion layer caused by the stress overshoots. Surprisingly, in the limit of very large samples the hysteresis loops vanish.  Although complicated crossovers complicate the analysis, we argue that the underlying universality class governing this pseudohysteresis and avalanches is again that of the apparently-very-different dissipative limit.  But there are history dependent amplitudes --- associated with the depletion layer --- that cause striking differences over wide ranges of length scales.  Consequences of this picture for the statistics and dynamics of earthquakes on geological faults are briefly discussed.  
\end{abstract}

\section{Introduction}

Extended elastic manifolds  
 pulled through a {\em quenched} random medium by an 
applied force $F$ exhibit, in the absence of thermal fluctuations, a sharp transition from a  pinned
phase to a  
moving phase as $F$ is increased through a  critical 
value $F_c$ \cite{fisher1}.   Examples
 include interfaces between two fluids in porous media \cite{levine} or between oppositely magnetized ferromagnetic domains,  
vortex lines and lattices in type II superconductors \cite{dahmen},
\cite{blatter}, charge density waves \cite{narayan1}, and planar crack fronts in solids \cite{ramanathan1}.  

Although the {\it depinning} transitions of interest are driven non-equilibrium transitions, it is instructive
to draw an analogy with equilibrium phase transitions with the average velocity  $\bar{v}$  playing the role of an order parameter, $F$ a tuning parameter,
and the quenched variations of the random potential loosely analogous --- at least as giving rise to an ensemble --- to 
thermal fluctuations.  The character of depinning transitions can, one might expect, 
be either discontinuous transitions with hysteresis --- loosely like first-order transitions ---, or critical --- analogous to  second-order ---  transitions depending on the system and, perhaps, on its {\it history}; such history dependence is an effect that cannot occur in true equilibrium.
Theoretical analysis has shown that a broad class of realistic models undergo a {\it critical} depinning transition with a unique, history independent critical force in the limit of a large system and     
\begin{equation}
\bar{v}\sim (F-F_c)^{\beta}
\end{equation}
for $F$ just above $F_c$. Several different universality classes have been studied, including both short and long-range interactions, and random forces with or without periodicity --- the former arising for manifolds with a periodic structure in the direction in which they move \cite{narayan1},\cite{narayan2}, 
\cite{ertas}, \cite{nattermann}, \cite{fisher2}.
 
But most of the theoretical analysis has 
focused on dissipative dynamics for which   both inertia and any wave or other non-local stress propagation effects are ignored. The purpose of this paper is to study some of the consequences of these and other effects which we shall generically refer to, for reasons to be explained shortly, as {\it stress overshoot} effects. 

In order to understand the potential of these effects to significantly change the nature of depinning, it is necessary to consider the nature of the irregular local motion that underlies the critical depinning phenomena.  The 
elasticity of the manifold
mediates between
two competing types of  forces: the applied driving force and the random local pinning forces. For small $F$, the pinning dominates and 
the system  relaxes to one of many static locally-stable configurations. But as the force is slowly increased, there will be local instabilities  when the driving force exceeds the random pinning in some small region. A segment of the manifold will then move forward rapidly and there will be some transient motion, limited in spatial extent, until a new static configuration is reached.  
For large $F$, in contrast,  the 
applied force will dominate and the system will approach a nonequilibrium statistically steady state with a non-zero mean velocity. Nevertheless, especially  if the force is not too large, the motion on short length and time scales will be very irregular with instantaneous local velocities that far exceed the macroscopic average velocity $\bar v$.    

In the absence of inertia or wave propagation, each segment of the manifold will move in response to the total force applied to it: from the applied drive, from the random pinning, and from the other segments via elasticity.  As long as the applied force is non-decreasing in time, the motion (at least after initial transients have decayed away) will be only in the ``forward" direction in which the system is driven. This, combined with the convexity properties of elasticity, means that the static configuration the system will settle into after it is disturbed by an increase --- global or local --- in the applied force does not depend on the details of its dynamics.  The motion can thus be considered as quasistatic, {\it in spite of} the rapid local motion that occurs.
 
Now consider what can happen in the presence of either inertia  or elastic waves that carry stress from one region to another in response to motion of one segment. The local dynamics  would then appear to be crucial:  If the local motion is rapid enough that the relaxation to a new static configuration is underdamped, a moving segment can overshoot one or more potential static configurations before settling, if at all, in another.  Even if the inertia is small enough that such local prolonged jumps do not occur, as long as the motion is sometimes underdamped, a segment can temporarily overshoot a static configuration before relaxing back into it; this will produce a temporary overshoot of the stress  --- above its eventual static value --- that this motion induces on neighboring segments.  Any arbitrary small overshoot in the stress has the potential to dislodge another segment if there is one nearby that was sufficiently close to being destabilized in the absence of the overshoot; again, the effects of this will be to cause the system to skip through a potentially static configuration without stopping.   Elastic waves, just like their electrodynamic cousins, carry with them pulses in stress that are larger than the eventual static stress that will obtain long after the waves have passed by.  These stress overshoots, like those from the inertia of local motion, have the potential to cause overjumping.  

Very generally, overjumping of any kind means that which configuration a pinned manifold stops  in depends on details of its local dynamics and on its history, in a way that cannot occur in the absence of inertial effects.  One particularly interesting consequence of this is the coexistence in two identical samples at the same value of the driving force of a static locally stable configuration, and a moving configuration  that will ``overtake" the static configuration.  What the {\it macroscopic} consequences of this are is the primary subject of this paper.  As stress overshoots can occur more readily than local overjumps caused by  inertia, we will generally refer to both types of effects as stress overshoots, although, with slight inconsistency that we trust will not be confusing, we will characterize  their strength by a parameter that we denote $M$.

 What is the nature of the depinning transition as $M$ is
increased from zero?  There are various scenarios one can readily envisage. For large enough $M$, in the model we introduce in the next section,  an infinitesimal increase in $F$ from {\it any} pinned state  
will result in a non-zero average velocity at long times.  This is 
because sufficiently large stress overshoots always
induce other segments  to move when triggered by an initial segment that
moves in response to an increase in $F$. The increased
triggering will spawn further motion, 
despite the fact that the stress overshoot is only 
temporary. This  process will run away; and the manifold will acquire a non-zero average velocity.    
While it might thus seem likely that the 
transition will become 
``first-order" for large enough $M$, does
it do so for arbitrary small  $M$?   If there are regimes in which the depinning is indeed discontinuous in some way, is  there macroscopic hysteresis? In what sense?  

More generally, what happens to the
depinning transition beyond the dissipative limit?  If it remains critical --- at least in some respects --- for a range of $M$, what is its nature?  Can the quasistatic behavior persist macroscopically for small $M$ in spite of the presence of additional microscopic hysteresis?  If so, what is the size dependence of the hysteresis and related phenomena? The system size dependence is particularly relevant for geological faults for which the statistics of the earthquakes are affected both by the nature of the drive and the distribution of the ``sizes" of faults.

Several recent papers have undertaken some preliminary studies of the effects of stress overshoots in models of depinning of elastic manifolds.  Reference \cite{ramanathan2}  studied
several one-dimensional models with long-range 
elasticity and stress overshoots  motivated by planar crack fronts driven by applied  loads. 
Reference \cite{schwarz} introduced a particularly simple model 
with short-range static elasticity and
 analyzed its 
infinite-range limit  in which all segments of the manifold are equally coupled to 
each other.  In this limit, the 
spatial properties of the manifold are averaged
away;  only time dependence remains and   mean
field theory becomes exact. Such mean field models were the starting point for theoretical understanding of the 
finite-dimensional physics in the quasistatic limit \cite{narayan1}.  Whether or not they provide a useful starting point beyond the dissipative limit, is one of the questions that we must address.

In this paper, we investigate numerically and phenomenologically  the
finite-range version of the stress overshoot  
model introduced in \cite{schwarz}.

\subsection{Outline}

The remainder of this paper is organized as follows:  In the next section we introduce the basic lattice model on which we focus.  In Section III 
the general scaling picture is introduced and known results for the dissipative case are summarized. In Section IV the critical behavior in the moving phase is studied, initially for the dissipative case, and then in the presence of stress overshoots.  We summarize a variety of evidence that  the critical behavior  which occurs as the driving force is decreased until the system stops is in the {\it same} universality class as the dissipative case. 

 Section V turns to the key aspect of overshoots: hysteresis. We analyze the hysteresis loops that occur when  the system is stopped from  the moving phase and restarted by gradually increasing the force.   Various puzzling aspects of the data are discussed and some understanding of the hysteresis loops in terms of a low density of segments that can be readily triggered by an increase in $F$ is reached.  In the following section, studies of the dynamics and statistics of avalanches that occur as the force is gradually increased are presented.  These again lead to puzzling dependence on overshoot and pinning strengths although some aspects of the avalanches appear very similar to those in the dissipative limit.  The data suggest various subtle crossovers may be occurring.  In Section VII the dynamics of the nucleation of restarting after a system has been stopped from the moving phase are analyzed.  It is found that over a substantial range of sizes, bubble-like nucleation can occur.  

In Section VIII the puzzling aspects of the various sets of data are tentatively resolved in terms of a crossover as a function of length scale and system size that manifests itself in different ways as the various parameters are varied.
Finally, in Section IX, the conclusions are summarized and applications to the dynamics of earthquakes are discussed briefly,

In the main body  of this paper, we restrict consideration to weak enough overshoots that they do not totally change the local dynamics.  But for sufficiently large $M$, the overshoots cause dramatic changes in the macroscopic behavior.  Although these are interesting, they are probably peculiar to certain aspects of the model; results on these will be presented elsewhere \cite{thesis}.

\section{Model}

Near the depinning transition, the dynamics is very
jerky with segments of the manifold spending most of their time stationary or almost so, but occasionally  getting unpinned by the forces from other segments and moving forward only  to get pinned again by a combination of the newly explored random forces and the elasticity. The inherent discreteness of these {\it local jumps} suggests that   we model the manifold as a large number of 
segments  that can jump discontinuously from one pinning position to another; this is also convenient for numerical studies.  We define
$h({\bf x},t)$ to be the {\em single-valued} scalar
displacement of the manifold from some undeformed reference configuration with both the  position,  
${\bf x}$ and the time, $t$,  taken to be discrete.  Note that by constraining the displacement field to 
be single-valued, we exclude ``overhangs" as well as defects such as dislocations that could otherwise occur in periodic systems.
The forces on a segment of the manifold consist of three terms: the applied force $F$, a static random pinning force $\eta({\bf x},h({\bf x,t}))$, and the stress caused by the elasticity $\sigma({\bf x},t)$. 

The stress depends linearly on the displacements of other parts of the manifold via
\begin{equation}
\sigma({\bf x},t)=\sum_{\bf y}\sum_{\tau>0} J_{{\bf xy}}(\tau)h(
{\bf y},t-\tau)-\hat{J}h({\bf x},t),
\end{equation}
where 
\begin{equation}
\hat{J}=\sum_{{\bf y},\tau}J_{{\bf xy}}(\tau)
\end{equation}
and the sum is over nearest neighbors of ${\bf x}$.  To model stress overshoots, we assume the simplest possible form:   that 
the overshoot only applies to neighbors and only lasts for one time step so that \begin{equation}
J_{{\bf xy}}(\tau=0)=
\frac{1}{Z}(1+M)
\end{equation}
 and 
\begin{equation}
J_{{\bf xy}}(1)=-\frac{M}{Z}, 
\end{equation}
with $Z$ 
the number of nearest neighbors.  With this stress transfer, the 
jump
of any nearest neighbor of ${\bf x}$ induces an extra temporary stress on the ${\bf x}$th 
 segment.
When $M=0$, 
the stress transferred to the 
${\bf x}$th segment 
is simply proportional to the static curvature at ${\bf x}$. This stress will  not decrease  with time as long as the ${\bf x}$th segment does not move and the other segments only move forward or remain at rest; this limit is thus the dissipative dynamics already studied extensively.  However, 
for positive $M$ the  stress on  the ${\bf x}$th segment caused by a jump forward of one of its neighboring segments will
 first increase by a larger amount than in the absence of $M$ and then {\it decrease} at the next time step to reach its quasistatic value. 

Modeling of the local pinning forces also involves substantial arbitrariness.    In
Ref. \cite{schwarz} we chose randomly spaced pinning positions for each segment with  uniform pinning strengths, but this choice did not affect substantially the mean field behavior \cite{thesis}.  For our present purposes, it is more convenient to choose the pinning  positions for each segment to be uniformly spaced but with their {\it yield strengths}, the maximum force they can sustain, randomly distributed.  In particular, we take the distribution
of these yield strengths to be uniformly distributed from $[0,
f_p^{max}]$.

Because these random forces pin, or hold back, the manifold the corresponding forces $\eta({\bf x},h({\bf x}))$, take on negative values in the range $\big(-f_p({\bf x},h({\bf x})),0\big)$. 
  With this form of the pinning forces, the equation of motion is simply given by
\begin{equation}
h({\bf x},t+1)=h({\bf x},t) + \Theta[\sigma({\bf x},t)+F -f_p({\bf x},h({\bf x},t))],
\end{equation}
where $\Theta$ is the unit step function. The 
theta function is imposed so that a segment can  move only forward and does so when the net force $f(\bx)$ on it (the argument of the theta function), 
is positive; otherwise it remains stationary.  
With this dynamics, when a  segment jumps
its displacement always increases by one.  
As long as $\bar{v}\ll 1$, the (artificial) upper limit on the velocity,  this automaton dynamics
mimics the continuous time motion reasonably well.  

Note that in the absence of elasticity, with $M=0$, 
$F_c= 
f_p^{max}$ with each segment becoming stuck on an anomalously strong pinning site.  In the presence of the elasticity, not all of the segments can be simultaneously pinned on strong pinning sites, and the critical force will decrease. For weak pinning forces, however, the definition of the applied force in this model is somewhat pathological: when released from a pinning segment, a segment can jump forward far enough that the total force on it becomes negative and it can then pull forward other segments resulting in overall motion even if $F$ is negative.  To make it more realistic, one could replace $F$ by $F+1$; so that there is always enough force to make the forces at pinning segments non-negative. With non-zero $M$, another adjustment should really be made as more realistic forms of stress overshoot involve a concomitant negative force on the segment that has moved.  Since the zero of the applied force is entirely a convention, we will not make these adjustments; this will mean, however, that in some regimes the critical ``force"  will be negative.
 
Several additional  aspects of the model need to be  specified: the initial conditions and the order of the updating.  To avoid lock-step or other ``faceting" like behavior \cite{cieplak}, we choose the pinning positions of each segment to be offset from one another by random amounts in the interval $[0,1]$.  Because of the integer character of the jumps, the fractional part of the displacement of a given segment does not evolve with time. It might be thought that the displacement-independent randomness induced by this constraint could dominate over the randomness of interest, especially as far as determining the variations of critical forces, etc. in finite size systems.  [Indeed, just such an effect does occur for systems with periodic randomness such as charge density waves \cite{narayan1}.   But in the present case it can be shown that the additional randomness is analogous to a spatially random force that is the derivative of a random function. This, combined with the statistical tilt symmetry,  mean that its  effects are subdominant for large systems (although they could give rise to additional corrections to scaling.)  The fact that the variations in the critical forces in finite size systems decrease substantially faster than the inverse square root of their area supports this assertion. 

The updating of the displacements are done in {\it parallel} after computing all of the stresses.
While there are alternate sequential methods of updating,
this parallel method requires the least amount of computation and does not appear to introduce any troublesome artifacts in the regime of smallish $M$ of primary interest here. 

Finally, to limit boundary effects, we impose periodic 
boundary conditions on $h({\bf x},t)$. This is especially important as we will use finite size scaling to analyze much of the data; this  is substantially more straightforward with periodic boundary conditions. Our simulations are restricted to two dimensions which we chose because of the availability of the widest range of system sizes without running into the complications associated with very large
stresses that arise in one dimensional depinning with short range interactions.  We study systems of size $L\times L$ up to $256
\times 256$ with most of our ``large" system data on $128\times 128$ samples.

\section{Scaling and Dissipative Dynamics}

Before presenting new results for the systems of interest with stress overshoots, we briefly summarize the scaling behavior that obtains near the  critical force in the absence of stress overshoots; i.e. for $M=0$. 
As the force is adiabatically increased from zero, local instabilities lead to a succession of {\it avalanches}, most of which will be small, but which can occasionally become large as the unique critical force $F_{c0}$ is approached.  Above $F_{c0}$ the mean velocity in the statistical steady state rises continuously with an exponent $\beta$.  The motion is jerky out to length scales of order the velocity {\it correlation length} $\xi_v$ which diverges at the critical force as 
\begin{equation}
\xi_v\sim \frac{1}{(F-F_{c0})^\nu}\ .
\end{equation}
The characteristic time for relaxation on scales of order $\xi_v$ is 
\begin{equation}
\tau \sim \xi_v^z
\end{equation}
and in this time the manifold typically moves  forward by an amount 
\begin{equation}
\Delta h \sim \xi_v^\zeta .
\end{equation}
These three exponents, $\nu$, $z$ and $\zeta$ characterize the scaling behavior near the transition.  The velocity exponent is related to these via the observation that the mean velocity is of order the characteristic displacement per characteristic time so that 
\begin{equation}
\bar v \sim \xi_v^{\zeta-z} \sim (F-F_{c0})^\beta
\end{equation}
with 
\begin{equation}
\beta=(z-\zeta)\nu .
\end{equation}

In the pinned phase, the critical behavior as the force is adiabatically increased can,  in the absence of stress overshoots, be related to that in the moving phase reviewed above. In particular, the scaling of the dynamics and shape of the avalanches, the probability that they will be large, the divergence as $F_{c0}$ is approached of the cutoff size in their distribution, and the ``roughness" of the manifold at the critical point are all given in terms of the same three exponents.  In Section VI we will discuss the avalanches in  detail, but for now we focus on the macroscopic behavior such as the velocity in the moving phase and the  mean displacements in the pinned phase. 

The mean displacement in response to a spatially varying applied force yields, via a statistical symmetry of the system, a scaling law that relates two of the exponents.
This
relation can be derived from the average static polarizability 
\begin{equation}
\chi(q,\omega=0)
\equiv \frac{\partial \overline{h(q,\omega=0)} }{\partial \epsilon(q)}
\end{equation}
 to a  perturbing force $\epsilon(q)\cos({\bf q\cdot x})$.  A change of variables to  $h({\bf x},t)=h'({\bf x},t) -
\nabla^{-2}\epsilon(x)$, yields an equation of motion for $h'$ that is statistically identical to the original one for $h$,  {\it independent}
of the perturbing force.  Therefore 
\begin{equation}
\frac{\partial \overline{h(q,\omega=0)} }{\partial \epsilon(q)}
= \frac{1}{1-J(q)} \sim \frac{1}{q^2} .
\end{equation}
  Since the polarizability should scale as $\Delta h/\Delta F \sim  \xi^{\zeta+1/\nu}X(q\xi)$, with $X$ a scaling function, this yields the scaling law
\begin{equation}
 \zeta+\frac{1}{\nu}=2.
\end{equation}

\subsection{Dissipative  exponents}

The critical exponents for $M=0$ take simple mean field values of $\nu=1/2$, $z=2$, characteristic of diffusive dynamics, and $\zeta=0$ above the critical dimension of $d_c=4$ for short range elasticity.  The velocity scales with $\beta=1$ \cite{narayan1}. 

Below four dimensions, renormalization group expansions have been performed that justify the scaling laws and claims of universality as well as yielding results for the exponents as expansions in powers of 
\begin{equation}
\epsilon\equiv 4-d \ :
\end{equation}
\begin{equation}
\zeta\approx\frac{\epsilon}{3}
\end{equation}
and 
\begin{equation}
z\approx 2-\frac{2}{9}\epsilon .
\end{equation} 
These yield 
\begin{equation}
\beta\approx 1-\frac{1}{9}\epsilon .
\end{equation}
Recently Chauve et al \cite{chauve} have computed these exponents to 
second-order in $\epsilon$, obtaining 
\be
\zeta\approx\frac{\epsilon}{3}(1+0.14331\epsilon) \ 
\ee  
and
\be 
z\approx2-\frac{2}{9}\epsilon - 0.04321 \epsilon^2 \ ,
\ee
although there are some doubts about the validity of these second-order results \cite{narayan2}. 

\section{Critical Behavior in Moving Phase}

 We will shortly turn to presentation of our numerical results for the
critical behavior  in the moving phase.  But first, it is instructive to summarize the behavior found in the mean field limit for small $M$ and to consider several possible scenarios that might obtain in short-range systems.  We can then determine which scenario is most consistent with the data.  

\subsection{Mean field limit and scaling scenarios}

In the mean field limit all of the sites are coupled to all of the others so that the number of ``nearest neighbors" $Z$ is equal to the number of segments, $N$ (more precisely, $Z=N-1$).  In this limit, the critical force is found to be {\it unchanged} for $M$ less than a critical value, $M_c=1$.  The velocity versus force curve is modified for any non-zero $M$, however, but the exponent $\beta$ remains at its quasistatic value of $\beta=1$ for $M<M_c$.
The other universal properties of the transition are also unchanged for small $M$, including the lack of hysteresis in steady state and the asymptotics of the distribution of large avalanches as  the critical force is approached from below.

The simplest scenario for the short range systems of interest would be like that of the mean field limit: unchanged critical behavior and no macroscopic hysteresis for small $M$.  But previous work has shown that this {\it cannot} be the case: As shown in references 
\cite{ramanathan2}, \cite{fisher3}, {\it any} stress overshoot will cause the critical force to be shifted downwards, in the sense that dynamic behavior can persist for some ($M$-dependent) range of forces below the quasistatic critical force $F_{c0}$. This implies that some form of macroscopic hysteresis can exist since  locally stable --- at least linearly stable --- static configurations  exist up to $F_{c0}$.  
But whether such configurations are non-linearly stable to, for example, an arbitrarily small increase in $F$, is a question of substantial importance to which we will return later.  
For now, we focus on the moving states and how they stop as the force is lowered.

The {\it simplest scenario} that cannot immediately be ruled out is a modified version of the mean field scenario: a velocity versus force curve with a well defined critical force, $F_c^\downarrow$, that is non-hysteretic as the force is decreased; history independent steady states above $\fd$; and critical exponents, $\beta,\nu^\downarrow,z^\downarrow$ unchanged from their quasistatic values. 
In renormalization group language, this would correspond to $M$ being an {\it irrelevant} perturbation at least as far as behavior in the {\it moving phase}.

If this mean-field-like scenario indeed  applies for small $M$ to the finite-range 
model, we expect the following scaling behavior for both $M$ and the proximity to the $M$-dependent critical force, 
\be
f\equiv F-\fd (M).
\ee
small:
\begin{equation}
\bar{v}(M,F)\sim f^{\beta}B(\frac{M}{f^{\phi}})
\label{cross-scaling}
\end{equation}
with the crossover exponent $\phi<0$ indicating the irrelevance of $M$, and $B$ a scaling function. In analogous situations in equilibrium statistical mechanics, if a parameter such as $M$ does not change the nature
of the transition, the effect of $M$ on $F^{\downarrow}_c$ can be taken into
account perturbatively.  Because 
of the singular nature of the critical fixed point that describes the quasistatic depinning --- resulting, in part, from the absence of thermal fluctuations but, more essentially, from the jerky nature of the motion --- the ``analytic parts" might not be smooth functions of $M$, but under the assumption  that they are, we expect that 
\be
F^{\downarrow}_c(M)=F_{c0}-aM-bM^2+ \dots \ .
\ee 
Nevertheless, there can be singular {\it corrections to scaling} determined by the form of $B$ and the crossover exponent $\phi$.

This scenario, in which $M$ is irrelevant when it is small, we call the {\it dissipative scenario}. 
We note, however, that this scenario is compatible with  a change in behavior at a critical value of $M$ as occurs in mean field theory.  This would give rise, for $M$ close to its critical value, to a crossover to some kind of multicritical behavior emerging at larger velocities.

More interesting behavior would occur if the qualitative mean field results on the effects of small $M$  do {\it not} simply carry over to the finite-range
case.  This would be the case if  $M$ is a {\it relevant} perturbation and would correspond to a crossover scaling function, such as $B$ in Eq. (22) with $\phi$ positive. 
A relevant perturbation $M$ would yield a singular correction to the critical force of the form
\be
 [F_{c0}-F_{c}^\downarrow(M)]_{\rm sing} \sim M^{1/\phi},
\ee
which would dominate over the leading (or subdominant) analytic shift if $\phi>1$ (or $\phi>\frac{1}{2}$). Earlier numerical results by Ramanathan and Fisher suggested that this might be the case \cite{ramanathan2}. 

There are two simple scenarios for the velocity versus force if $M$ is relevant. 
One possibility is that the depinning transition is driven {\it discontinuous
immediately} for any non-zero $M$.  The average velocity would then have a discontinuity  of 
\be
\bar v_{min} \sim M^{\frac{\beta}{\phi}}
\ee
We call this the {\it first-order scenario}.

If $M$ is relevant but the transition is still {\it continuous}, one would expect it to be in a  different critical universality class. In this case,  the scaling function  
$B(y\to\infty)\sim y^{\rho}$ 
so that  $\bar{v}(f)$ remains continuous
but with a new exponent 
\be
\beta=\beta_0-\phi\rho
\ee
with $\beta_0$ the quasistatic value.  Asymptotically close to the critical force the new critical behavior would obtain but for $f>>M^{\frac{1}{\phi}}$,
the average velocity curve would crossover to the dissipative behavior.  We refer to this as the {\it new universality class scenario}.  

While more exotic scenarios may be possible, we will limit our consideration
to  the three scenarios  enumerated above.   

Note that we have explicitly {\it not} considered scenarios in which the velocity in the moving phase is hysteretic.  Although we cannot rule this out entirely, the fact that the random environment through which the manifold moves acts, to some extent, like thermal noise, suggests that if there were more than one possible moving phase for 
the same applied force, there would be some stochastic process by which the system could jump from one to the other.  The result of this would be that, as in equilibrium transitions, true ``coexistence" would not be possible over a range of parameters.
This argument does {\it not} apply to coexistence between static and moving phases, as the former are not subject to time dependent ``noise".

\subsection{Numerical results in moving phase}

 Our primary
numerical results in the moving phase were carried out on square two-dimensional samples of linear dimension  $L=128$.  The maximum pinning force $f_p^{max}$, was chosen to be either $0.5$ or $1.0$ so that the typical pinning force is comparable to the change in elastic forces caused one neighbor of a segment jumping.

\begin{figure}[h]
\begin{center}
\epsfxsize=8cm
\epsfysize=8cm
\epsfbox{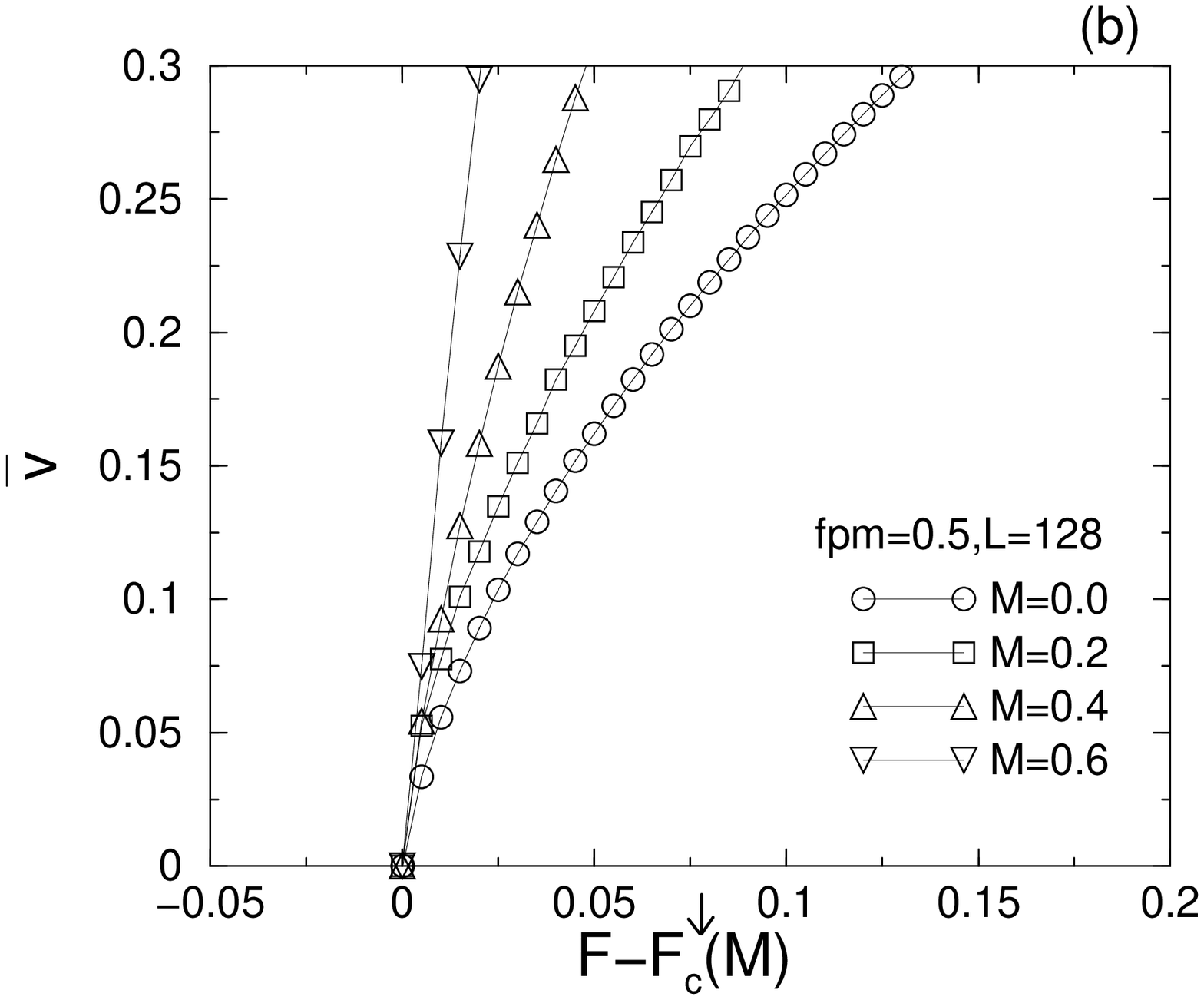}
\caption{\label {vfplot}(a) Steady state velocity, $\bar{v}(F)$, for $L=128$ and 
$\fpm=0.5$.  On the actual plot, as well as on all remaining 
ones, $f_p^{max}$ is denoted by ${\rm fpm}$.  The applied force is
changed by an increment of $0.005$ after the manifold  has been
equilibrated for $12,000$ time steps (see text). Initially,  
$F>F^{\downarrow}_c(M)$ and then it is decreased until $\bar{v}=0$ before 
being increased back up again to its starting value.   The filled arrows 
indicate this cycling of $F$.  The errors bars 
are smaller than the symbols.\label {vfshiftedplot}(b) The same as (a) but with the curves shifted by an approximate 
$F^{\downarrow}_c(M)$.}
\end{center}
\end{figure}

We focus on the results for $M$ small enough that sublattice effects are not too important --- for $f_p^{max}=0.5$, we study $M\leq 0.8$.  
Fig. \ref {vfplot}(a) shows $\bar{v}(F)$ for $0\leq M \leq 0.6$ in
increments of 0.2.  To generate these curves we start, for each $M$ at $F>\fd$, 
with almost flat initial conditions --- segment displacements random in $(0,1)$ ---
 and then decrease the applied
force very slowly until $F^{\downarrow}_c(M)$ is reached.

\vspace{-1.0cm}
\begin{figure}[h]
\begin{center}
\epsfxsize=8cm
\epsfysize=8cm
\epsfbox{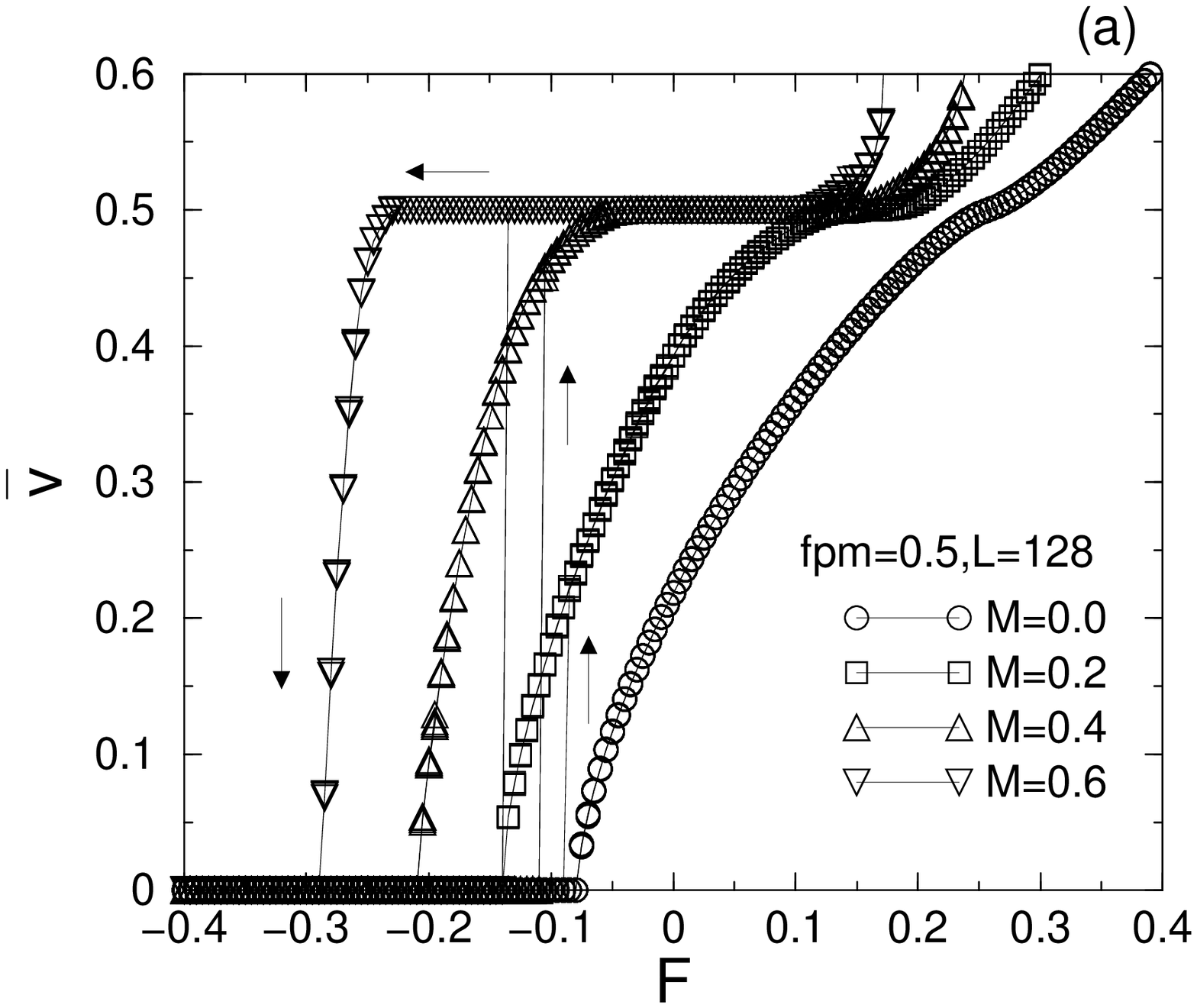}
\end{center}
\end{figure}
\vspace{-2.5cm}

For our finite systems with $L=128$,  $F^{\downarrow}_c(M)$ is defined as being the force below which  the system halts
after $12,000$ time steps, a value chosen so as to be long enough for transients to decay, but not so long that rare configurations of the randomness with anomalously strong pinning forces can dominate. 
  
For $M=0$, there is a finite-size crossover regime in which the system 
may stop due to an anomalously   strongly pinned region 
and the infinite system behavior will  no longer be observed.  This tends to occur when the  
average velocity while it is still moving is of order  $L^{-\frac{\beta}{\nu}}$.  For $L=128$, 
this value is about $0.013$ and with this average velocity,
the manifold will typically travel a distance several times the characteristic displacement $L^{\zeta}$
within $12,000$ time steps.  As we shall see, our estimates are that at least one measure of the characteristic time scale, which grows as $L^z$, is only weakly dependent on $M$. 
Thus, we use
this same criterion for non-zero $M$ while being aware that it may bias our scaling results slightly such that if time scales do change substantially with $M$ we may 
be observing more of either finite-size effects or non-equilibrium effects, the former if the time scale decreases with $M$ and the latter if it increases with $M$.   
For other system sizes,  the equilibration time is decreased correspondingly, roughly with $L^{z_0}$, i.e., according to the dynamic scaling found in the dissipative limit. 

\begin{figure}[h]
\begin{center}
\epsfxsize=8cm
\epsfysize=8cm
\epsfbox{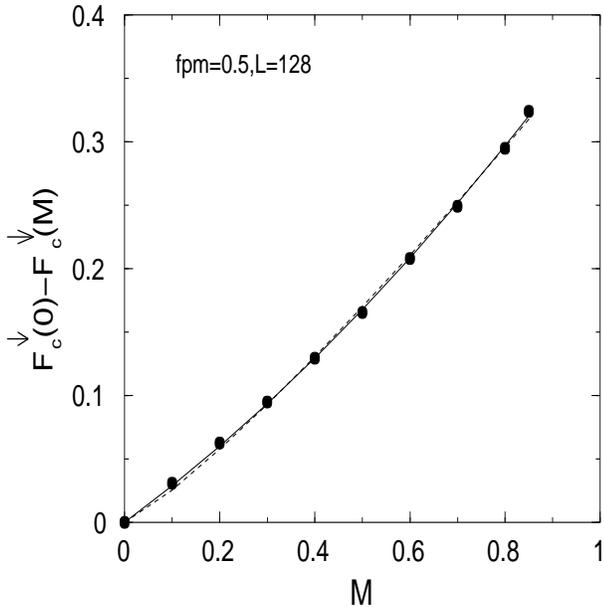}
\caption{\label {fcshiftplot} Critical force, $F^{\downarrow}_c(M=0)-F^{\downarrow}_c(M)$, as a function of $M$ indicated  by filled circles; the solid line is 
the result of a quadratic fit $bM+cM^2$ with  
$b=0.27\pm 0.01$, and $c=0.12 \pm 0.01$ while the dotted line is  the
result of a fit to $aM^{1+D}$ with $a=0.384 \pm 0.003$ and
$D=0.18 \pm 0.02$.  The error bars are smaller than the symbols. }
\end{center}
\end{figure}

 We first study the dependence on $M$ of the critical force, $\fd$, below which the steady state motion ceases, in particular to test whether
$F_c^{\downarrow}(M)$ is a singular or smooth function of $M$.   See Fig. \ref {fcshiftplot}  
The results, along with
a quadratic fit, 
\be 
F_c^{\downarrow}(0)-F_c^{\downarrow}(M)=0.27(\pm 0.01)M+
0.012(\pm 0.01)M^2
\ee
are shown in Fig. \ref {fcshiftplot}.  [Note that if we had included a constant in the fit, 
the constant would have vanished within one standard deviation, as it should,  and
the linear coefficient would have been only slightly modified to $0.26\pm 0.01$.]  A natural expectation --- although overly naive ---  is a linear decrease of $\fd$ by an amount $M/Z$.  For small $M$, this appears to work rather well. The reason for this linear shift and the corrections to it will be discussed later.

The analytic fit should be compared to a fit --- with the same number of parameters --- to 
an 
arbitrary power-law $M$ dependence of the shift in $\fd$ such as would obtain if $M$ were a relevant perturbation. The best fit to the $\fd$ data
 yields an exponent  $\frac{1}{\phi} =1.18\pm0.02$; note, however, that by 
eye the quadratic fit looks slightly better than the power-law fit. 
Although with a weakly relevant $M$ with a crossover exponent less than unity, as the power law fit suggests, one would presumably have a linear analytic term as well and thus the inferred $\phi\approx 0.85$ should not be taken too seriously, In any case, one  must ask whether it is consistent with our other data. It does not appear to be: if we use this value of 
$\phi$ to try and find a  scaling function $B(\frac{M}{f^{\phi}})$ for the velocity data of Fig. \ref {vfplot}(a), 
the curves do not collapse.  This suggests that either $M$ is irrelevant, or that it is sufficiently weakly relevant that the crossover exponent $\phi$ is small enough that it would not dominate the shift in $\fd$.  Other data, as summarized below, suggests that, in fact, $M$ is irrelevant, at least for the steady state moving phase. 
\vspace{-2.0cm}

\begin{figure}[h]
\begin{center}
\epsfxsize=8cm
\epsfysize=8cm
\epsfbox{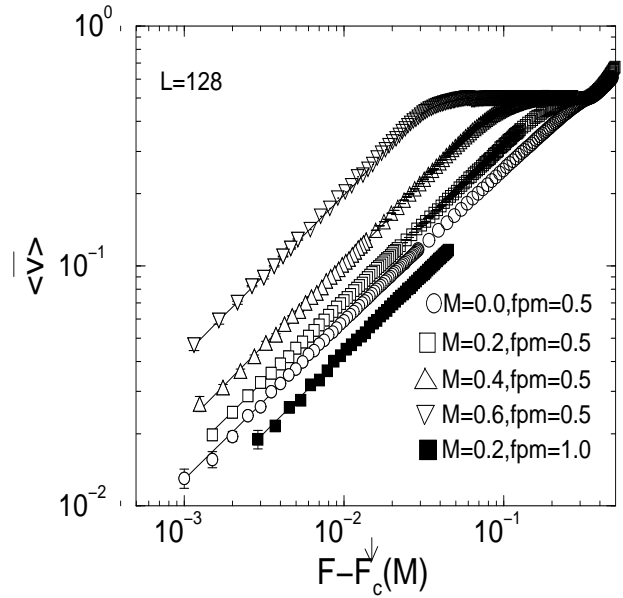}
\caption{ \label {loglogvfplot} Log-log plot of the average steady state $\bar{v}$ as a function of $f=F-\fd(M)$, 
where $F^{\downarrow}_c(M)$ is varied to optimize the linearity of each curve within the 
scaling regime.  The 
symbols used
are the same as in Fig. \ref {vfplot}(a), with the filled 
squares representing $M=0.2$, 
$\fpm=1.0$.  Each curve represents an average over 10 samples 
with the error
bars representing the rms sample-to-sample variations  in the 
velocity.  The applied force increment  is 
$0.0005$ for approximately $\bar{v}<0.1$, while for $\bar{v}>0.1$  the force increment is $0.005$.  
For the weaker randomness, the best fit slope for $M=0.0$ is $0.66 \pm 0.03$; for 
$M=0.2$ it is $0.66 \pm 0.02$; for $M=0.4$ it 
is $0.68 \pm 0.05$; and for $M=0.6$ the slope is 
$0.67 \pm 0.10$.  For the stronger randomness at $M=0.2$,
$\beta=0.66 \pm 0.02$.}
\end{center}
\end{figure}

The mean velocity data can be used to  obtain the critical exponent $\beta$ and see whether it depends on $M$.  Figure \ref {loglogvfplot} shows a  log-log plot of the 
$\bar{v}(F)$ curves with the best $F^{\downarrow}_c(M)$ value, that which  makes  the curve the  most linear, determined by
hand for each $M$.   The error bars indicated are the rms variations in the average 
velocity over $10$ samples.
The values of the applied forces used are separated by an  interval of $5 \cdot 10^{-4}$, one-tenth of those used in  Fig. \ref {vfplot}(a).

The velocity critical exponent $\beta$ inferred from these data is, 
for $M=0$,  
\be
\beta_0=0.66 \pm 0.03 .
\ee
Surprisingly, this value appears to be consistent within one 
standard deviation with the  data   for all the $M$ values shown.  
Even for $M=0.6$, we find 
\be
\beta=0.67 \pm 0.1 \ ,
\ee
although the straight line fit is only over one and a half decades in the reduced force, $f$, substantially less than the three decades of the fit for $M=0$.  From Fig. \ref {vfplot}(a), it is apparent that this reduced range of scaling is primarily due to a larger amplitude for the singular velocity for larger $M$. See Fig. \ref {vfshiftedplot}(b).  

The data suggest that the 
evidence is at least consistent with the dissipative critical behavior obtaining asymptotically for small $M$; i.e., with $M$ being an irrelevant perturbation.

\begin{figure}[h]
\begin{center}
\epsfxsize=8cm
\epsfysize=8cm
\epsfbox{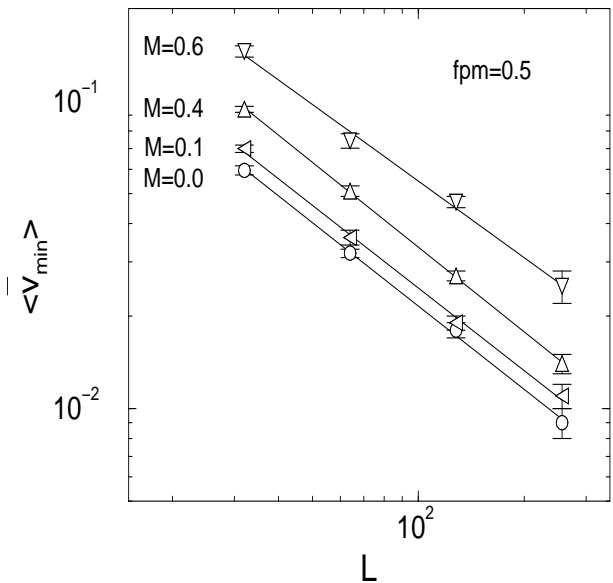}
\caption{\label {vminplot} Plot of the minimum spatially averaged velocity 
$\bar{v}_{\rm min}$ as a function of both
$M$, $L$ and $\fpm$. For the determination of $\bar{v}_{\rm min}$, see the text.  The symbols 
for the various $M$'s are the same as in Fig. \ref {vfplot}(a), except for the left triangles
which represent $M=0.1$.  The best fit slope for the $M=0.0$ curve is $-0.90 \pm 0.02$,
for $M=0.1$ it is $-0.89 \pm 0.03$, for $M=0.4$ the slope is
$-0.91 \pm 0.01$, and for $M=0.6$ it is $-0.83\pm0.04$. Each set of data are 
averaged over 10 samples.}   
\end{center}
\end{figure}

We must be careful, however, especially as Figures \ref {vfplot}(a) and (b), 
and \ref {loglogvfplot}  indicate that the minimum average velocity is
increasing with increasing $M$.  Might this observation suggest that
there is a discontinuity opening up as a power of $M$, perhaps suggesting that the
transition is driven discontinuous immediately?  We can check the data against the {\it finite-size crossover} behavior expected in the   $M=0$ quasistatic limit; this would yield
\be
\bar{v}_{\rm min}   \sim L^{-\frac{\beta}{\nu}} .
\ee
Figure \ref {vminplot} tests for this scaling and finds it to be consistent with the data: even for
$M=0.4$, the slope of the log-log plot is $-0.91 \pm 0.01$, which agrees within one standard 
deviation with the $M=0.0$ slope.  While we do observe an increase in
the minimum average velocity with increasing $M$ at fixed size, it is {\em not} the size-independent  
power-law increase with $M$ that would have been expected if the transition became discontinuous.  Instead, there is an $M$-dependent coefficient

associated with the average velocity, just as occurs for small $M$ in mean field theory.  Similar minimum velocity data was also obtained for stronger randomness, with $\fpm=1$.    
We thus see that the finite-size  data are  consistent with the dissipative scenario, with {\it small stress overshoots being  irrelevant for the velocity versus force curves}.

But it is still possible that a new critical universality class is emerging for small $M$  if
$\rho<<1$, so that  the emergence of the new universality class would be
difficult to detect by simply measuring the velocity 
exponent as this would be little changed from its dissipative-limit value. We therefore look more closely at the finite-size crossover regime to 
investigate whether other aspects of the behavior really look similar to the $M=0$ quasistatic depinning.
    
\begin{figure}[h]
\begin{center}
\epsfxsize=8cm
\epsfysize=8cm
\epsfbox{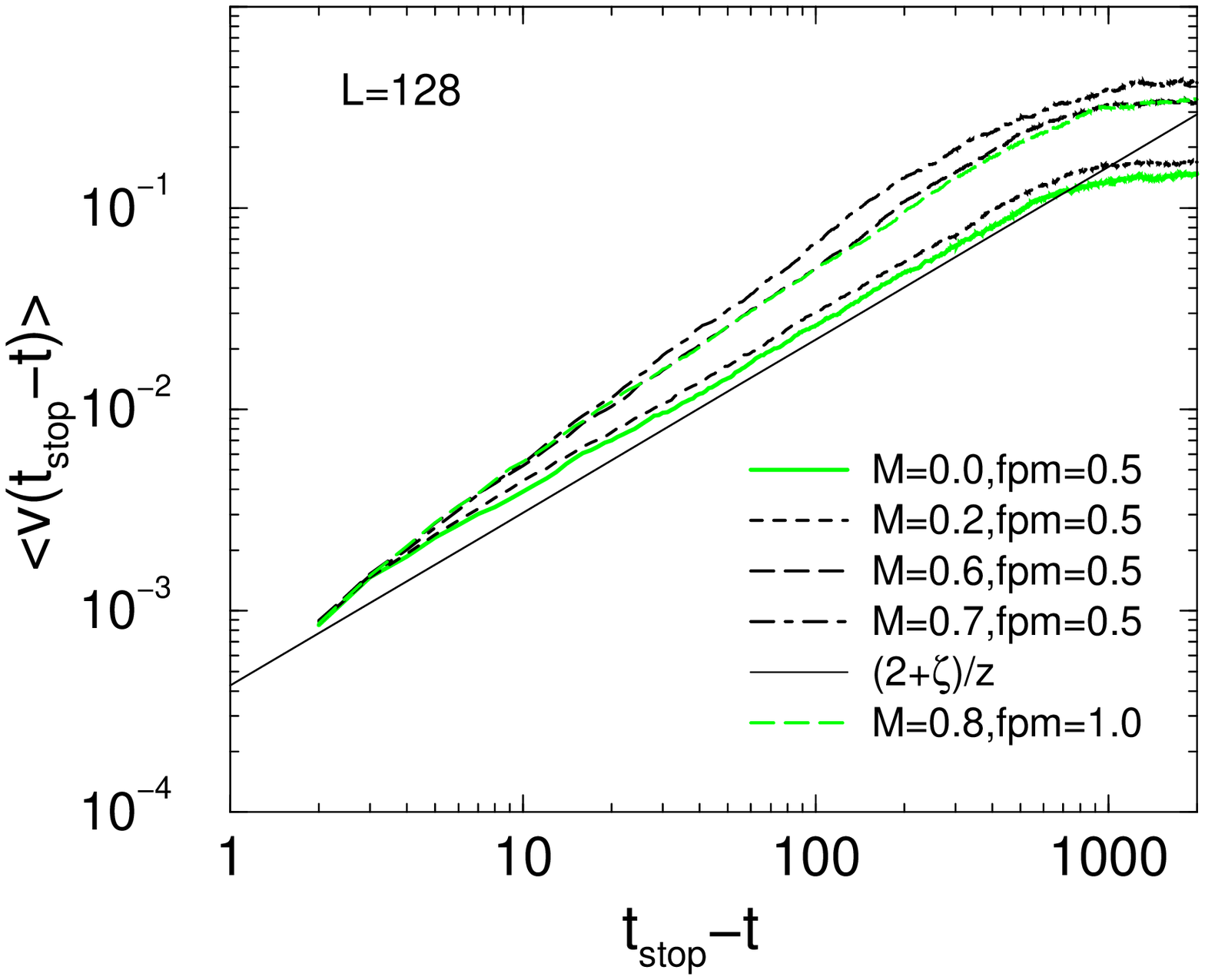}
\caption{\label {vstopplot} Log-log plot 
of $v(t_{\rm stop}-t)$ as a function of the time before the motion ceases at $t_{{\rm stop}}$. The data were obtained 
from a uniformly random initial condition on the interval $[0,1]$ and 
equilibrated at an applied force higher than $F_c^{\downarrow}$ by 
three times the standard deviation of this critical force.  After equilibration, the applied force 
was lowered to the average $F_c^{\downarrow}$, whereupon $v(t)$ is recorded until the manifold 
eventually stops. The solid line is the theoretical expectation within the scaling 
regime. The slopes of the numerical data, in order of increasing $M$, are
$0.831\pm0.001$, $0.846\pm0.002$, $0.981\pm0.003$, $1.085\pm0.004$, and 
$0.975\pm0.004$  
respectively.  Note that for the $M=0.8$ curve, $\fpm=1.0$. Each curve represents an average over those of approximately 
$500$ samples.}
\end{center}
\end{figure}

Anticipating that it might be the stopping behavior that would distinguish between quasistatic and overshoot dynamics,  we have explored some of the 
dynamics of the stopping process.  Given that there is a distribution of $\fd$'s, we 
let the manifold equilibrate at an $F$ three standard deviations above the average of 
$\fd(M)$.  We then lowered $F$ to the average of $\fd(M)$ and waited for the manifold to 
come to a stop.    
Figure \ref {vstopplot} plots the instantaneous spatially averaged velocity  $v(t)$ averaged over many samples as they come to a stop at time 
 $t_{stop}$.  Here we explicitly see that  the various $M$
samples come to a stop in a similar gradual manner; only the amplitudes
vary. The differences between the curves are presumably due to the $M$ dependence of the amplitude of the steady state $\bar v(F)$ curves.    
The final stages of the stopping process can be analyzed --- at least for the dissipative ($M=0$) case --- by a simple scaling argument.  Once only a small fraction of the system is still moving, the average velocity will be inversely proportional to the area $L^d$.  Since the velocity scales as $\xi^{\zeta-z}$ and lengths as $t^\frac{1}{z}$, we expect
\be
v(t)\sim \frac{ (t_{stop}-t)^{(d+\zeta-z)/z}}{L^d} \ ;
\ee 
the data on a log-log plot, Fig. \ref {vstopplot}, are 
reasonably consistent with this for all values of $M$ tested. 

\begin{figure}[h]
\begin{center}
\epsfxsize=8cm
\epsfysize=8cm
\epsfbox{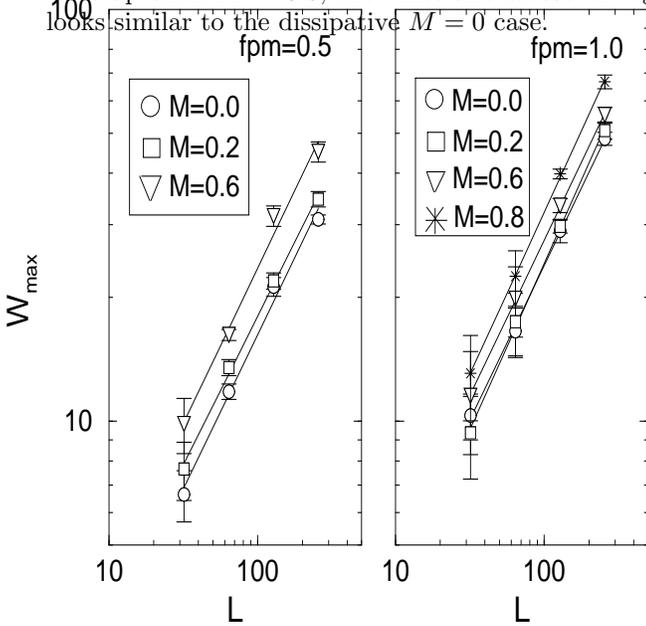}
\vspace{1.0cm}
\caption{\label {maxwidplot} The average maximum width $W_{max}$ of  stopped manifolds as a 
function of $L$ for various $M$'s and $\fpm$'s. The data are taken after the manifold has stopped moving following a gradual decrease of  the driving force. Each $W_{max}$ is an average over 10 
samples.  The slopes can be used to extract the exponent $\zeta$. For the left part of the figure, $\fpm=0.5$; the
$M=0.0$ data are best fit by  $0.75\pm 0.05$; $M=0.2$ yields a
slope of $0.72\pm 0.03$; and for $M=0.6$ the slope is $0.75 \pm 0.05$.  For the 
the right plot, $\fpm=1.0$; for the $M=0.0$ curve, the slope is
$0.75 \pm 0.02$; for $M=0.2$ it is $0.81 \pm 0.02$; the $M=0.6$ slope 
is $0.75 \pm 0.01$; while for $M=0.8$, the slope is $0.79\pm0.01$.  
Note that the overall magnitude of $W_{max}$ is slightly 
larger for the stronger pinning.}
\end{center}
\end{figure}

We can also probe the finite-size crossover regime  in terms of the
 {\it roughness}.  We define the {\it maximum width} of the manifold 
$W_{\rm max}$ as the absolute value of the maximum deviation of the displacement from its spatially averaged value, $\overline{h}$: $W_{\rm max}=<\max|h({\bf x})-
\overline{h}|>$.   This should scale as $L^{\zeta}$ with 
$\zeta\approx 2/3$ for $M=0$.  If $M$ is  irrelevant, then the same
should hold true for all small $M$.  Figure \ref {maxwidplot} demonstrates that the
maximum width does indeed obey this scaling with system size, but with   apparent values 
of $\zeta$ that are somewhat larger than $\frac{2}{3} $ even for $M=0.0$ for which 
$\zeta=0.75 \pm 0.05$ is inferred from the data with $f_p^{max}=0.5$.  
Although the exponent appears to be $M$-independent,  there is  an $M$-dependent coefficient 
with the overall width  increasing --- albeit only slightly --- with increasing $M$.  
For the stronger randomness data, $\fpm=1$, this tendency is less strong  but that
is because the value of $M$ at which the behavior changes character is larger for stronger pinning.
 Overall it appears that, at least up until 
$M=0.6$, the finite-size crossover regime looks similar to  the dissipative $M=0$ case. 
\vspace{-1.0cm}

\begin{figure}[h]
\begin{center}
\epsfxsize=8cm
\epsfysize=8cm
\epsfbox{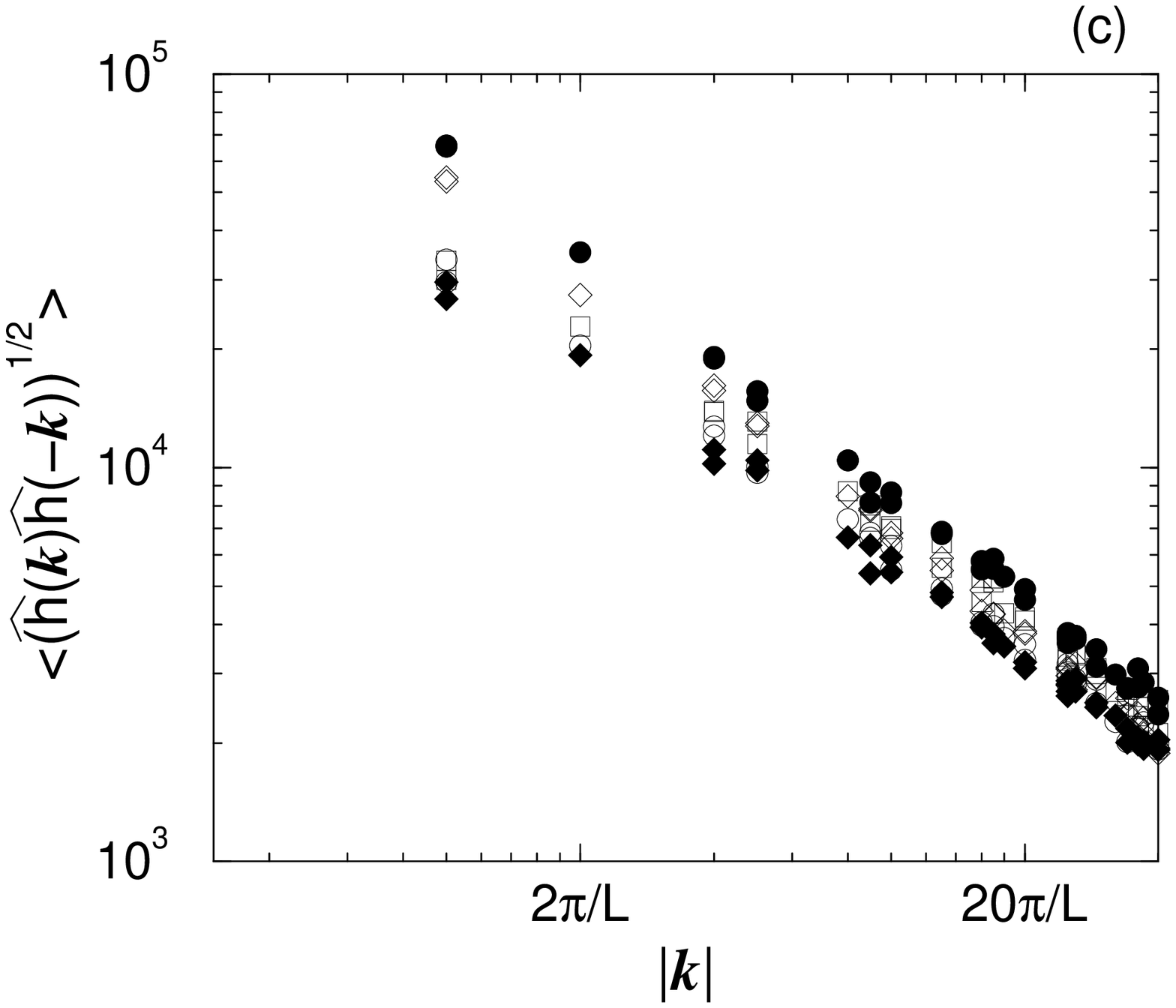}
\caption{\label {spatialpowspecplot} \label {spatialpowspecampplot}(a) Log-log plot of the square root of the power spectrum 
of a manifold in steady state both just above the depinning transition, open symbols, and just after the manifold has stopped after the force is decreased gradually, filled symbols.  The various sets of data  have been shifted along the vertical axis for clarity.  For the moving configurations, $F=-0.0775$ for
$M=0.0$, $F=-0.1389$ for $M=0.2$, and $F=-0.2436$ for $M=0.5$.  For each 
$M$ 
the data are averaged over 100 samples of size of $L=256$.
   The roughness exponent $\zeta$ can 
be extracted from the linear fits over the appropriate wave vector 
interval: the expected slope is $-(1+\zeta)$.  The quoted error bars are a measure of the dependence of the apparent slope on the range over which the data are fit.  The slopes of the moving data
are $-1.65 \pm 0.01$, $-1.65 \pm 0.02$, $-1.69 \pm 0.01$ for $M=0.0$, 
$M=0.2$, and $M=0.5$ respectively.  While the slopes for the stopped data
are $-1.72 \pm 0.02$ 
and $-1.74 \pm 0.01$ for $M=0.0$ and $M=0.5$. Note that at the smallest wavevectors flattening is evident in the moving data but not the stopped data, as expected.  Also note the appearance of a peak at the zone corner in the moving data; this represents the tendency for alternating sublattice motion that occurs, especially for larger $M$. (b) We plot data for $M=0.8,{\rm fpm}=1.0$, and 
$L=256$ for both stopped (light dots)  
and moving (dark dots) configurations  
We extract a $\zeta$ of $0.75\pm0.01$ for both curves.  These 
curves have not been shifted 
with respect to each other.  We note that the data in both (a) and (b) has been thinned out for clarity. (c)
Same plot as (a) without any shifting of the data and over a narrower 
region in ${\bf k}$-space so that the amplitudes can be compared. }
\end{center}
\end{figure}

\begin{figure}[h]
\begin{center}
\epsfxsize=8cm
\epsfysize=8cm
\epsfbox{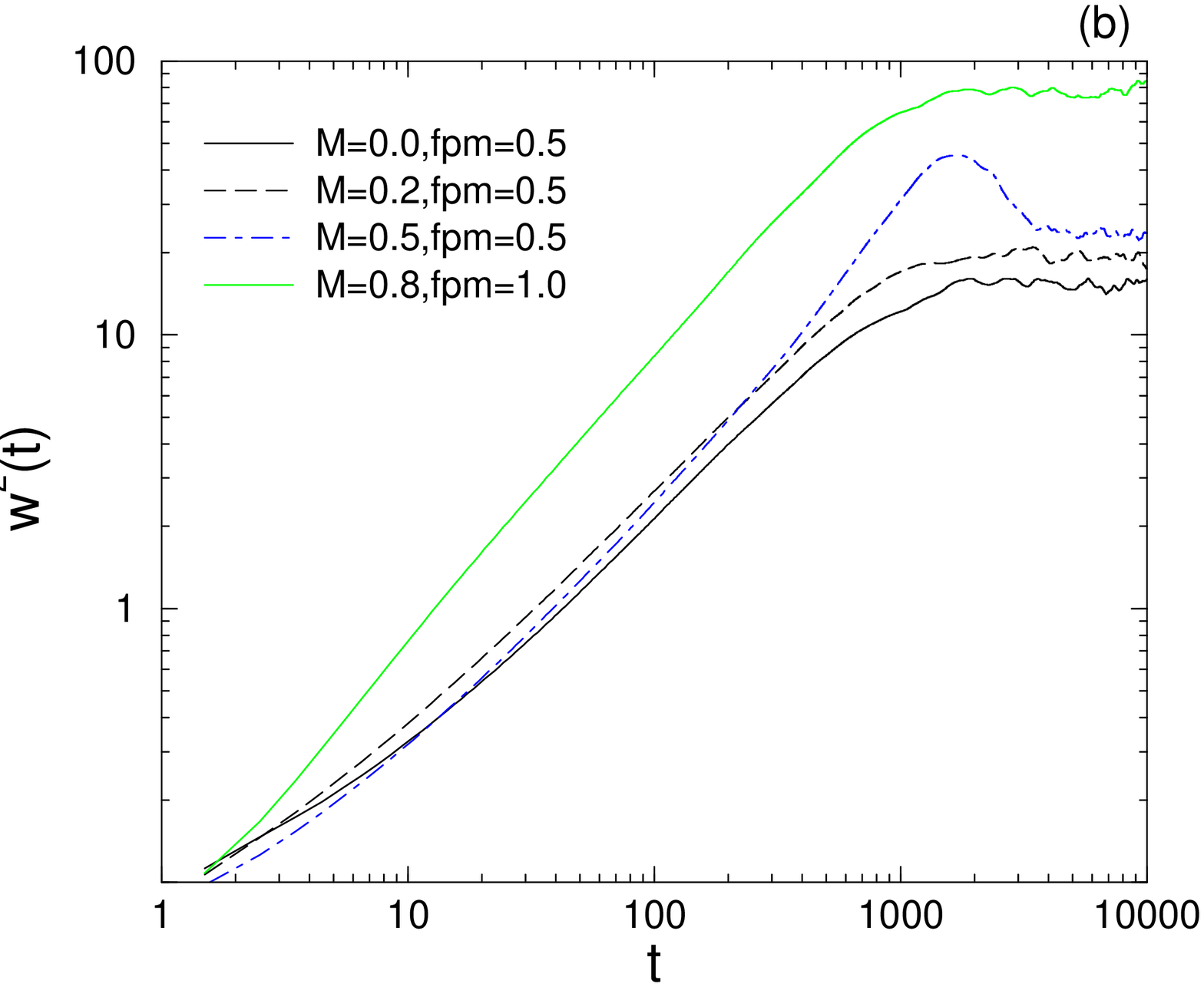}
\caption{\label {wtplot} \label {wtampplot} (a)Dynamical roughening as the manifold starts moving just above $\fd$ from almost flat initial conditions. The mean-square width of the manifold as a function of time is shown on a log-log plot.The curves has been shifted along the vertical 
axis for clarity. 
  The expected slope in  
the scaling regime of these curves is $\frac{2\zeta}{z}$ so that
we can extract the dynamic exponent $z$ after measuring $\zeta$ 
independently (Fig. \ref {spatialpowspecplot}(a)).  The linear fits yields a slope 
of $0.86 \pm 0.01$ for $M=0.0$ and $0.86 \pm 0.01$ for $M=0.2$. The scaling regime should obtain only 
for $t<<\xi^z$; since we are above the transition $\xi<<L$ so that a 
crossover to time-independent roughness should occur when $t\sim \xi^z$.  The crossover scale is in rough agreement with theoretical estimates.  For $M=0.5$ the behavior is rather different that smaller $M$: they exhibit
too much curvature to be readily explained by corrections to scaling and a substantial  overshoot which we believe is due to the merging of many nucleating bubbles as the steady state is 
approached.  The apparent slope in this regime is $1.20 \pm 
0.02$, although this is most likely just an effective exponent.  For the stronger randomness, $\fpm=1.0$, at $M=0.8$ a fit would yield $\frac{2\zeta}
{z}=0.99\pm0.01$.   Despite this larger dynamic roughening exponent, the 
static roughness exponent is also larger and the inferred $z$ is not much changed from the $M=0$ data.    
(b) Same as (a) but without any shifting of the curves. }
\end{center}
\end{figure}

We can also determine $\zeta$ from 
the power spectrum  of the displacement just after the motion has stopped from the
moving phase; we use  the
same equilibration time and applied force increments as in Figs. \ref 
{spatialpowspecplot}(a) and 
\ref {spatialpowspecplot} (b).
In two dimensions, 
\be
<|\hat{h}(k)|^2> =  \int d^2x\  e^{i{\bf k\cdot x}}  <h(x)h(0)>\ \sim k^{-2\zeta-2},
\ee
 where the brackets denote averaging over samples.  From Fig. \ref {wtplot}(a) we 
see that fitting to this form over a range of one and a half decades for samples with $L=256$ yields $\zeta=0.72 \pm 0.02$ and $0.74 \pm 0.01$
for $M=0.0$ and $M=0.5$ respectively, again in mutual agreement and similar to the values from the maximum width discussed above.  But again, apparently slightly larger than the $\zeta=\frac{2}{3}$ value from the first-order epsilon expansion. 
Again, the cumulative  evidence
suggests that at least up until $M>0.7$, 
it is not likely that
a new universality class is emerging in the ``equilibrium" moving or stopped phase as $\bar{v}\rightarrow 0$.

It is useful to compare our values of the exponent $\zeta$
with those previously  obtained. Leschhorn {\it et. al.}
found, in numerical simulations,  $\zeta=0.75\pm 0.02$ for $M=0.0$. \cite{leschhorn}
The most solid theoretical result is that $\frac{\epsilon}{3}$ is a {\it lower bound} for $\zeta$.  This comes from application of finite size scaling to the connections between the variations of the critical force and the correlation length exponent --- see discussion below.  While Narayan and Fisher \cite{narayan2} had 
argued that the value of $\frac{\epsilon}{3}$ is exact to all orders in $\epsilon$,
Chauve {\it et. al.} \cite{chauve} have computed the exponents to second-order in 
$\epsilon$ and found that the roughness exponent is increased to $\zeta=\frac{\epsilon}{3}
(1 +0.14331\epsilon) + {\cal O}(\epsilon^3)$, which naively extrapolated to two dimensions yields $\zeta \approx 0.86$ substantially higher than the value inferred numerically.   Whether or not this 
discrepancy is due to the neglect of terms higher order in $\epsilon$, to some problem with the expansion, or to  corrections to scaling  
remains in doubt.   It is worth noting in this context that for one-dimensional systems with the long range interactions appropriate for crack fronts, Ramanathan and Fisher \cite{ramanathan2} found that to obtain reliable and universal values of $\zeta$ analysis of corrections to scaling were needed.  With these included, they found a value of $\zeta$ very close to that predicted from the first-order $\epsilon$-expansion {\it without} any higher order corrections.

\vspace{4.0cm}
\begin{figure}[h]
\begin{center}
\epsfxsize=8cm
\epsfysize=8cm
\epsfbox{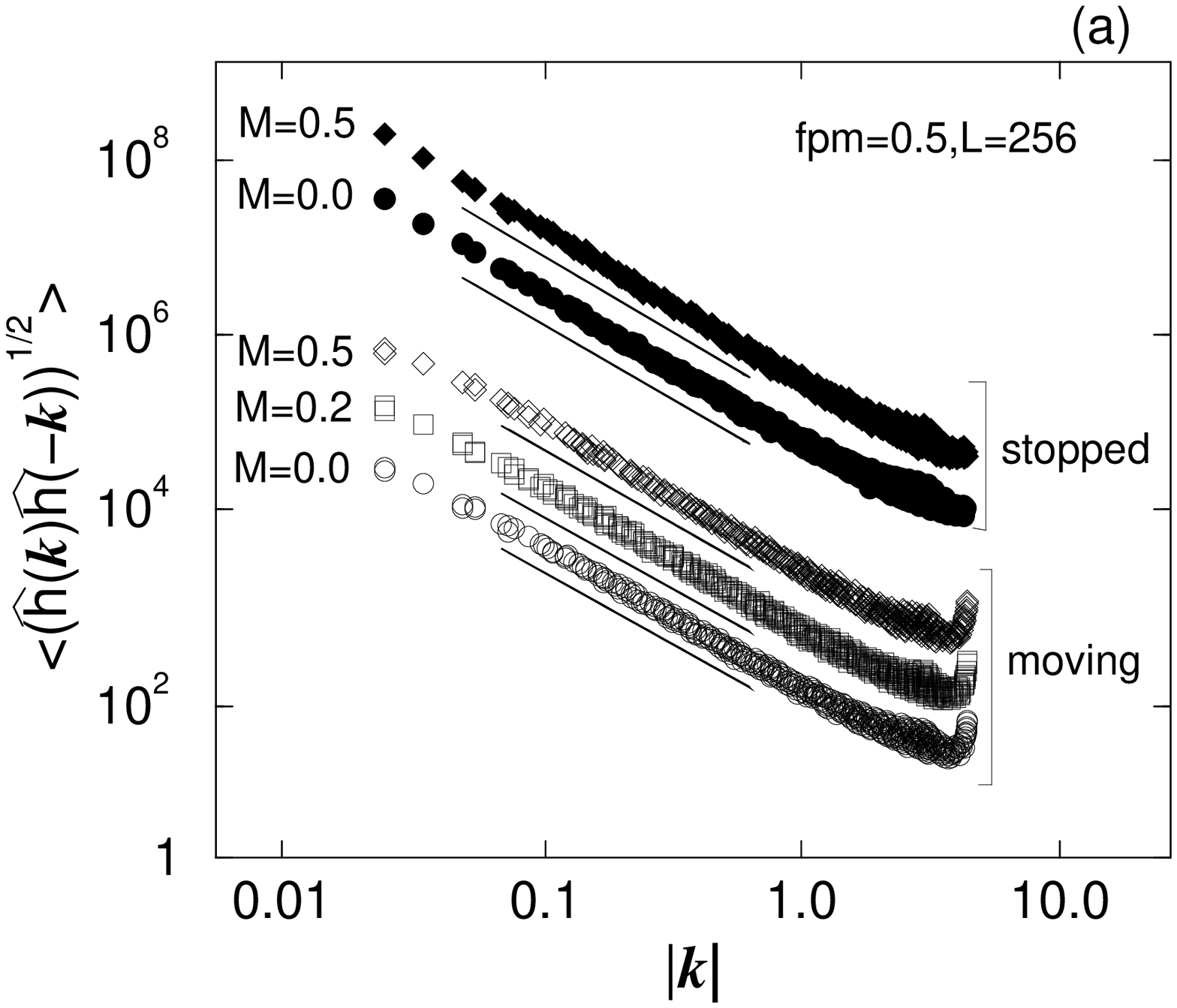}
\end{center}
\end{figure}
\vspace{-2cm}
\begin{figure}[h]
\begin{center}
\epsfxsize=8.0cm
\epsfysize=8.0cm
\epsfbox{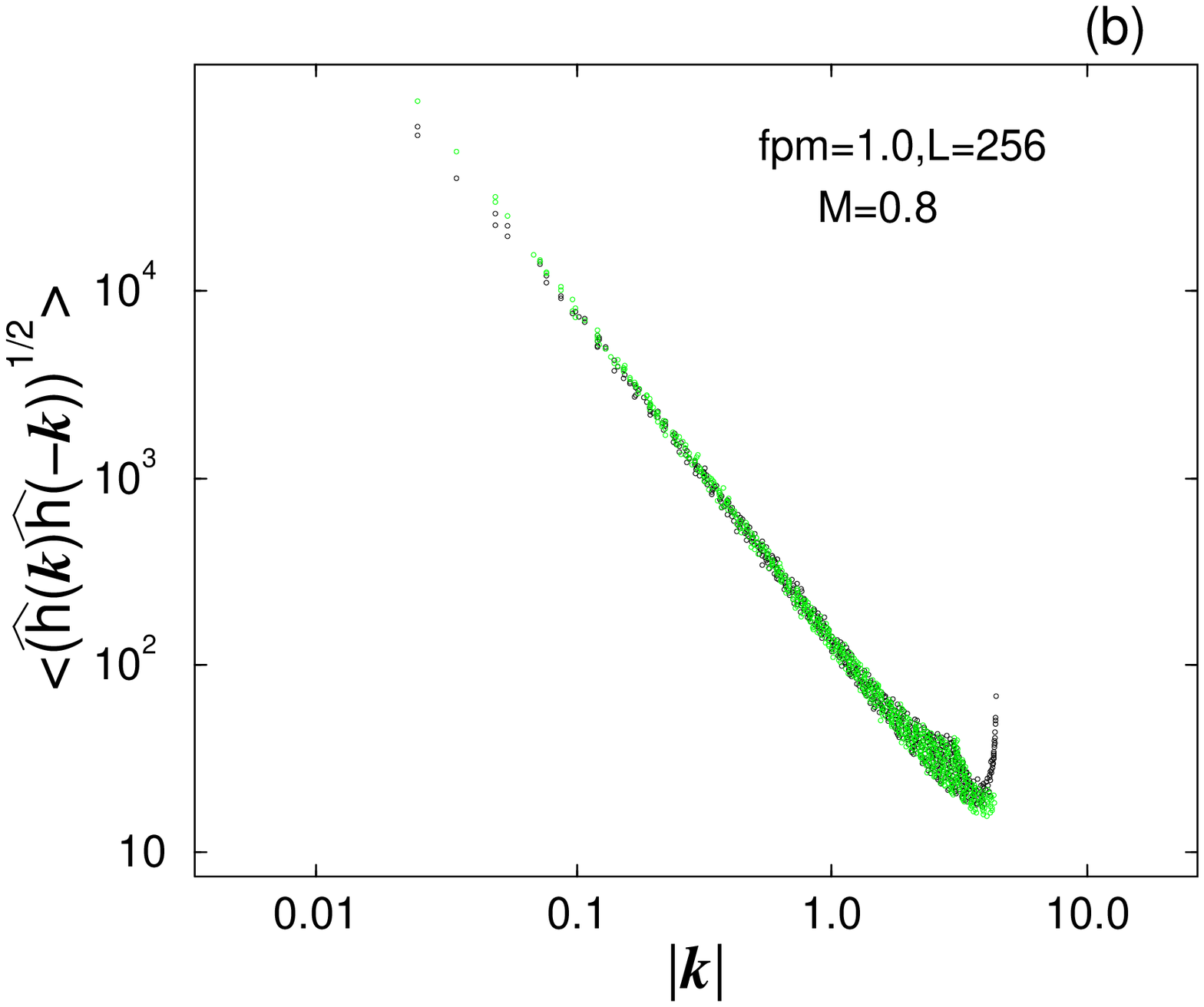}
\end{center}
\end{figure}
\vspace{-2cm}

For completeness, we have also studied the spatial power spectrum of the displacements just above the depinning transition. Figure \ref {spatialpowspecplot}(a) and (b) shows the square root of the averaged power spectrum 
 for $\bar{v}<<1$.   The roughness exponents
obtained for $M=0.0,0.2, 0.5$ are again roughly in agreement with each other ---
$\zeta=0.65\pm 0.01$, $0.65\pm 0.02$, and $0.69\pm 0.01$ respectively --- but slightly smaller than those found for the stopped manifold.  
In the moving phase, however,   there should be a crossover at
long wavelengths, $k\xi_v\sim 1$, to the Edwards-Wilkinson universality class with only logarithmic roughness \cite{edwards}. This arises because on scales larger than $\xi_v$, the motion makes the randomness appear like white noise in both space and time and
the displacement correlation function becomes 
\begin{equation}
<\hat{h}(k,\omega)\hat{h}(-k,-\omega)> \sim \frac{1}{|-i\omega +Dk^2|^2},
\end{equation}
with  $D(F)$ an effective  diffusion constant.  
This crossover is observed for the smallest $k$ in 
Fig. \ref {spatialpowspecplot} (a) where the slope of the 
 power spectrum
decreases in contrast to  the data taken after the motion has stopped.  We note that this 
difference in slope between the moving and stopped spectra is not as 
prominent when the randomness is stronger as shown in Fig. \ref {spatialpowspecplot} (b).

For the moving configurations we  observe a peak in the power spectrum 
at $\bf k=(\pi,\pi)$.  This peak is 
caused by the tendency of one segment's motion to trigger jumps of its neighbors at the {\it next} time step.   The structure and amplitude of the peak looks similar  for all of the  $M$'s shown, although the wavevector dependence indicates that there are somewhat more segments participating in the 
sublattice behavior at $M=0.5$ than at $M=0$; but only about  a third more.

The dynamic exponent in the moving phase,
$z$,  can be determined in various ways from data near to the critical force.  We first study the {\it non-equilibrium  roughening} of the manifold starting with almost flat initial conditions. We define 
\begin{equation}
w^{2}(t)=\langle \overline{(h({\bf x},t)-\overline{h(t)})^2}\rangle 
\sim t^{\frac{2\zeta}{z}},
\end{equation}
with the overbar denoting spatial averaging over the sample. The scaling behavior is  expected for times short compared to the critical correlation time $\tau$ which diverges at $\fd$.
Once the roughness exponent has been calculated independently from the 
same set of configurations in Fig. \ref {spatialpowspecplot}(a),
$z$ can be extracted from the log-log plot of
$w^2(t)$ vs. $t$ as is done in Fig. \ref {wtplot}(a)(b)
For  $M=0.0$ and 
$M=0.2$ we see that the dynamic 
exponents $z$ are very similar: 
$z=1.51 \pm 0.03$ for both.  However, for
$M=0.5$, the data are somewhat further above the transition as they correspond to an  equilibrium average velocity of $\bar{v}=0.08$.  While these data could be fit with the same $z$ over a limited range of times, there is clearly some new physics emerging: an upward curvature on the log-log plot and a substantial --- more than a factor of two --- overshoot in the velocity before it settles down to its steady state value.
 This effect is related to 
the change in the dynamical onset of the motion to which we will turn in Section VII. 

 Aside from the transient effects associated with approach to steady state,  all our  measurements in the moving phase suggest that the critical behavior is  most consistent with the
dissipative universality class obtaining for all sufficiently small  $M$ as the transition is approached from above; this {\it despite}
the $M$-dependent 
shift in $F^{\downarrow}_c$ and the concomitant hysteresis that is possible because of the existence of linearly stable static configurations up to $F_{c0}$ which is larger than $\fd$. We will analyze this paradox later.  For now, 
it appears that the mean-field-like scenario has won out over the 
first-order scenario and the new universality class scenario.  

\vspace{-2.5cm}
\begin{figure}[h]
\begin{center}
\epsfxsize=8cm
\epsfysize=8cm
\epsfbox{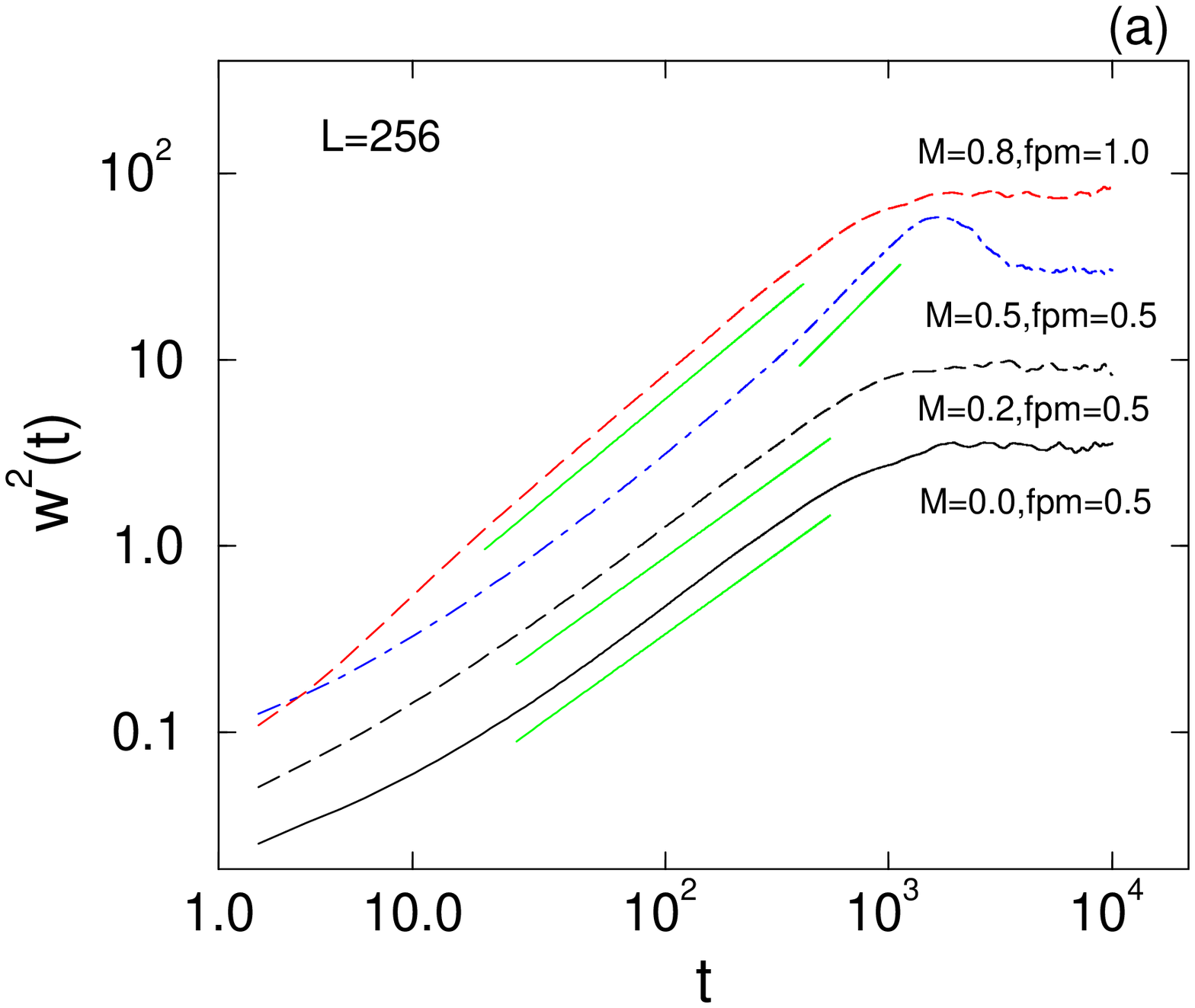}
\end{center}
\end{figure}
\vspace{-2.5cm}

\subsection{Amplitude ratios}

If we accept that $M$ is irrelevant for the steady state moving critical behavior,  the observed changes as $M$ is increased in  the moving phase for $F$ just above $\fd$ appear to be primarily attributable to   the increased amplitude $A_v(M)$ of the velocity near $\fd$, $\bar{v}\sim A_v(M) f^{\beta}$.  As seen in Figs. \ref {vfshiftedplot}(b) and \ref {loglogvfplot}: For larger $M$, the velocity rises more rapidly with increasing $F$ until it becomes close to $\frac{1}{2}$ at which point sublattice effects set in.   A similar increase in the amplitude of the velocity  is observed in mean field theory, with the exponent $\beta$ unchanged for small $M$ but the amplitude of the velocity  growing with $M$. 
 In both cases, this implies that  the width in $F$ of the depinning transition
narrows with increasing $M$ --- naively just by the narrowing of the range of $F-\fd$ over which $\bar v$ is small.

If the universality class of the critical behavior in the moving phase is independent of $M$ over a range of $M$, then there should be various universal relations between non-universal coefficients: {\it universal amplitude ratios}. A priori, we would expect three non-universal scale factors associated with the scaling relationships between length and, respectively, deviation from criticality, $F-\fd$; displacement, and time.  We can define these, for example, by the scaling of the correlation length,
\be
\xi_v \approx \bigg(\frac{A_F}{F-\fd}\bigg)^\nu ;
\ee
mean square displacements at separations smaller than $\xi$, 
\be
\la[h(\bx,t)-h(0,t)]^2\ra \approx A_h^2 |\bx|^{2\zeta} ;
\ee
and velocity
\ba
\bar{v}& \approx& \frac{A_h}{A_t} \xi_v^{-\beta/\nu} \nonumber \\
&\approx&\frac{A_h}{A_t}\bigg(\frac{F-\fd}{A_F}\bigg)^\beta \nonumber \\
&=& A_v (F-\fd)^\beta
\ea
where we measure all lengths in units of the lattice constant.  
But in the absence of stress overshoots the statistical ``tilt" symmetry of the system that relates the exponents $\zeta$ and $\nu$ via the triviality of the averaged response to a static spatially varying additional applied force, also relates the associated coefficients:
\be
\frac{A_F}{A_h} = C_K K
\label{AFH-scaling}
\ee
where $C_K$ is a universal dimensionless coefficient and $K$ is the long-wavelength elastic constant, which, in our model, is simply the inverse of the coordination number $Z$. 

 For $M=0$, we thus expect that amplitudes of scaling laws that only involve the three exponents $z,\nu$ and $\zeta$ should  be expressible in terms of the two amplitudes $A_h$ and $A_t$ only.  For example, the rms variations in the critical force $\fd$, in finite size systems should be expressible in terms of $A_F$ or, via 
Eq. (36), $A_h$:
\be
\sqrt{{\rm var}(\fd(L))} \approx C_\Delta C_F L^{-\frac{1}{\nu}}
\ee
with $C_\Delta$ universal. 

It is not clear, {\it a priori}, whether the statistical tilt symmetry argument can be applied in the presence of stress overshoots and the concomitant local hysteresis. This is because, in essence, it relies on the history independence of linear response of at least some quantities. In the moving phase, on which we are currently focusing, it seems reasonable that the argument should apply and the amplitude ratios hence be related as in the dissipative case.  But we should remain alert to the possibility that apparent failure of expected scaling laws may be due to this assumption.

\begin{figure}[h]
\begin{center}
\epsfxsize=8cm
\epsfysize=8cm
\epsfbox{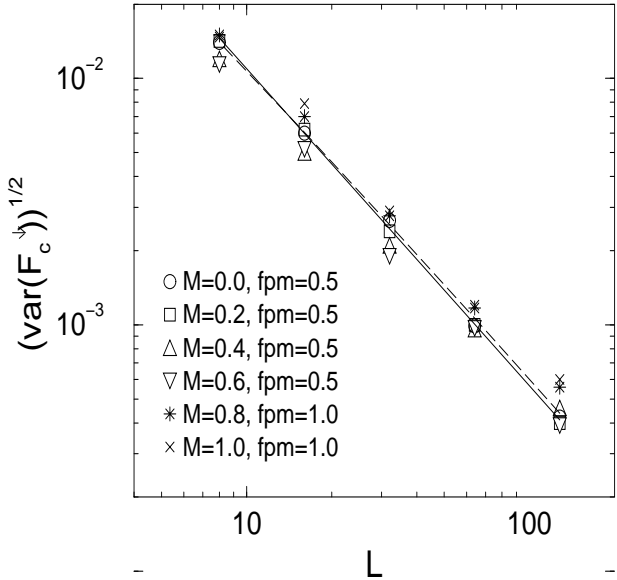}
\caption{\label {varfcdownplot}Log-log plot of the rms variations of $F_c^
{\downarrow}(M)$
as a function of the system length $L$. The inverse correlation exponents $1/\nu$ can be inferred from the slopes. For the $M=0.0$ curve the 
slope is $-1.29\pm 0.03$; for $M=0.2$, it is $-1.29 \pm 0.02$; 
for $M=0.4$, the slope is $-1.18\pm 0.06$; and for $M=0.6$, it is 
$-1.22\pm 0.03$.  For the stronger randomness data, when $M=0.8$,  
 the slope is $1.21\pm 0.03$, while for $M=1.0$, it is $1.19\pm 0.05$.    
The dotted line along the $M=0.0$ curve is the result of a fit that includes a correction 
to scaling with a correction exponent $\frac{2}{3}$ and correction amplitude  $B_{\nu}=0.78\pm 0.01$; the inferred modified 
correlation length exponent
is $-1.30\pm 0.03$.}
\end{center}
\end{figure}

Qualitative examination of the numerical data for the roughness (e.g. 
Fig. \ref {spatialpowspecplot}(a) and the variations in the critical force $\fd$, (see Fig. \ref {varfcdownplot})) 
suggest that neither of the amplitudes $C_h$ and $C_F$ are strongly dependent on $M$ in the range studied.  One would then guess that the 
dependence of the amplitude of the velocity, $A_v(M)$, on $M$ is primarily caused by a {\it decrease} in the characteristic {\it time scale} as $M$ increases.  This will have consequences for the behavior of other quantities as will now be discussed.  
\vspace{1.0cm}

A useful quantity to study is the temporal fluctuations in the instantaneous spatially averaged velocity $v(t)$ in finite-size samples: this has information about both length and time scales.  For a fixed average 
velocity $\bar{v}$, the magnitude of the fluctuations should increase with 
increasing $M$ because the system is effectively   closer to the transition.
Since regions of size of order the velocity correlation length will fluctuate roughly independently, the  variance of the instantaneous velocity should, in a system much larger than $\xi$, be 
\be
{\rm var}(v) \equiv \la (v(t)-\bar{v})^2)\ra \approx \bar{v}^2\frac{\xi_v^d}{L^d} .\label{xiv-def}
\ee
Since we have not yet defined the correlation length precisely, this could well serve as its definition, thereby fixing the definition of the amplitude $A_F$.
The universality of the amplitude ratios can then be checked by comparing the  $M$ dependence of the amplitude of ${\rm var}[v(t)]$ with those of the variations in $\fd$ and the roughness. These results are presented in Table I.

\clearpage
\begin{table}
\caption{Universal amplitude ratio data.   The parentheses  
denote the uncertainty in the last digit of the quoted amplitudes. 
The first row is $M$; all the data shown have $\fpm=0.5$.  The second 
row is obtained from the amplitude of the Fourier transform of the 
$h-h$ correlation function for $L=128$ slightly above 
$F_c^{\downarrow}$ like the data shown in Fig. \ref 
{spatialpowspecplot}(c) for $L=256$.   The 
third row is the combination of amplitudes $A_h/A_t$ determined from the combination $\bar{v}(\frac{L^{d/2}(\Gamma(0))^{1/2}}{\bar{v}})^{\frac{2\beta}{d\nu}}$, where $\Gamma(0)$, $\bar{v}$, $\beta$ and $\nu$ are all measured.  The fourth 
row is obtained from the ratio of the second to the third rows  
The fifth row is obtained from the combination  $\frac{\bar{v}}{A_h}\frac{K^{1/4}}
{(F-F_c^{\downarrow}(M))^\beta}$.  The sixth row is the expected universal ratio $C_K$ determined from the
second, the fourth, and the fifth rows: note the agreement within error bars of all the columns.  The seventh row is determined 
from $A_F=C_K K A_h$, with the $M=0.0$ value for $C_K$ used.  The 
eighth row, the velocity amplitude $A_v$, is obtained from the relation 
$\frac{A_h}{A_t}\frac{1}{A_F^{\beta}}$.  The ninth row shows the 
$A_v$ obtained from the best power law fits of the velocity data of Fig. \ref {loglogvfplot}.   Note the agreement within the errors with the eighth row, with the exception of the largest $M$ for which there is a small apparent inconsistency.
The tenth row is obtained, via Eq. (\ref{gamma-int}) 
from the integral of the  ratio of the time-dependent velocity-velocity correlations $\int \Gamma(t) dt$ to $\Gamma(0)$ for particular choices of the forces close to the critical force.  The eleventh row, the velocity correlation time, $\tau_v$, is 
obtained from the decay constant of 
exponential fits to the same $\Gamma(t)$ as the tenth row.  The twelfth row 
is the expected universal ratio $C_{vv}$ obtained from the previous two rows; note that is is consistent with being constant except for the largest $M$ data. The thirteenth row is the correlation length $\xi'_v$  obtained from the velocity autocorrelations $\xi'_v = 
(\frac{\tau_v}{A_t})^{1/z}$, with $\tau_v$ taken from the eleventh row. 
$\xi_v$ can also be obtained from $\Gamma(0)$ and $\overline{v}$ 
via Eq (\ref{xiv-def}).  This data is presented in row 14.  The fifteenth row is the force amplitude obtained from $A_F\approx 
(F-F_c^{\downarrow}(M))(\xi')_v^{1/\nu}$. The sixteenth row is obtained from the finite-size variations in the critical force shown in Fig. \ref {varfcdownplot}, 
and the seventeenth row is the putative universal ratio $C_\Delta$ inferred from the previous two rows; note it is consistent with being constant within the (admittedly large error bars). In the eighteenth row, the velocity amplitude $A_v$ is obtained from the fifteenth row and from the third row.   }
\vspace{8cm}
\begin{tabular}{|c||c||c|c|c|c|}
$1$ & $M$ & $0.0$ & $0.2$ & $0.4$ & $0.6$ \\ \hline \hline 
$2$ & $A_h$ & $0.98(1)$ & $1.03(1)$ & $1.13(1)$ & $1.23(1)$ \\ \hline 
$3$ & $A_h/A_t$ & $0.39(2)$ & $0.47(4)$ & $0.60(7)$ & $1.0(1)$ \\ \hline
$4$ & $A_t$ & $2.5(3)$ & $2.2(4)$ & $1.9(6)$ & $1.2(6)$ \\ \hline 
$5$ & $\frac{1}{C_K^\beta A_t A_h^\beta}$ & $0.51(6)$ & $0.6(1)$ & $0.7(1)$ & 
$1.0(2)$ \\ \hline
$6$ & $C_K$ & $0.70(8)$ & $0.69(8)$ & $0.7(1)$ & $0.6(2)$ \\ \hline 
$7$ & $A_F$ & $0.16(2)$ & $0.17(2)$ & $0.19(4)$ & $0.21(5)$ \\ \hline
$8$ & $A_v$ & $1.3(2)$ & $1.53(3)$ & $1.8(6)$ & $3.0(9)$ \\ \hline
$9$ & $A_v$ (sim.) & $1.20(1)$ & $1.49(2)$ & $2.29(2)$ & $4.26(3)$ \\ \hline 
$10$ & $C_{vv}\tau_v$ & $158(5)$ & $146(10)$ & $145(16)$ & $210(20)$ \\ \hline
$11$ & $\tau_v$ & $83(1)$ & $88(2)$ & $94(4)$ & $222(6)$ \\ \hline
$12$ & $C_{vv}$ & $1.90(6)$ & $1.7(1)$ & $1.5(2)$ & $1.1(1)$ \\ \hline
$13$ & $\xi'_v$ & $10(1)$ & $11(2)$ & $13(3)$ & $31(10)$ \\ \hline
$14$ & $\xi_v$ & $14(2)$ & $21(3)$ & $23(4)$ & $42(11)$ \\ \hline
$15$ & $A_F$ & $0.11(2)$ & $0.10(2)$ & $0.08(3)$ & $0.08(5)$ \\ \hline
$16$ & $C_\Delta A_F$ & $0.21(2)$ & $0.22(1)$ & $0.13(1)$ & $0.14(2)$ \\ \hline
$17$ & $C_\Delta$ & $1.9(1)$ & $2.20(9)$ & $1.6(2)$ & $1.8(4)$ \\ \hline
$18$ & $A_v$ & $1.7(2)$ & $2.3(4)$ & $3(1)$ & $5(2)$    \\ \hline 
\end{tabular}
\end{table}
The correlation time
$\tau_v$ can also be obtained from the {\it truncated} velocity fluctuations,
\be
\Gamma(t) \equiv \la v(t)v(0)\ra -\bar{v}^2 .
\ee
 Integrating over time yields, 
\be 
\int  \Gamma(t) dt \approx C_{vv}\tau_v \bar{v}^2 \frac{\xi^d}{L^d} \label{gamma-int}
\ee
with $C_{vv}$  a universal coefficient.  The resulting $\tau_v$ can, together with $\xi_v$ be used to check other scaling relations. See Table I. 
Note however that there are difficulties associated with the subtraction needed to obtain the truncated correlations in the most interesting regime in which the fluctuations are large and an accurate extraction of $\bar{v}$ problematic. 
 
The velocity-velocity correlations can also be used to probe the nature of the dynamics; in particular by studying the power spectrum which is the  Fourier transform, $\hat{\Gamma}(\omega)$ of $\Gamma(t)$. To understand how this is expected to behave in the dissipative limit, it is useful to consider the {\it local} velocity-velocity correlation function, $\la v(\bx,t) v(0,0)\ra$. At long distances, $|\bx|\gg \xi_v$ or time separations, $t\gg\tau_v$, this will approach $\bar{v}^2$.  But within a correlation space-time volume,  the local velocities will be characteristic of avalanche events and hence be fractal.  The correlations will be proportional to $\bar{v}$ times a conditional expectation of $v(\bx,t)$ given that there is motion --- i.e. a jump --- at $(0,0)$.   These conditional correlations within the  space-time correlation  volume will reflect the fractal structure, being of order $1/|\bx|^{z-\zeta}$ or $1/t^{1-\frac{\zeta}{z}}$ whichever is smaller.  Integrating the associated scaling forms over $\bx$ and Fourier transforming in time, one finds that for $\omega \gg\frac{1}{\tau}$, 
\be
\hat{\Gamma}(\omega) \approx C_{PS} \frac{\bar{v}^2}{\omega^{\frac{d+\zeta}{z}}} \frac{\xi^d}{L^d}\frac{1}{\xi^{d-z+\zeta} A_t^{\frac{d-z+\zeta}{z}}} ,
\ee 
with $C_{PS}$ a universal coefficient. 

\begin{figure}[h]
\begin{center}
\epsfxsize=8cm
\epsfysize=8cm
\epsfbox{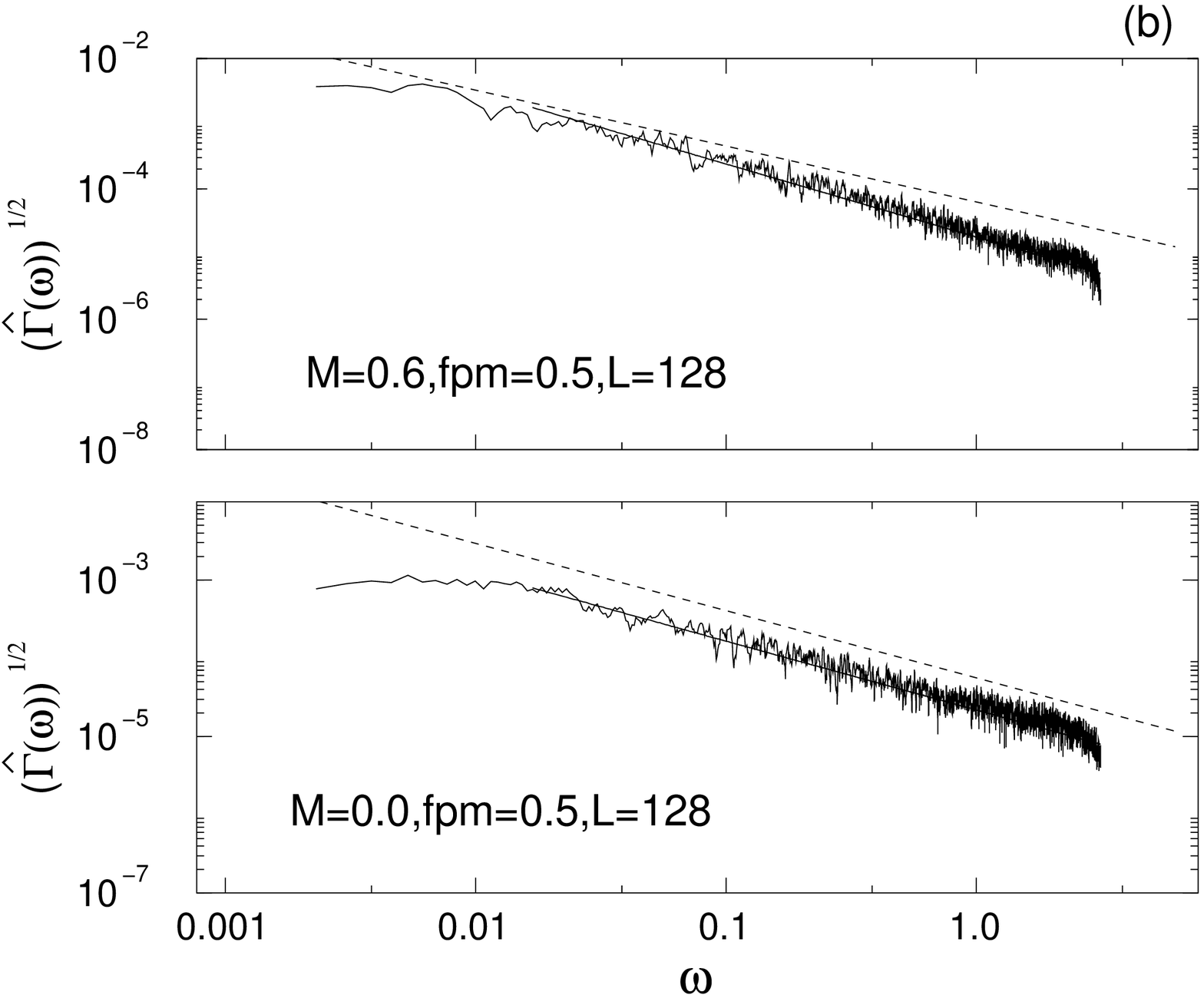}
\caption{\label {vtplot} \label {vtpowspecplot}(a)Time dependent fluctuations of the instantaneous spatially averaged velocity $v(t)$ for $M=0.0,
0.2, 0.4,$ and $0.6$.  For all data, $L=128$ and $\fpm=0.5$. 
Each sample is first equilibrated at some applied force which 
is then subsequently lowered at a rate of $4.2\,\, \times \,\, 10^{-8}$ 
(see text),
until the applied force is such that the steady state 
average velocity
is approximately $0.04$. (b) Log-log plot of the square-root of the 
velocity power spectrum for $M=0.0$ (lower plot) and $M=0.6$ (upper plot).   The data have been smoothed by averaging over 
groups of five frequencies.The dashed 
line in both plots is the theoretical expectation in the scaling regime: a slope of $-\frac{d+\zeta}{2z}\approx -0.86$ for $M=0$.  
A fit yields a slope of $-0.88 \pm 0.01$ for $M=0$.  But for $M=0.6$, the 
slope is rather larger than expected: $-1.12 \pm 0.01$. }
\end{center}
\end{figure}

The log-log plot of the square root of the velocity power spectrum (Fig. \ref {vtpowspecplot}(b)) 
appears to exhibit power-law behavior  for large $\omega$ for both the values of $M$ shown.  For $M=0$, the observed exponent is close to the expected value of  $\frac{d+\zeta}{2z}\approx 0.86$ over two and a half decades in frequency.  For $M=0.6$, the best fit slope is somewhat larger, but some curvature is evident and consistency with the dissipative result is not ruled out. 
In both of these sets of data, there is a crossover at low frequencies to a flat spectrum.  This is more pronounced in the $M=0$ data which are effectively further from the critical regime --- $\xi/L$ smaller --- than the $M=0.6$ data because the two sets of  data were taken at the approximately the same $\bar{v}$ which is closer to the corresponding $\bar{v}_{min}(L,M)$ for $M=0.6$.

Not only does the value of $F^{\downarrow}_c(M)$ provide us useful information, but
its variations do as well. 
In the finite-size limited scaling regime in which $L\ll \xi$, $\xi$ is replaced with $L$ in scaling laws and we expect
\be
[{\rm var}(F^{\downarrow}_c(M))]^{1/2}\sim L^{-1/\nu}\ .
\ee
 Figure \ref {varfcdownplot} demonstrates 
this scaling with the system length,  the obtained correlation 
length exponents being $0.78 \pm 0.02$, for $M=0$ and for $M=0.6$, 
$\nu=0.81\pm 0.02$.  Calculations of Narayan \cite{narayanunpub} yield the leading 
irrelevant eigenvalue at the quasistatic fixed point as approximately 
$-\frac{\epsilon}{3}$.  This suggests a fit of the data of  Fig. \ref {varfcdownplot} with the 
form ${\rm var}(F_c^{\downarrow}(M))=CL^{-\frac{1}{\nu}}/(1+B_{\nu}L^{-\frac{2}
{3}})$ yielding $\nu=0.77\pm 0.03$ with $B_{\nu}=0.78\pm 0.01.$
   
We have found that a variety of properties of the steady state moving phase as well as the variations of the critical force at which the system stops on decreasing the drive are all consistent with critical behavior that is independent of the magnitude of the stress overshoots.  Given the local hysteresis that is intrinsic  with stress overshoots, this universality is more than a little surprising.  In the next section, we consider to what extent this applies more generally, in particular as far as   macroscopic hysteresis.
\vspace{-1cm}
\begin{figure}[h]
\begin{center}
\epsfxsize=8cm
\epsfysize=8cm
\epsfbox{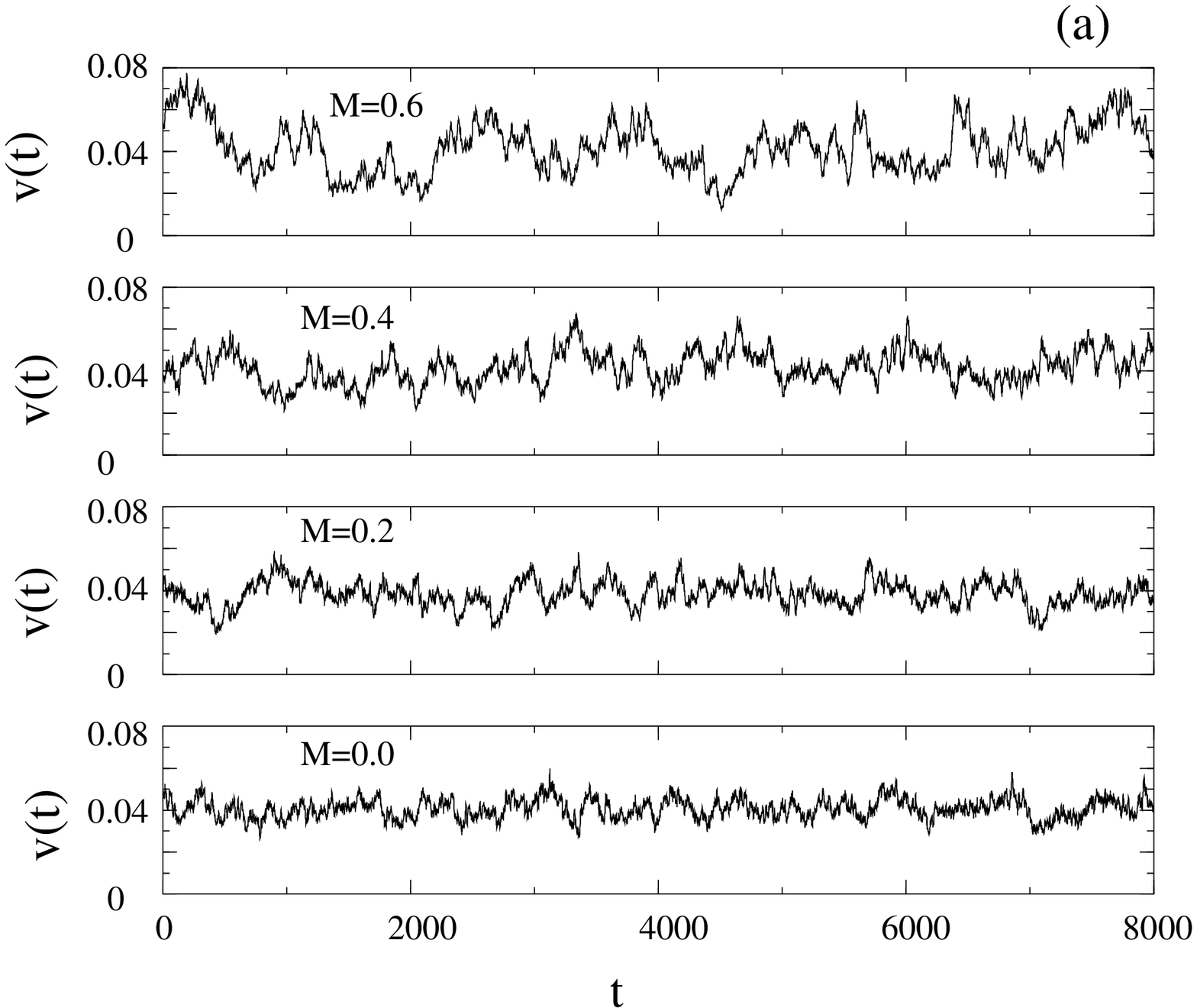}
\end{center}
\end{figure}
\vspace{-2cm}


\section{Hysteresis}

We now turn to an analysis of the hysteretic phenomena that are  implied by the coexistence of moving and stationary solutions at the same force in the presence of stress overshoots.  A crucial question which we must address is whether hysteresis persists in {\it macroscopic} systems that are not prepared in special ways.  In particular, are there hysteresis loops with a width that is non-zero in the limit of large systems?  If not, as we shall see is the case, how does the hysteresis depend on system size? Can one understand this in terms of the purely dissipative dynamics that appear to control the properties of the steady state moving phase?  Or is new physics needed?

In Fig. \ref {vfplot}(a), hysteresis loops are shown for typical samples of size $128^2$ with $\fpm=0.5$ and $M$ from $0$ to $0.6$.  An upwards arrow indicates the force $\fh$, at which the system starts moving again {\it after} it has been stopped at $\fd$ by a gradual decrease in the force. For $M=0$ no hysteresis is apparent while for positive $M$, the difference $\fh-\fd$ appears to be close to $M/4=M/Z$, the magnitude of the stress overshoot.  On the basis of these data, it would appear that the situation is rather simple:  on decreasing the force 
the steady state moving phase and the stopping process are not qualitatively dependent on $M$, but once the system has stopped, an increase of the force by $M/Z$ is required to start it up again.  Once restarted, the velocity rapidly increases to that  of the apparently unique moving ``state".  The reason for this macroscopic hysteresis would appear to be simple: If the force on each of the segments caused by the last motion of its neighbors before the system stopped was not enough to cause it to move, then at later times the force will be less than that needed to make a segment move  by at least $M/Z$ as the stress overshoots from its neighbors jumping will no longer be in effect.  If this applies to all of the segments, it should  be necessary to increase the force back up again by at least $M/Z$ before anything can start to move.   This would imply truly macroscopic hysteresis that is independent of size for large systems. 

A more careful examination of both the data and the argument above shows that it is fallacious: even with $\fpm=0.5$, a small fraction of samples have substantially narrower hysteresis loops. There must thus be some segments that can be restarted by an increase in the force from the stopped state by less than $M/Z$.  In the next subsection, we discuss the origin of this effect,  but first, we present and analyze the numerical data. 

\subsection{Distributions of $\fh$}

For reasons that will become clear later, we can obtain more useful data on the hysteresis by increasing the strength of the randomness.  Most of our detailed hysteresis data is for $\fpm=1.0$ and $M=0.8$, the latter being sufficiently large that the effects of overshoots are strong, but not so large (for this larger value of $\fpm$) that sublattice effects start to play a role.  For these parameter values, the mean force at which the system stops on decreasing $F$ under the procedure discussed in Section IV is 
\be
\la \fd \ra \approx -0.0665
\ee
with  the 
rms variations about this of
\be
\sqrt{{\rm var}(\fd(\fpm=1.0,M=0.8))} \approx  0.0006 \ .
\ee

\begin{figure}[h]
\begin{center}
\epsfxsize=8cm
\epsfysize=8cm
\epsfbox{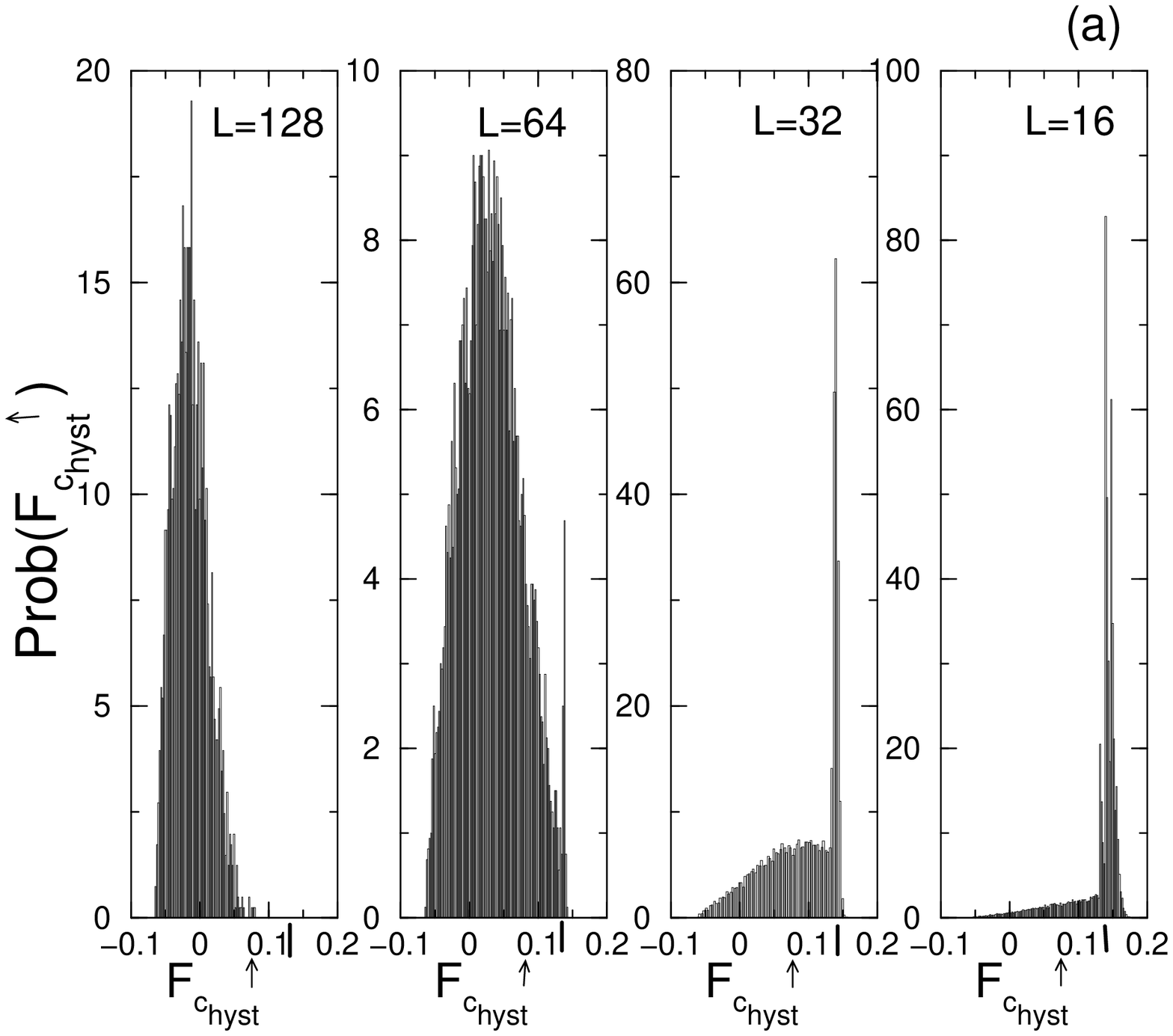}
\end{center}
\end{figure}

\begin{figure}[h]
\begin{center}
\epsfxsize=8cm
\epsfysize=8cm
\epsfbox{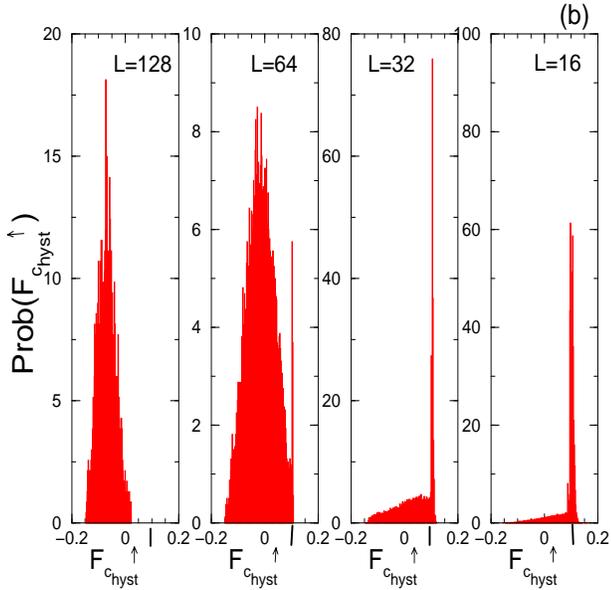}
\caption{\label {loopwidth0.8plot} \label {loopwidth1.0plot} 
(a)Probability distribution of 
$F^{\uparrow}_c$,  $Prob(F^{\uparrow}_{c_{hyst}})$ 
for $M=0.8$ and $fpm=1.0$ for different system sizes.   (b) Same as (a) but with $M=1.0$.
The vertical bar on each horizontal axis indicates $<F_c^{\downarrow}>+M/Z$.}
\end{center}
\end{figure}
\begin{figure}[h]
\begin{center}
\epsfxsize=8cm
\epsfysize=8cm
\epsfbox{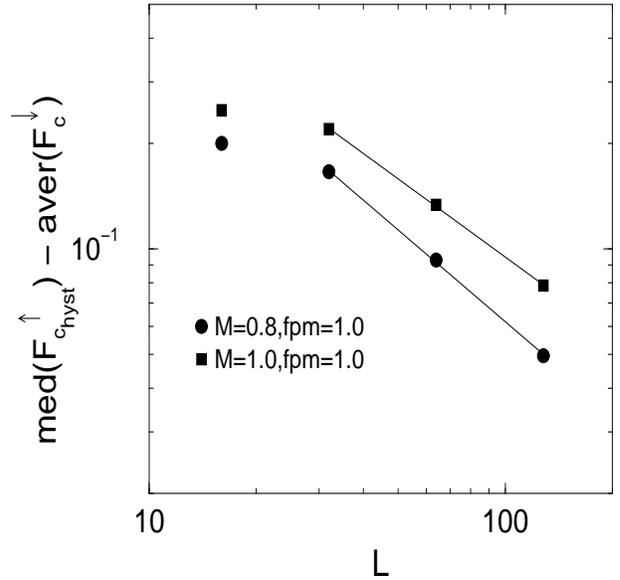}
\caption{\label {loopvanishplot}Log-log plot of the media width of the hysteresis loop for 
$M=0.8$ and $M=1.0$ with ${\rm fpm}=1.0$.  The slope of the large $L$ data for  $M=0.8$  is
$-0.87\pm0.02$ with an amplitude of $3.4\pm0.2$.  For the $M=1.0$ 
data, the slope is $-0.74\pm0.01$ with an amplitude of $2.9\pm 
0.2$.  We did not include the $L=16$ data points in the linear 
regression as there is a clear crossover at small scales.}
\end{center}
\end{figure}

In Fig. \ref {loopwidth0.8plot}(a), the distributions of $\fh$ are shown for various system sizes. It can be seen that they are much broader --- by almost two orders of magnitude for $L=128$ --- than the distributions of $\fd$. The shapes of the distributions at first appears rather strange: For the smaller system sizes, a substantial fraction of the weight is in a narrow peak that has similar width to that of the distribution of $\fd$, but is shifted up from this by an amount $\fh-\fd\approx M/Z$.  In the largest samples, this peak has completely disappeared and we see that the width of the hysteresis loop has narrowed considerably.    
The narrowing with size of the median width of the hysteresis loops is shown in a log-log plot in Figure \ref{loopvanishplot} .
As 
was evident in the shape of the distributions, a crossover length of around $L=20$ is seen in these data.  For small sizes the hysteresis loops have width that is typically close to $M/Z$.  But for the large sizes, the typical width appears to  decrease as a power of $L$:
\be
\fh-\fd \sim \frac{1}{L^\frac{1}{\mu}}
\ee
with
\be
\mu \approx  \ 1.15 \pm 0.03
\ee
for $M=0.8$  and
\be
\mu \approx  \ 1.35 \pm 0.02.
\ee
for $M=1.0$.  Note that these exponents are obtained from a narrow range of length scales and  crossover behavior is likely to be playing a role; we will return to the issue of crossovers later. 
Nevertheless, in spite of uncertainties in the asymptotic size dependence of the hysteresis loops, the overall trend is clear: in the limit of large systems, the {\it width of the hysteresis loops vanish}! This in spite of the fact that there are many linearly stable static configurations that coexist with the moving state up to the force $F_{c0}$ which is substantially greater than $\fh$.

Before discussing this result, it is instructive to consider what  happens in the dissipative limit, $M=0$. Although it might appear that there would be no hysteresis in this case, this is not strictly correct for finite size systems that have been stopped from a moving state.  At the force at which the system  stops, it gets stuck in a somewhat anomalously 
strong pinning region (how anomalous depends on the rate of decrease of the force). Getting it unstuck from such a region, in the sense that all parts of the system move for at least some distance, can require an increase in the force that is comparable to the width of the distribution of $\fd$.  Thus we expect ``hysteresis loops" in the dissipative limit to have a width of order $L^{-1/\nu}$.

If the systems with stress overshoots behaved like the dissipative case in all universal  aspects, one would expect the asymptotic large system-size dependence of the width of the hysteresis loops to have the same exponent as the dissipative case, i.e. that $\mu=\nu$.
Up to questions about estimation of  uncertainties in the presence of complicated crossovers, it {\it appears} that this is {\it not} the case: we seem to find
\be 
\mu > \nu
\ee
corresponding to system size dependence of $\fh$ being slower than that of $\fd$. If this were indeed the case, we would expect that it would most likely hold asymptotically  for any $M>0$.  Unfortunately, the range of data is not so large as to conclusively rule out equality rather than inequality, although if $\mu=\nu$ one would probably need either a large dimensionless amplitude ratio between the coefficients of the size dependence of the two critical forces, or strongly nonmonotonic behavior; we will later explore  such scenarios.
But for now we focus on the scaling behavior that seems to be emerging for system sizes larger than of order 20 or so for  ($\fpm=1.0,\ M=0.8$).

It appears that the larger system sizes {\it do} exhibit scaling behavior of the {\it distributions}, a more stringent test than exponents.  Indeed, the size dependence of the distributions provides a useful way to understand the causes of the size dependence of the hysteresis.

\begin{figure}[h]
\begin{center}
\epsfxsize=8cm
\epsfysize=8cm
\epsfbox{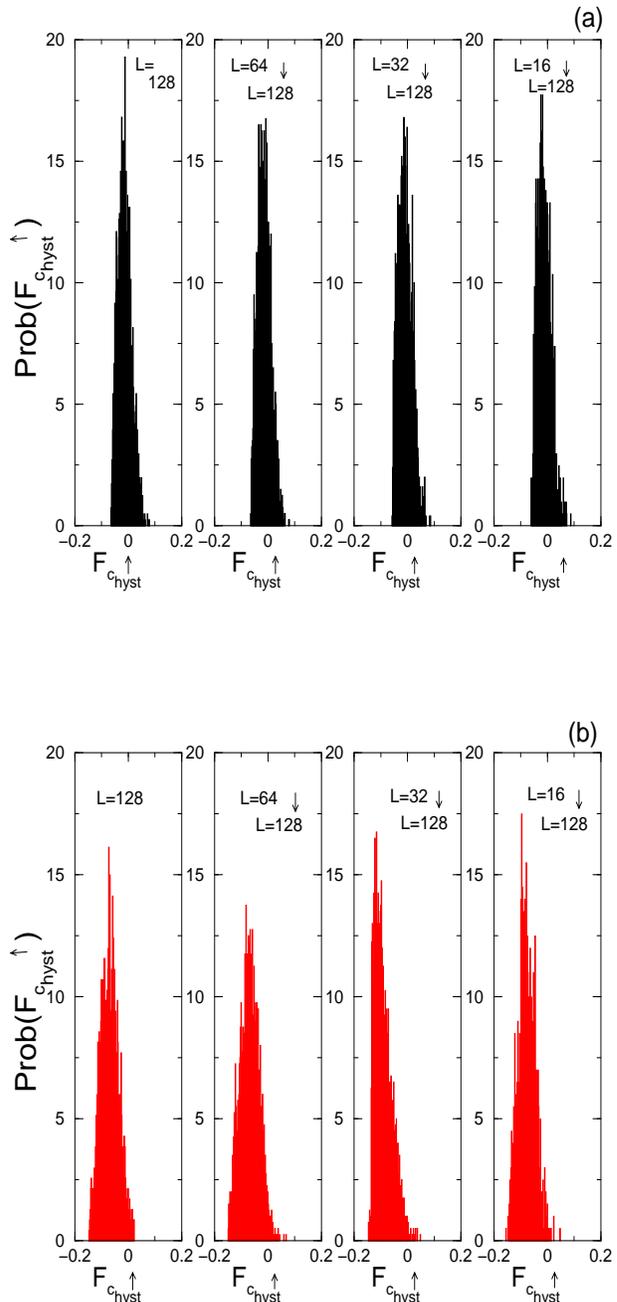}
\caption{\label {lwidth0.8rescaledplot} \label {lwidth1.0rescaledplot} (a) Distributions,  $Prob(\fh)$, 
for $M=0.8$, $fpm=1.0$ and $L=128$ inferred from data for smaller sizes.  The $L=128$ curve is the same as in Fig. (\ref{loopwidth0.8plot}) while the other distributions are obtained from the $L=64,\ 32$ and $16$ data from Fig. ( \ref{loopwidth0.8plot})  rescaled as described in the text. (b) Same as (a) except with $M=1.0$.}
\end{center}
\end{figure}

Let us assume that restarting on increasing the force after stopping occurs via some kind of nucleation process whose occurrence  is dominated by scales
that are much smaller than the system size. Then we expect a density of nucleation segments --- or ``seeds" --- with a distribution of values of the local critical forces $F_s$ needed to restart. If this distribution extends down to $\fd$, larger systems are more likely than smaller ones to have a seed with a small $F_s$.  As the lowest $F_s$ in a given stopped configuration will be the one that determines $\fh$, this will yield a distribution of critical forces that becomes squeezed down to $\fd$ as $L\to \infty$. 
A simple check on the assumption of locality of the seeds is accomplished by estimating the distribution of $\fh$ for a sample of size $L$, by considering it as being made of $b^d$ {\it independent} samples of size $L/b$ whose $\fd$s are drawn independently from the {\it observed} distribution for these smaller size samples.  In Figs. \ref {lwidth0.8rescaledplot}(a) and \ref {lwidth1.0rescaledplot}(b), 
this is carried out for $L=128$ and $b=2,4$, and $8$.  As can be seen, the distributions obtained agree very well with those measured for the $L=128$ samples directly.  This agreement is particularly striking given that the data for $\fh$ for $L=16$ in Figs. \ref {loopwidth0.8plot}(a) and \ref {loopwidth1.0plot}(b) have a very different form than those for the large samples. (This difference in form for the distributions is the source of the crossover observed in the size dependence of the width of the hysteresis loop.) 

The agreement of the actual distribution of $\fh$ for $L=128$ with that obtained from the distribution with $L=16$ suggests that the nucleation process in a sample of size 128 is typically dominated by regions whose diameter is less than 16. Once such a  small nucleation region gets going, it will typically expand to make the whole system restart, independent of the existence or lack thereof of seeds in other regions, or of other stochastic properties of the rest of the system.   Before analyzing the consequences of this, we must caution that a crucial question is whether such a relation between the distributions of $\fh$ for systems of size $L$ and $L/b$ holds, in the limit of large $L$, for only a limited  range of $b$,  for any $b\ll L$, or for $b$ up to some (subdominant) power of $L$. 

For now, we will extract the shape of the distribution from the observation that there is at least a substantial range of $b$ over which the relationship between the distributions of $\fh$ for size $L$ and size $L/b$ does hold.  

\begin{figure}[h]
\begin{center}
\epsfxsize=8cm
\epsfysize=8cm
\epsfbox{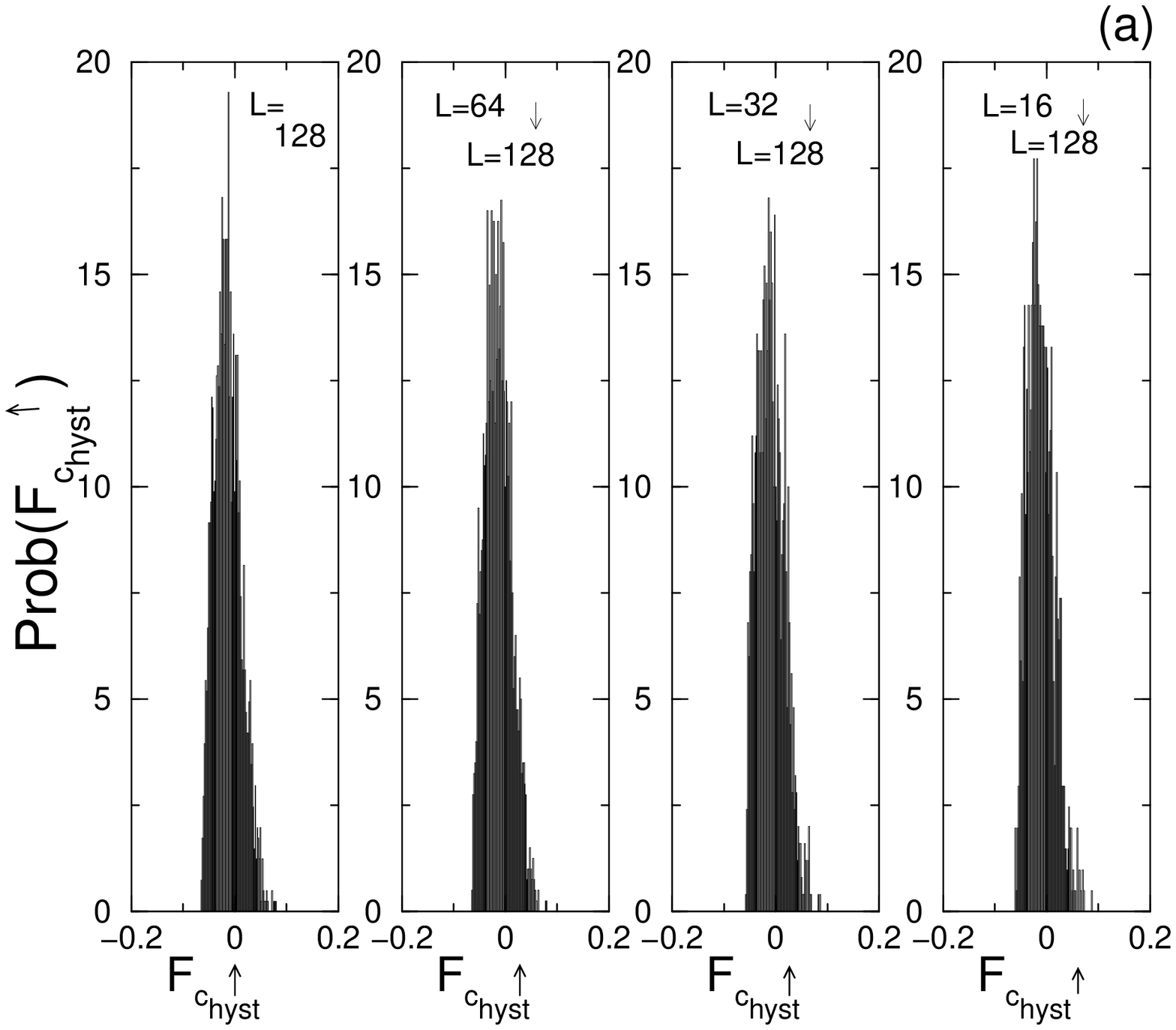}
\end{center}
\end{figure}

\begin{figure}[h]
\begin{center}
\epsfxsize=8cm
\epsfysize=8cm
\epsfbox{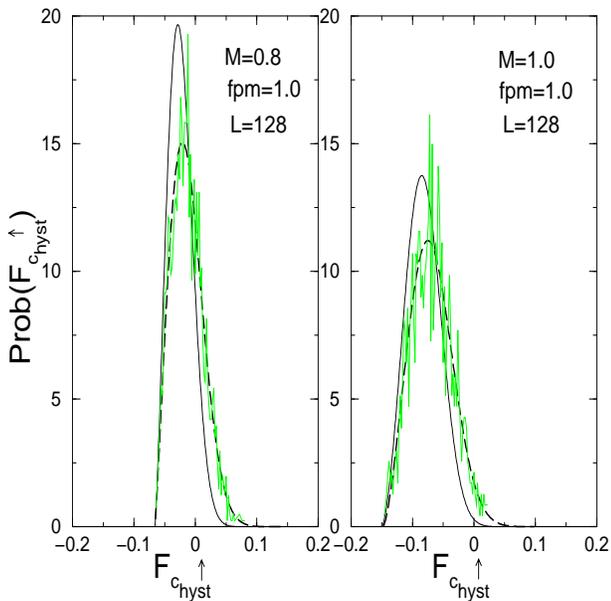}
\caption{\label {weibullplot} Left plot shows distribution of $\fh$ with dark solid line a Weibull distribution, Eq. (\ref{weibull}), with
scale parameter of $0.048$ and $\mu=1.15$ obtained 
from Fig. \ref {loopvanishplot}, and dashed line a best fit to 
the Weibull form with two  fitting parameters. The light solid line is the data for $L=128$ and $M=0.8$.  
The right plot is, similarly, 
$M=1.0$ with the scale 
parameter for the solid line Weibull distribution $0.078$ with  $\mu=1.35$
from Fig. \ref {loopvanishplot}, and the dashed line the best fit Weibull distribution. }
\end{center}
\end{figure}

The basic picture of the restarting being controlled by the least pinned of many independent seeds enables one to relate the size dependence of the median $\fh$ to the form of the distribution. Picking the minimum of the $\fh$'s from  $S=b^d$ subsystems each of which has a distribution of $\fh$ that vanishes as $(\fh-\langle\fd\rangle)^{d\mu-1}$ for small $\fh-\langle\fd\rangle$, yields a power law   decrease with $L$ of the width of the distribution, and, indeed,  the actual {\it form} of the distribution:  We expect a Weibull distribution with one non-universal scale parameter, $f_h$, \cite{bouchaud}. 
\begin{eqnarray}\label{weibull}
\pr[d\fh]\approx d\fh\ \frac{d\mu}{f_h} 
\left(\frac{\fh-\fd}{f_h}\right)^{
d\mu-1} \nonumber \\ 
\exp\left(-\left(\frac{\fh-\fd}{f_h}\right)^{d\mu}\right) \ . 
\end{eqnarray}
As can be seen in Fig. \ref {weibullplot}, this 
yields a rather good fit to the data for $L=128$ with the value of $\mu$ extracted from the size dependence of the median. If, instead, we do a best fit to the shape of the distribution for the largest size, we find, $\mu\approx 1.08$ for $M=0.8$ and for $M=1.0,\ \fpm=1.0$, $\mu\approx 1.25$.  Note that these values are slightly smaller than those obtained from the size dependence and thus somewhat closer to $\nu$; this may well be a sign of slow crossover to asymptotic behavior that is like the dissipative limit.

\subsection{Origins of seeds for restarting}

We next develop an understanding of  the origins of the unusual hysteretic behavior that was found in the numerical studies: in particular, the origins of the seeds for nucleation of motion when the force is increased back up again after the motion has stopped.
To do  this, we need to understand how the  manifold
stops moving as $F$ is decreased to $F^{\downarrow}_c(M)$ 
as it is this that sets up the configurations in which the seeds exist.
We first analyze the basic role of the stress overshoots in the steady state moving phase.  

A very crude approximation to the effects of the stress overshoots is to ignore their local and transient natures. We thus consider an artificial 
model in which if {\it any} segment has moved on the previous time step, the force on {\it all} the segments is  increased by $M/Z$ above what it would be with purely dissipative dynamics. As long as something is always moving, this is identical to merely increasing the applied force by $M/Z$.  But once the system has stopped --- because of a decrease in the applied force or because of running into a strongly pinned region --- no segment can move again until the force is increased by $M/Z$; this is because any segment that could move with less of an increase in $F$, should, {\it a fortiori} have moved already because of the stress overshoot that was present before the motion stopped. In terms of the distribution of the total force, $\{f(\bx)\}$, on a segment, which must be positive for it to jump, the stopped configuration will have a {\it depletion layer}: no segments with $f(\bx)$ in the interval $(-M/Z,0)$ (and of course none with $f(\bx)>0$). The behavior of this crude model is thus very simple: the $v(F)$ curve is shifted down (in $F$) by $M/Z$, the steadily moving states are {\it identical} to those at $M=0$ with $F\to F+M/Z$. When the force is increased after stopping, no motion will occur until the depletion layer disappears, therefore $\fh=\fd+M/Z$ for each sample.  

Before we return to the model or primary interest, it is worth noting that a model which  is much less pathological than the crude model discussed above nevertheless has much of  the same behavior.  This  {\it non-additive stress overshoot model} has nearest 
neighbor stress overshoots that last for one time step and a ``self overshoot"  (analogous to inertia) that likewise lasts for one time step.  But the stress overshoots are  non-additive so that any site has a stress overshoot that is either zero or $\frac{M}{Z}$.  For example, if two nearest neighbors 
of a segment jump, that segment feels a stress overshoot of 
$\frac{M}{Z}$, in contrast to the $\frac{2M}{Z}$ overshoot it would feel in our primary model.  The partial equivalence between this non-additive stress overshoot model and the crude model can be simply understood:  If the total force on a segment does not change 
from the previous time step  it cannot jump at the next time step. Therefore   
{\em changes} in the total force on a segment are what  determines whether a segment jumps  or not at the next time step.  
At fixed $F$, changes in the total force on a segment arise from nearest neighbors  jumping at the previous 
time step and from the segment itself jumping at the previous time step. In the non-additive model, each of these will involve an extra $\frac{M}{Z}$ stress. Thus as long as motion has existed somewhere in the systems for more than one time step, given a configuration of the crude model and which segments have jumped on the previous time step, there is an exactly equivalent configuration of the non-additive model which will have the same dynamics at all future times as long as $F$ remains fixed. But the dynamics are {\it not} fully equivalent: when the first segment moves in the crude model, it can trigger others far away; this cannot happen in the non-additive model, 
Nevertheless, the steady state velocity as a function of force will be the same in these two models, with the critical force in the infinite system limit shifted down by exactly $\frac{M}{Z}$ from the dissipative case.  The hysteresis loops will also be similar, but not identical: in both cases there will be a depletion layer of width $\frac{M}{Z}$ after the system has stopped and the force will have to be increased by this for motion to start again,  But in finite size samples the behavior will be slightly different as it is much more likely in the non-additive model that motion could start in one region but die out: the actual critical force for restarting would then be slightly higher.  The dynamics of the transient motion on restarting would also differ due to the locality of the non-additive model.

The decrease of $\fd$ with $M$ in the crude and non-additive models 
is the underlying cause for the linear decrease of $\fd(M)$ in the primary model: As long as segments only move in response to their neighbors moving, what is crucial in determining $\fd$ is how the system gets through potential sticking points. Some of these are likely to involve only one neighbor of a segment moving at the previous time step; if they do, then the critical force at which they can proceed will, in the absence of other changes of the dynamics due to $M$, be just $M/Z$ lower than it would be with $M=0$. In the limit of small $M$, we expect that the sequence of jumps at $F$ will be very close to that at $F+M/Z$ in the absence of overshoots.

The non-linear part of the  dependence of $\fd$ on $M$ is of a different origin. As $M$ grows, the equivalence between the sequences of jumps at different values of $M$ no longer obtains because of, for example,  the effects of two neighbors jumping at the same time which increases the stress on a segment by $2M/Z$.  This will tend to make stopping less likely as  a region can be restarted by motion in other regions that is caused by such multiple-neighbor jumps. As one would thus expect, $\fd$ decreases {\it faster} than linearly as $M$ is increased.  

The focus on particular sites and whether they can be retriggered by a given increase in $F$ is also useful for understanding the hysteresis in the
model of primary interest.  A crucial  
question about the local dynamics is: How close can a segment be to moving {\it without} one of its neighbors having moved on the previous time step? We must consider the most recent time in the past, say time one, at which a neighbor of the segment $\bx$ of interest moved.  For simplicity, let us assume that none of the neighbors moved at time zero. At time zero, the total force on $\bx$ is then
\be
f(\bx,t=0)=F-f_p(0)+\sigma(0) <0 \label{s0}
\ee
with $\sigma(0)=\sigma(\bx,0)$ and $f_p(0)=f_p({\bf x}(0),h({\bf x}(0)))$ the initial elastic force and initial pinning strength, respectively, at $\bx$.
If $n$ out of the $Z$  neighbors jump at time one, 
\be
f(\bx,1)= F-f_p(0)+\sigma(0) + n(1+M)/Z  \ . \label{s1}
\ee
If this is negative, then $\bx$ will not move and, at later times, the force $f(\bx, t>1)< -nM/Z$ as the stress overshoot will no longer apply; thus  segment $\bx$ will not be in the depletion layer. If, however, $\bx$ does jump at time two in response to its neighbors jumping, i.e., if $f(\bx,1)>0$, then the total force on it at later times will be
\be
f(\bx,2)=F-f_p(1)+\sigma(0) -1 +n/Z  \label{s2}
\ee
with  a new random pinning force $f_p(1)$.  As long as $f(\bx,2)<0$, then this segment will not move further unless one of its neighbors does. Thus this represents a possible local configuration when the system has just stopped. 

The condition  $f(\bx,0)<0$  implies, from Eq. (\ref{s0}) 
and Eq. (\ref{s1})  that  $f(\bx,1)<n(1+M)/Z$.
The maximum of $f(\bx,2)$ is then obtained when $f_p(1)$ is minimal (i.e. zero) and $f_p(0)$ maximal (i.e. $\fpm$); this yields 
$f(\bx,2)<\fpm-1+n/Z$.   If the force is now  increased back up by $\Delta F$ so that $f(\bx,2)+\Delta F=0$, the segment $\bx$ will jump and could trigger restarting of the overall motion.  We must thus ask how close to zero $f(\bx,2)$ can be.

If the last motion in any region were always via single neighbors triggering each other, then there would be a depletion layer on stopping and macroscopic hysteresis for $\fpm<1-\frac{1}{Z}=\frac{3}{4}$ in our case with coordination number $Z=4$.  [This would obtain if the stress overshoots were not additive but instead such that any number of neighbors jumping yielded the same value of the stress overshoot as if only one had.] But in reality, it is possible that any number of neighboring segment on a segment could jump at one time and then not again; some results on the simultaneous hopping of a number of neighbors are presented in Table II.  As such sets of simultaneous jumps can occur for any $n\le Z$ --- including for $n=Z$ --- even as the system is stopping, there will be {\it no depletion layer} even for arbitrarily weak randomness  and no hysteresis in the infinite system limit.  Nevertheless, for weak pinning,  the depletion layer will only be filled by simultaneous jumps of multiple neighbors followed by a jump of the central segment that does not trigger further jumps of any of its neighbors,  Although we expect that a local condition such as this will always occur for some finite fraction of the segments, it appears likely that the potential seeds for restarting with small increases in $F$ will be very rare for weak pinning; this, indeed, turns out to be the case. It is the proximate cause of the long crossover lengths apparent in the size-dependent distributions of $\fh$.

\begin{table}
\caption{The  
probability of $n=0,1,2,3$ or $4$ of the nearest neighbors of a segment having
jumped at the previous time step given that the segment jumps.   For $n=0$,  there are large fluctuations so we record a range of values. The size is $L=128$ and average 
velocity  approximately $0.1$.}
\begin{tabular}{ccccccc}   
$f_p^{max}$ &$M$  &0\  (range) &$1$    &$2$    &$3$    &$4$ \\ \hline
$0.5$  &$0.0$        &$(0.0005,0.002)$   &$0.55$ &$0.30$  &$0.10$  &$0.05$ \\
$0.5$  &$0.2$        &$(0.0,0.001)$      &$0.45$ &$0.35$  &$0.15$  &$0.05$ \\
$0.5$  &$0.4$        &$(0.0,0.001)$      &$0.30$ &$0.35$  &$0.25$  &$0.10$ \\
$0.5$  &$0.6$        &$(0.0,0.001)$      &$0.20$ &$0.30$  &$0.30$  &$0.20$ \\
$0.5$  &$0.8$        &$(0.0,0.00025)$    &$0.10$ &$0.20$  &$0.30$  &$0.40$ \\
$1.0$  &$0.2$        &$(0.01,0.02)$      &$0.45$ &$0.35$  &$0.15$  &$0.05$ \\
$1.0$  &$0.8$        &$(0.002,0.008)$    &$0.20$ &$0.30$  &$0.30$  &$0.20$ \\
$1.5$  &$0.2$        &$(0.05,0.06)$      &$0.50$ &$0.30$  &$0.10$  &$0.02$ \\

\end{tabular}
\end{table}
The above analysis gives a qualitative explanation for the lack of macroscopic hysteresis.  But to explain the observed dependence of the widths of the hysteresis loops on system size, we must understand the density of states of segments with small negative total force on them in the stopped configurations, and what happens after motion is triggered by one of these seeds as the force is increased.      
Quite generally, the continuous nature of the distributions of yield strengths and the discrete nature of the stress transfer means that the density of states for local properties should either be zero or be strictly positive.  Specifically, from the above we expect that the density of states of the local forces will be positive at zero in the stopped state. To check this, we have computed the  probability density per site, $r(F)$, that some motion is triggered with a small increase in $F$ to $F+dF$ in stopped systems; this is normalized so as to include only those samples that have not yet restarted macroscopically (although they might have already had some transient local motion). This rate of triggering of jumps appears to go to a constant as $F$ decreases to $\fd$ and exhibits relatively weak dependence on $F$ as the force is increased; we will present the data later in the paper. 

From the data for the distribution of $\fh$, which {\it vanishes} approximately linearly at $\fd$, a constant density of states for triggering motion is perhaps surprising.  The reason for this must lie not in the seeds themselves, but in how they grow.  In particular, very close to $\fd$ local triggering must be less likely to induce restarting than it does at higher forces.
In order to understand this, it is necessary to investigate the {\it avalanche dynamics}, i.e., the transient motion in response to triggering of one segment. 
Before considering this in the context of the hysteresis loops and restarting, we analyze avalanches that occur in the approach to depinning from below.  

\section{Avalanche Dynamics}

In the previous section we have seen that to understand the hysteretic phenomena observed on cycling the force up and down, we need to understand how macroscopic motion starts once it has been triggered by a local instability that leads to one segment jumping. Before studying the case of interest for hysteresis loops, which involves initial conditions that are set by the stopping process, we analyze the behavior as the force is slowly increased from far below the depinning transition starting from more generic initial conditions. We will call this {\it initial depinning}. In particular, we are interested in the behavior as the depinning transition is approached from below. 
  
Even though there is no steady state motion  in this regime, there can be local, transient  
motion  in response 
to small increases in $F$.  Such {\it avalanches} will not persist indefinitely for small $F$ because  the pinning forces in other regions will eventually dominate as long as  
$F<\fa(M)$, the --- possibly history dependent --- critical force on increasing $F$.  

In this section we investigate the dynamics that result when $F$ is increased adiabatically: initially by
just enough that one segment 
moves. This can then trigger other segments to hop forward,
while $F$ is held fixed until the avalanche  stops.  The same procedure is then  
repeated until $F=\fa$. For an infinite system, this is defined as the force above which the motion persists indefinitely in the absence of any further increase.  In finite systems, there are some ambiguities in how it is defined; we choose to define it as the lowest force
at which all of the segments move during a single avalanche. 
The primary quantities of  interest  below the depinning are the sequence of avalanches and their statistical properties: numbers, sizes, durations, etc. More macroscopic quantities, such as macroscopic responses  to a small but non-infinitesimal increase in $F$ can be determined by integrating over the properties of the avalanches.

There are various measures of the size of an avalanche. Three of these will be of particular interest.  The {\it moment}, $m$,  of an avalanche is defined as   the total motion that occurs:
\be
m= \sum_\bx \left[h(\bx)_{\rm after} -h(\bx)_{\rm before}\right] \ ; 
\ee 
this is the quantity of primary interest for earthquakes.
Alternatively, one can consider the {\it area}, $a$,  (in the two dimensional case of interest): the total number of segments that move at least once during the avalanche.  Lastly, is the linear size, $\ell$, of an avalanche, one measure of which is its {\it diameter} defined, for example,  either via some weighted sum of distances of moving segments from its center, or as the diameter of the smallest circle that will enclose the avalanche. 
    
\subsection{Scaling}

For the purely  dissipative case, $M=0$, the scaling of the avalanches is related to that of the various quantities --- $h$, $x$, $F-\fa$ and $t$ --- discussed in the context of the moving phase.  In particular, if an avalanche has a diameter $\ell$, its area will scale as $\ell^{d_f}$, its duration as $\ell^z$, the typical maximum displacement --- change in $h$ ---  as $\ell^\zeta$, and its moment as 
\be
m\sim \ell^{d_f+\zeta} \ .
\ee
As long as the dimension is less than the upper critical dimension, $d_c=4$ for short range interactions, the avalanches will not be fractal and hence 
\be
d_f=d
\ee
so that the area is a good surrogate, which we will use,  for the length scale of an avalanche:
\be
\ell \sim a^\frac{1}{d} \ .
\ee

Well below the depinning transition, most avalanches are small.  But as the transition is approached from below, larger ones become possible, although the distribution of their diameters is cutoff by a correlation length $\xi$ that, 
in the dissipative limit,  diverges at $F_{c0}$ as
\be
\xi \sim \frac{1}{(\fa-F)^\nu}
\ee
with the same exponent $\nu$ as determines the scaling of the physically different characteristic length in the moving phase, the velocity correlation length. At the critical point in the dissipative limit, the distribution of avalanche sizes is a power law:
\be
\pr[{\rm diameter}\  >\ell] \sim \frac{1}{\ell^\kappa}
\ee
and a similar relation applies for other measures of size; for 
example, for the area, the exponent is simply changed to $\kappa/d$.
Near to the critical force, the distribution of the areas of large avalanches has the scaling form 
\begin{equation}
\pr[da]= p(a;F)da\sim \frac{1}{a^{\frac{\kappa}{d}}}{\cal P}(a/\xi^d)\frac{da}{a}
\end{equation}
where, of the avalanches   that occur within a small force interval around $F$, $p(a,F)da$, is the {\it fraction} that of these that whose area is between
$a$ and $a+da$\cite{narayan2}. The scaling function  ${\cal P}(y\rightarrow \infty)$ decays rapidly while for $y\rightarrow 0$, it goes to a constant. 

The statistical ``tilt" symmetry of the system that was used earlier to yield the scaling law, $\frac{1}{\nu}=2-\zeta$, can also be used, via relating the polarizability to the avalanche production rate and the distribution of their sizes, to show that 
for $M=0$, 
\be
\kappa=d-\frac{1}{\nu}= d-2+\zeta
\ee
 as derived in Ref. \cite{narayan2}.
The one crucial assumption is that the rate, $r(F)$,  of avalanche production, defined as $1/\Delta F$ times 
the number of avalanches per unit area of the system   as the force is increased by a small amount from $F$ to $F+\Delta F$, tends to a finite non-zero constant at the critical force.

We now turn to an analysis of the data for avalanche statistics and properties of the avalanches: for the dissipationless limit, to check the theoretical predictions outlined above; and for non-zero $M$, to investigate the effects of stress overshoots.

An easy quantity to  measure is the {\it cumulative distribution} of 
all the avalanches as the applied force is increased to 
$\fa(M)$. This is given by
\begin{equation}
p_{cum}(a)da \equiv \int_{-\infty}^{\fa} dF p(a;F)r(F)\frac{da}{a} \sim \frac{1}{a^{K_{cum}}} da\ .
\end{equation}
  Assuming that $r(F)$
approaches a constant as $F\to \fa$, the scaling laws for the dissipationless case yield a power law cumulative avalanche
area distribution with an exponent of unity, 
\be
K_{cum}(M=0)=1 
\ee  
{\it independent} of the values of the other exponents. This thus provides a good test of the general scaling theory that does not depend on particular predictions for exponents.
Although we do not expect the universal aspects of the avalanche statistics to depend on details of the initial conditions, to avoid effects that might arise from smoothening out  rough initial conditions, we take the initial configuration to be approximately flat: specifically, the initial $\{h(\bx)\}$ uniformly distributed in the interval $[0,1]$.

\subsection{Dissipative limit}

\begin{figure}[h]
\begin{center}
\epsfxsize=8cm
\epsfysize=8cm
\epsfbox{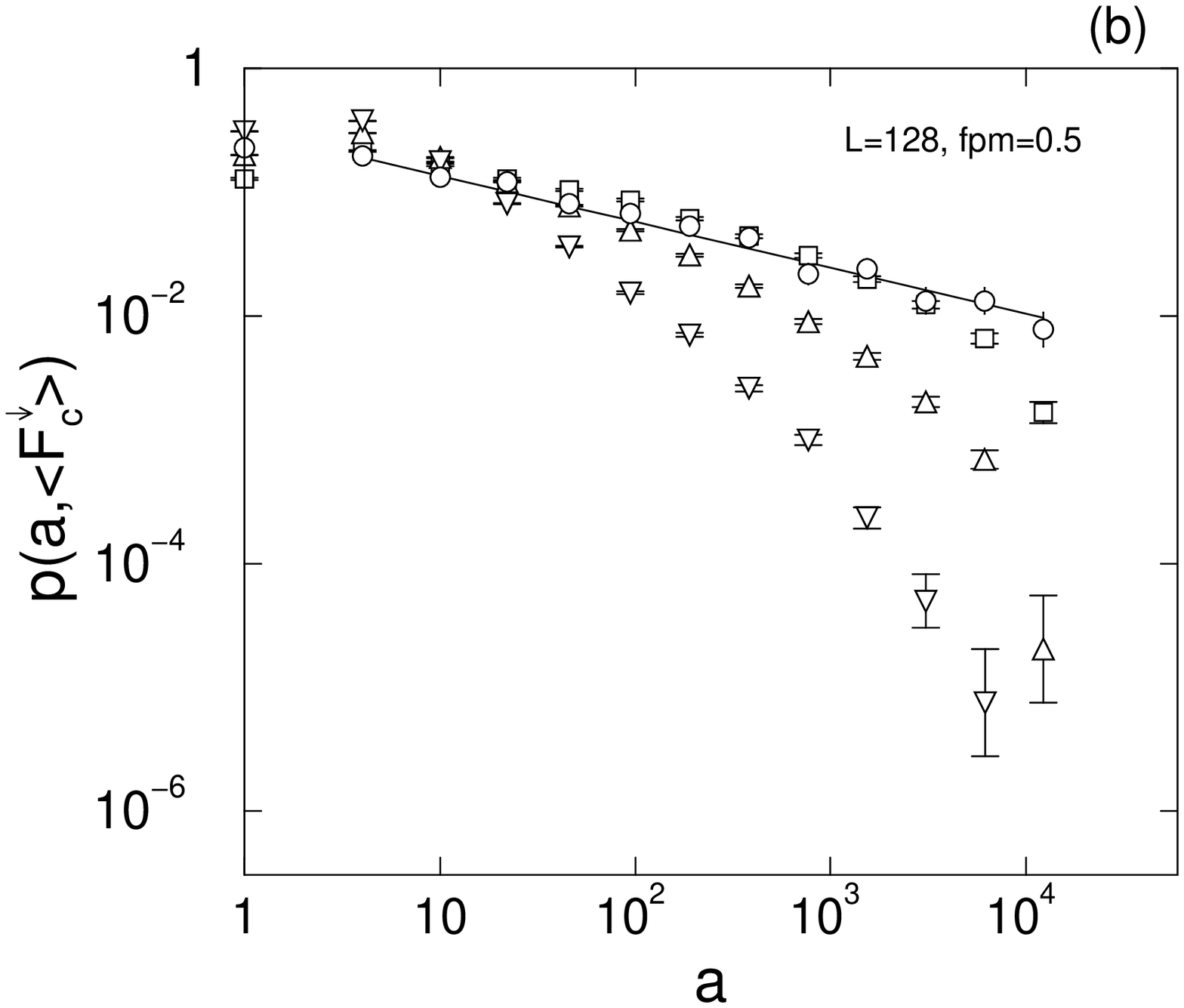}
\caption{\label {pcum0.5plot} \label {p0.5plot} (a)Log-log plot of the probability of avalanche area
$a$ occurring 
within the interval $[a/\sqrt{2},\sqrt{2}a)$ as $F$ is increased 
to $F^{\uparrow}_{c_{aval}}(M)$ 
for $\fpm=0.5$.  The 
circles are for $M=0.0$, squares $M=0.2$,  
triangles  $M=0.4$, and inverted triangles $M=0.6$.  The 
open symbols represent $L=128$ data, while the closed symbols represent 
$L=64$.    
The asymptotic slope should be $-K_{cum}/d$; fits yield: for $M=0.0$, $-0.89 \pm 0.02$; 
for $M=0.2$ , $-1.13 \pm 0.02$; for $M=0.4$, $-1.25 \pm
0.01$; and for $M=0.6$, $-1.53\pm 0.02$.  The dashed curve is the result 
of a two-parameter fit to the $M=0$ data: $C a^{-1}(1+B_{cum}a^{-1/3})$ with $B_{cum}=
-1.21\pm0.11$; with $K_{cum}$ also a
fitting parameter: $C a^{-K_{cum}}(1+B_{cum}a^{-1/3})$, these data yield $K_{cum}=
0.83 \pm 0.02$ and
$B_{cum}=0.53 \pm 0.01$. (b) Same  as 
in (a) and but including only the avalanches that were initiated with
an applied force greater than $F^{\downarrow}_c(M)$ --- well within the critical regime.   The expected slope is $-\kappa/d$; for
$M=0.0$, a best fit yields $-0.37\pm0.02$. For non-zero $M$, large crossover effects are evident. }
\end{center}
\end{figure}

We first analyze the data for $M=0.0$. In Fig. \ref {pcum0.5plot}(a), the  
cumulative avalanche area statistics are shown; a fit to the data on a log-log plot yields an exponent somewhat less than the theoretical expectation of $K_{cum}=1$,   If we restrict consideration to those avalanches that occur in the region close to the critical force in which most of the activity occurs, specifically, within 
the applied force region of $[\fa-0.1,\fa]$, the apparent exponent is roughly the same, $K_{cum} = 0.85\pm0.03$ as shown in Fig. \ref {pcum0.5plot}(a). 
Before trying to understand the apparent discrepancy of this with the scaling prediction, we consider the statistics  of avalanches that occur {\it in the critical regime}, specifically, only those that occur for  $F> \la F^{\downarrow}_c(M)\ra$. The distribution of these also decays as a power of the area, as shown in Fig. \ref {p0.5plot}(b). But the power is much smaller: $0.37\pm 0.02$.  Because the correlation length that would cutoff the avalanche distribution is of order the system size in this regime, the distribution should essentially be that of critical avalanches with an exponent $\kappa/d$.  We thus obtain an estimate
\be
\kappa(M=0) \approx 0.74\pm0.04
\ee
to be compared with the theoretical expectation of  $\kappa=d-2+\zeta=d-\frac{1}{\nu}$ equal to $\zeta$ in two dimensions.  We see that the agreement with our data for $\zeta$ is quite good, suggesting that the basic scaling scenario is correct.

There are various possible sources for the substantial discrepancy of the apparent cumulative exponent from unity. As can be seen in Fig. \ref {pcum0.5plot}(a),(b), 
the rate, $r(F)$  of avalanche production for $M=0$ increases sharply  as the critical force is approached.  Although it does appear to go to a finite constant, as it should on general grounds since $r(F)$ is a locally determined property, the precursor increase will bias the cumulative statistics as there are more avalanches produced near the critical point than further away, and these are the ones that have possibilities of being large. This will tend to put more weight in the large avalanche part of the cumulative distribution thereby decreasing the apparent $K_{cum}$. [The extreme limit of this weighting of those near the critical force would just yield the critical exponent $\kappa/d$ instead as in Fig. \ref {p0.5plot}(b).]   In particular, if there is a  cusp singularity in
the rate of avalanche production as $F\to \fa$ with an exponent $\alpha$, then  this would induce a multiplicative correction to the cumulative avalanche size distribution of the form  $1-C/a^\frac{\alpha}
{d\nu}$ which could complicate interpretation of data.

In general we expect there to be corrections to scaling arising from weakly irrelevant  operators at the RG fixed point that governs the depinning critical behavior.  In terms of areas, the leading irrelevant eigenvalue  $-\theta\approx\frac{4-d}{3}$ \cite{narayanunpub} would give corrections of the form $a^{-K_{ cum}}(1 + B a^{-\theta/d})$ with $\theta/d\approx \frac{1}{3}$ in two dimensions. Such corrections  would dominate over those from a cusp in the avalanche production rate unless $\alpha<0.5$. Indeed the simplest expectation is that the cusp in the avalanche production rate is controlled by exactly this correction exponent,
\begin{figure}[h]
\begin{center}
\epsfxsize=8cm
\epsfysize=8cm
\epsfbox{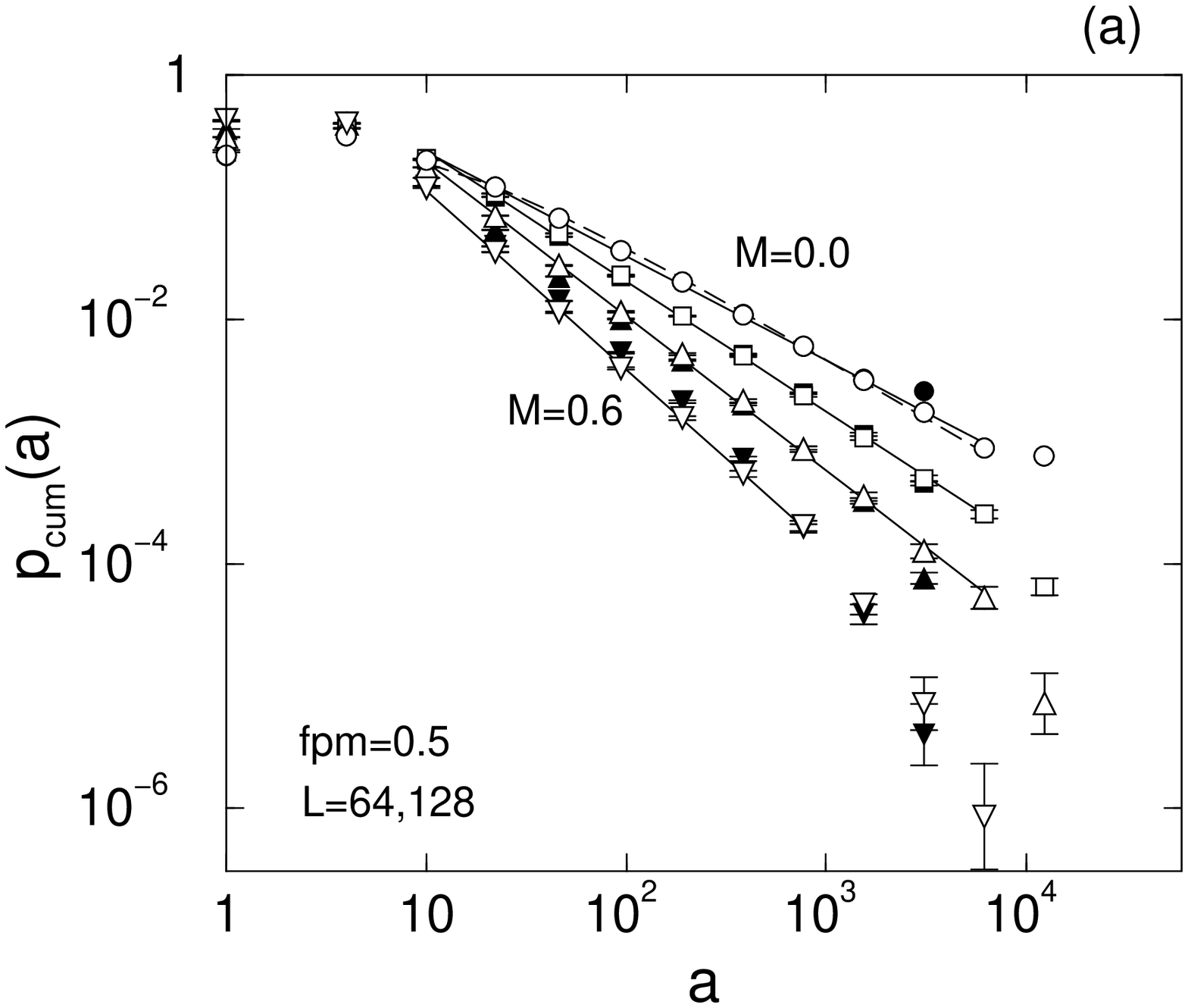}
\end{center}
\end{figure}

\be
r(F)\approx r(F_{c0})-C_r (F_{c0}-F)^{\theta\nu}
\ee
corresponding to $\alpha\approx 0.5$. A fit to this form with $r(F_{c0})$ and
$F_{c0}$ fixed, and $C_r$ and $\theta\nu$ as the free parameters, 
yields a lower 
value of $\alpha$.
If we take into account a correction to scaling by fitting the
log-log plot of the cumulative avalanche area distribution 
with a form that includes the leading correction (shown in Fig. \ref {pcum0.5plot}(a)), the
cumulative avalanche size exponent is changed from $0.89 \pm 0.02$ to 
$0.83 \pm 0.02$, i.e. in the wrong direction. But if the data are fit with the expected $K_{cum}=1$  and an $a^{-1/3}$ correction, --- a fit with the same number of parameters as an undetermined power law with no correction --- the inferred correction to scaling amplitude is $B\approx -1.2\pm0.1$, not unreasonably large. If this were the actual form of the distribution, some downward curvature at the largest 
sizes would  be expected as seen in the dashed fitting line in Fig. \ref {pcum0.5plot}(a).  A 
competing tendency, however, is the flattening of the distribution at areas of order the total system area; this is evident in Fig. \ref {pcum0.5plot}(a) 
in which data for $L=64$ and $L=128$ are shown.  These effects combine to make the  real uncertainties in the exponents here, and probably for other quantities, substantially larger than the apparent uncertainties. 

Note, however, that the useful range of length scales available for  avalanche data and other quantities for which one is trying to extract infinite system results from finite system data are, because of the crossover when the length scales approach  the system size, less than is available for quantities, such as the variance of the critical forces, that are {\it intrinsically} properties of finite size systems.  Thus we might hope to have more confidence in exponent estimates extracted from such intrinsic finite size properties.

\begin{figure}[h]
\begin{center}
\epsfxsize=8cm
\epsfysize=8cm
\epsfbox{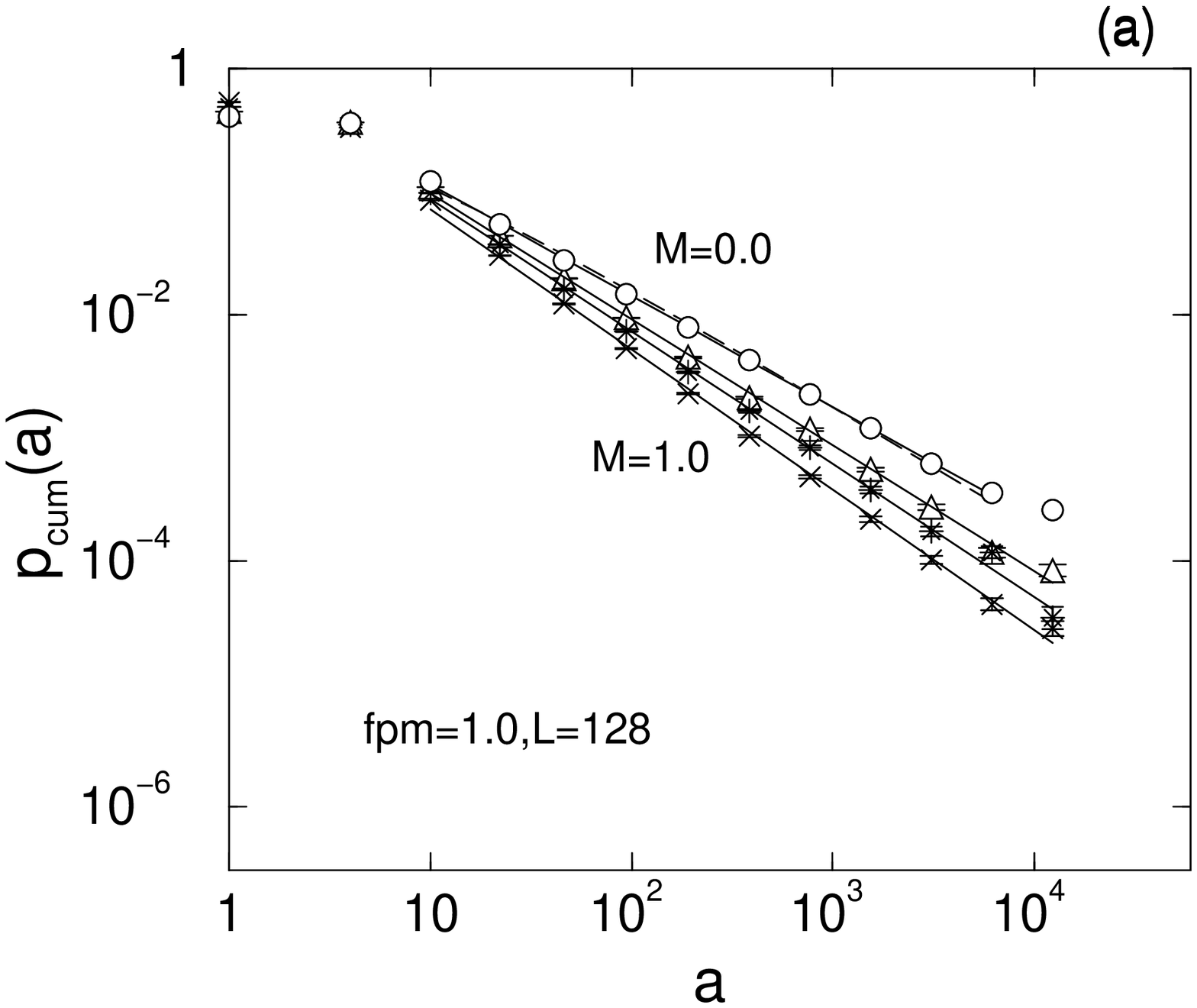}
\end{center}
\end{figure}
\vspace{-2.0cm}
\begin{figure}[h]
\begin{center}
\epsfxsize=8cm
\epsfysize=8cm
\epsfbox{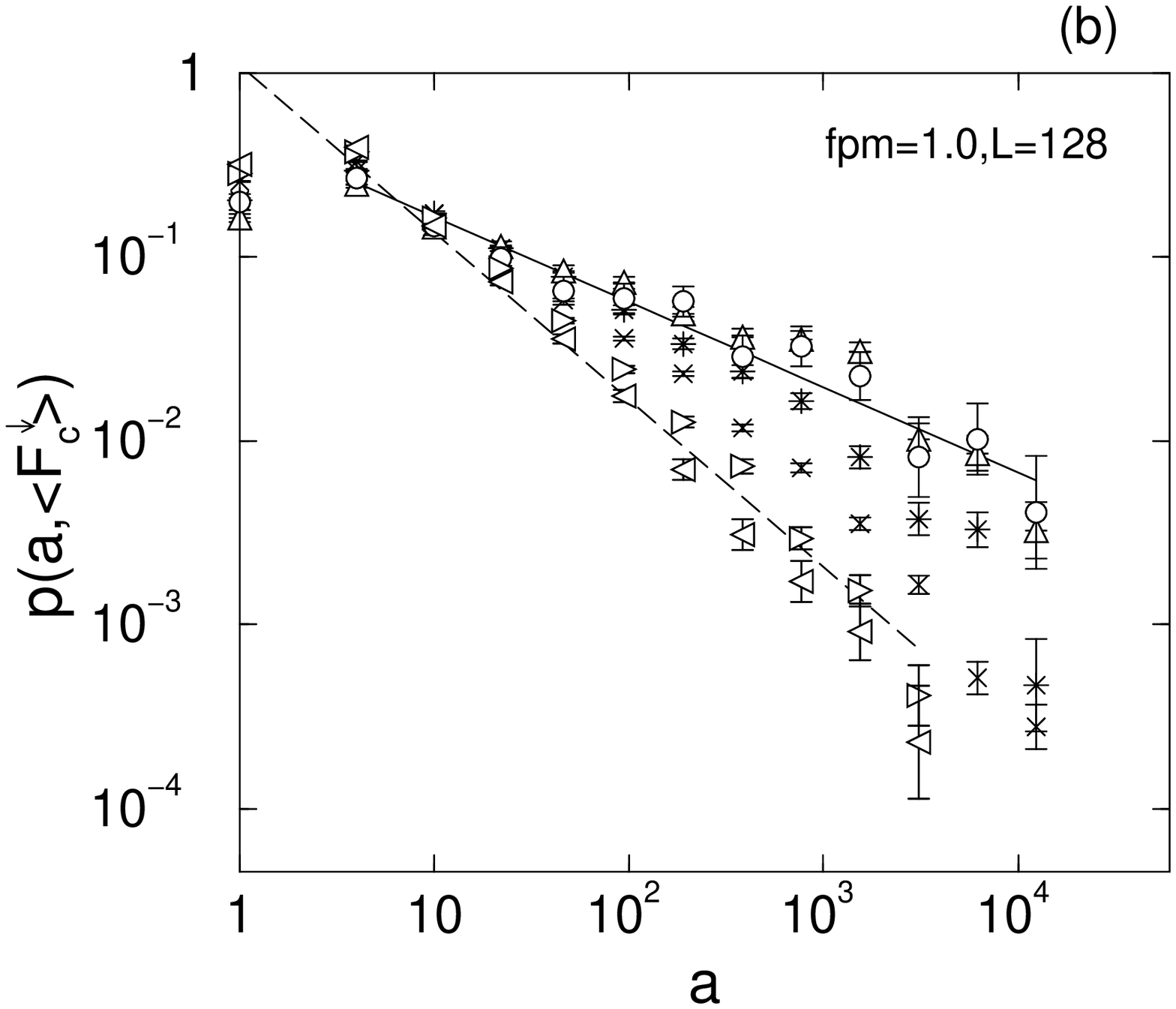}
\caption{\label {pcum1.0plot} \label {p1.0plot} (a)For initial depinning, log-log plot of the probability of an avalanche area
$a$ occurring in the interval $[a/\sqrt{2},\sqrt{2}a)$ as $F$ is 
increased to $\fa(M)$ for $\fpm=1.0$ and 
$L=128$.  The open 
circles are for $M=0.0$, the open triangles denote $M=0.4$, the 
stars represent $M=0.8$, and the $M=1.0$ data are denoted by  {\bf x}'s. 
The fitted slope for $M=0.0$ is $-0.90 \pm 0.01$; 
for $M=0.4$, $-1.02 \pm 0.02$; for $M=0.8$ , $-1.08 \pm
0.02$; and for $M=1.0$, $-1.14\pm0.02$.  The dashed curve is a fit to the 
$M=0$ data that includes corrections to scaling as in Fig. \ref {pcum0.5plot}(a); yields a correction to scaling coefficient 
of $B_{cum}=-0.97\pm 0.01$. (b)  Same as 
(a) but including only the avalanches that were initiated with
an applied force greater than $\fd(M)$, in the critical region. Comparisons with data taken on stopped samples  (starting with $F>\fd$ and 
lowering $F$ until the manifold stops before increasing again). For the initial depinning, the same 
symbols as in (a) are used. For the stopped samples,  the right triangles are for $M=0.8$ and 
the left triangles for $M=1.0$.  
For $M=0$, the slope is expected to be $-\kappa/d=-\zeta/d\approx-\frac{1}{3}$
 a fit yields $-0.39\pm0.02$.  For $M>0$, the data do not fit this and crossover effects are evident.  For comparison,  dashed lines are shown with a slope of $-0.9$, which corresponds to the measured 
{\it cumulative} avalanche area exponent $K_{cum}$ for the $M=0.0$ case.}
\end{center}
\end{figure}

\subsubsection{Durations, moments, and roughness}

\begin{figure}[h]
\begin{center}
\epsfxsize=8cm
\epsfysize=8cm
\epsfbox{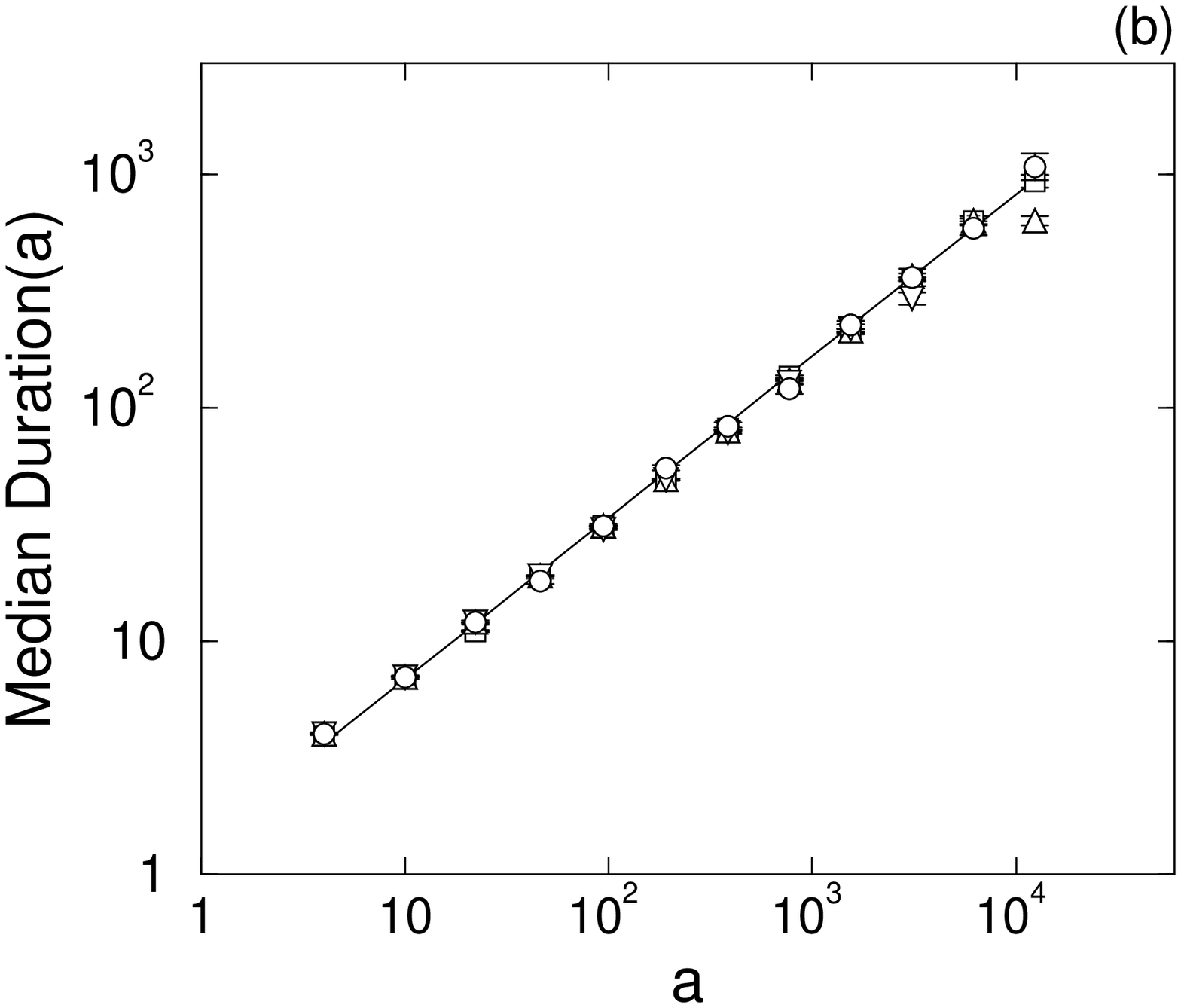}
\caption{\label {ameandurplot} \label {ameddurplot}(a) Log-log plot for the mean duration of  avalanches whose area is in the interval $[a/\sqrt{2},\sqrt{2}a)$. All data are for $\fpm=0.5$, $L=128$. The $M>0$ curves have 
been shifted along vertical axis for clarity by multiplying by successive factors of $3/2$. 
 The asymptotic  slopes should be $z/d$.  
Linear fits for both $M=0.0$ and $M=0.2$ (not shown) yield $z=1.40\pm 0.02$  .The 
dashed line is a non-linear fit  that includes 
corrections to scaling; for $M=0.0$ it yields  
$z=1.52 \pm 0.02$ with correction amplitude $B_z=-0.58 
\pm 0.05$;; see text. Linear fits for the $M=0.4$, 
$M=0.6$, $M=0.8$, $M=1.0$ data yield values of $z/2$ of $0.69\pm 0.02$, $0.70\pm 0.03$, 
$0.69\pm0.01$ and $0.68\pm0.02$ respectively. 
  (b) Log-log plot for the {\it median} duration avalanches with  area
in the interval $[a/\sqrt{2},\sqrt{2}a)$.}
\end{center}
\end{figure}

From the avalanche data, specifically for the {\it durations} of avalanches, one can  determine the dynamic exponent $z$: with non-fractal avalanches, the duration $\tau$ will scale as $a^\frac{z}{d}$.
In Fig. \ref {ameandurplot}(a) we plot the mean duration of avalanches whose area is within a factor of $\sqrt{2}$ of $a$ as function of $a$.  The slope of the
log-log plot yields 
\be
z(M=0)\approx 1.40\pm0.02 \ . 
\ee
As for the avalanche statistics, we can attempt to take into account corrections to scaling by fitting to the
form $\la \tau(a)\ra \approx \frac{Ca^{\frac{z}{2}}}{1+B_z a^{-1/3}}$. This three parameter fit yields $z=1.52\pm 0.02$ and  $B_z=-0.58\pm
0.05$; this value of $z$ is closer to the value of $\frac{14}{9}$ from a  naive extrapolation of the $4-\epsilon$ expansion, although $z$ is likely to have higher order corrections in $\epsilon$ even if $\zeta$ does not.  Ref. \cite{chauve} 
shows that $z=2-\frac{2}{9}\epsilon - 0.04321 \epsilon^2\approx 1.38$.  
The $z$ extracted from the avalanche dynamics agrees quite well with this 
value without any corrections to scaling.
  
The roughness
exponent $\zeta$ can be extracted from the {\it moment} as a function of the area: $m\sim a^\frac{d+\zeta}{d}$. Factor-of-two logarithmically binned histograms are shown in Fig. \ref{ameanmomplot}; for the dissipative case these appear to yield a rather 
small value of 
\be
\zeta_{aval}(M=0) \approx 0.44\pm0.02
\ee
however there is definitely upward  curvature observable at the larger sizes and a fit with $\zeta=\frac{2}{3}$ and an $a^{-\frac{1}{3}}$ correction to scaling (same number of parameters) is somewhat better and yields a correction to scaling amplitude of order unity, more precisely $-0.91$.  Note, however, that, as for the distribution of avalanche sizes, such fits suffer from finite size effects.
\vspace{-1cm}
\begin{figure}[h]
\begin{center}
\epsfxsize=8cm
\epsfysize=8cm
\epsfbox{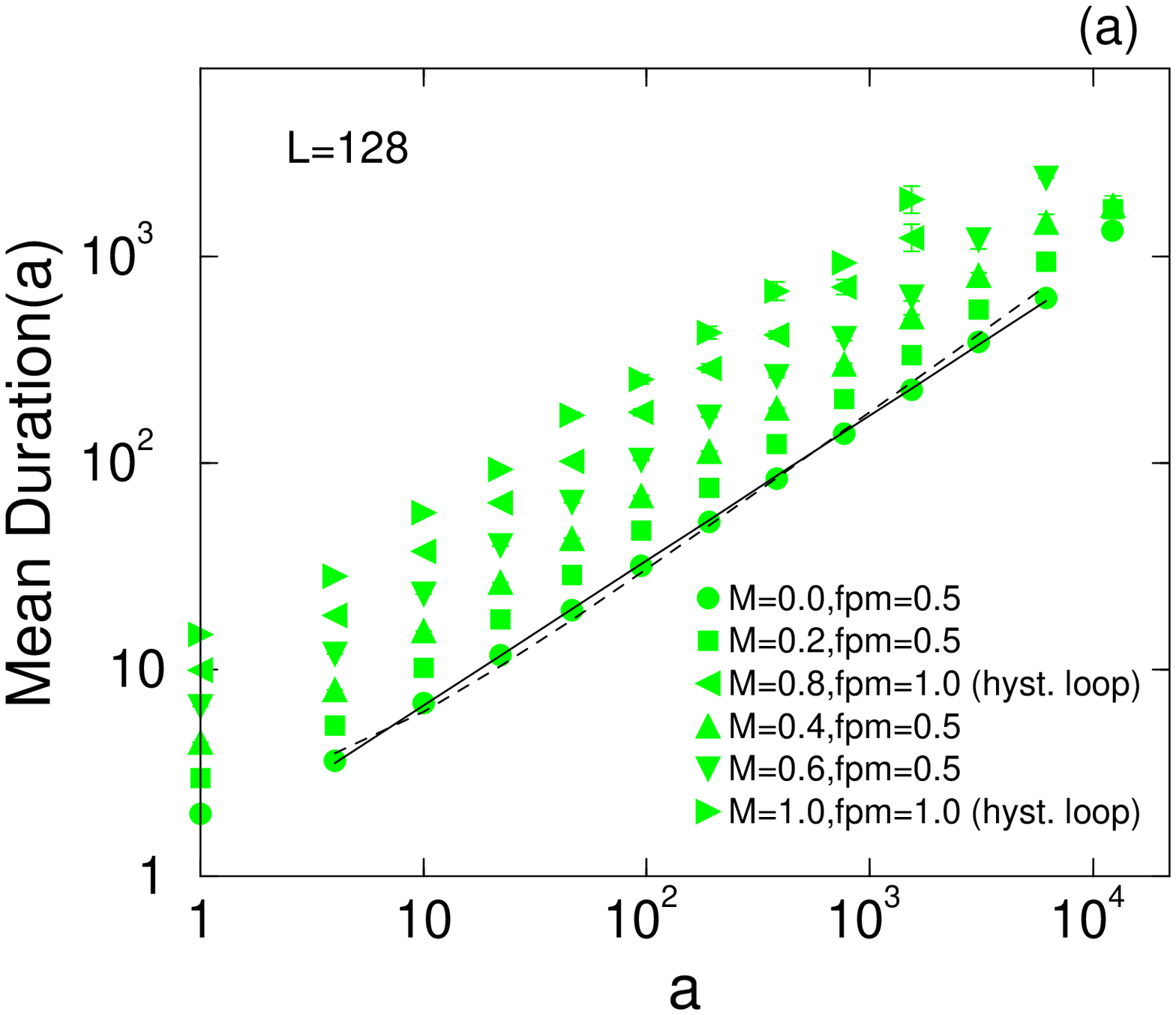}
\end{center}
\end{figure}
\vspace{-2cm}

\begin{figure}[h]
\begin{center}
\epsfxsize=8cm
\epsfysize=8cm
\epsfbox{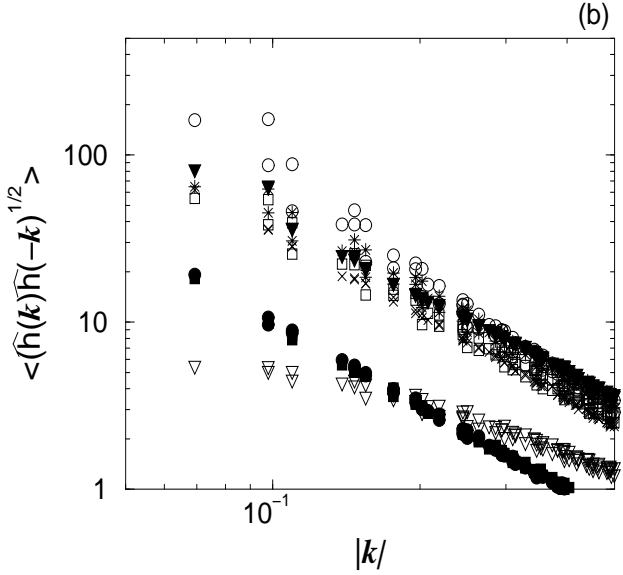}
\caption{\label {aspatialpowspecplot} (a)Log-log plot of the square root of the power spectrum contrasting the system just below the critical force on initial depinning, $\fa$,(open symbols); with its behavior when it has ``just started":immediately after every segment has moved during a
single avalanche event at $\fa$ (solid symbols).   
Each set of data represents an average over 1000 samples, and  the curves have been shifted along the vertical axis for 
clarity. The stars and
  ${\bf x}$'s denote $\fpm=1.0$, the other data $\fpm=0.5$.        
For the data just below $\fa$, the fitted roughness exponents, $\zeta$,  are indicated by the uppermost solid lines: with $\fpm=0.5$:  $0.71\pm0.05$ for $M=0.0$, 
$0.55\pm0.04$ for $M=0.2$, and $0.02\pm0.02$ for $M=0.6$; with $\fpm=1$:  $0.57\pm0.04$ for $M=0.8$ .  The lower solid curves are fits with corrections 
to scaling of the form $c_1 k^{-(1+\frac{2}{3})}(1+B_{\zeta}
k^{2/3})$, with  fitting parameters
$c_1$ and $B_{\zeta}=-0.006\pm0.002$ for $M=0.0$, $0.034\pm0.007$ for $M=0.2$, and $0.283\pm0.009$ for $M=0.8$. 
For the just started data, with $\fpm=0.5$, inferred $\zeta$'s are $0.65
\pm 0.02$ for $M=0.0$, $0.65 \pm 0.02$ for $M=0.2$, and $0.69 
\pm 0.02$ for $M=0.5$ (not shown); and  $0.72\pm0.03$ for $\fpm=1.0$, $M=0.8$. No fit was attempted for $M=0.6$, $\fpm=0.5$.  
The error bars are achieved in the same way as described in Fig. \ref {spatialpowspecplot}(a) and the data has been thinned out as in Fig. \ref {spatialpowspecplot} (a) as well.
(b) Same as  (a) for comparison of amplitudes: the curves have not been shifted and a narrower range of $|k|$ is shown.   }
\end{center}
\end{figure}

A roughness exponent can be more directly extracted from the statistics of the  roughness of the manifold  exactly at the critical force, $\fa$.  In Figs. 
\ref{aspatialpowspecplot}(a)(b), 
fits to the spatial power spectrum, $\langle |\hat{h}(k)|^2\rangle$, 
yield $\zeta(M=0)\approx 0.65\pm0.02$ and  $0.69\pm0.02$ for $\fpm=0.5$ and $\fpm=1.0$ respectively.  These 
values are close to those observed for systems that have stopped after a decrease of  $F$ from the moving phase.  

The apparent discrepancy between the scaling of the typical displacement of avalanches and the roughness at the critical point is somewhat troubling, although, as we have seen, it can readily be accounted for by corrections to scaling.  This sort of discrepancy has been seen previously in simulations of manifold depinning \cite{martys}.  In that case, the roughness exponent at the critical force was similar to ours, but the displacement of the avalanches scaled with a {\it larger} exponent.  The fact that the  discrepancy can be in either direction supports the belief that it is due to corrections to scaling rather than a difference between the two exponents, as had been conjectured in  
\cite{martys}.

\begin{figure}[h]
\begin{center}
\epsfxsize=8cm
\epsfysize=8cm
\epsfbox{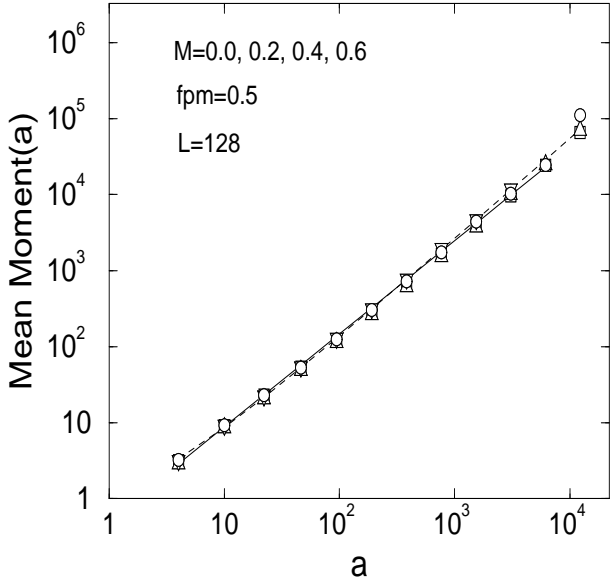}
\caption{\label {ameanmomplot} Log-log plot of the mean moment of avalanches with  area
in the interval $[a/\sqrt{2},\sqrt{2}a)$ for $L=128$ and
$\fpm=0.5$ using same symbols as in Fig. \ref{pcum0.5plot}(a).  
The expected slope is $(d+\zeta)/d$. 
A linear fit (solid line) for $M=0.0$  yields $1.22\pm-0.02$ 
For other 
$M$, the same exponent fits to within one standard deviation.  The dashed line is a fit  including corrections to scaling:  $C a^{\frac{\zeta+d}{2}}
(1+A_{\zeta} a^{-\frac{1}{3}})$ with $C$ and  $B_{\zeta}=-0.91\pm 0.02$ fitting 
parameters.  }
\end{center}
\end{figure}

As was done for the critical force defined from the moving phase, we can also extract the {\it correlation length exponent} from the finite-size scaling of the variance of the critical force.  In Fig. \ref{avarfcupplot}, the data are shown and exponent estimates  of $\frac{1}{\nu(M=0)}\approx 1.22\pm0.04$ and $1.28\pm0.02$ extracted from the data for, respectively, $\fpm=0.5$ and $\fpm=1.0$.  These yield 
\be
\nu(M=0)\approx 0.82\pm0.04 \  
\ee
for $\fpm=0.5$ and $\nu(M=0) = 0.78\pm 0.02$ for the stronger pinning.  
The scaling law $\zeta +\frac{1}{\nu}=2$ is consistent with the inferred exponents at the level of a few times the apparent error bars.

In spite of the discrepancies noted above for the typical displacement of avalanches and their cumulative statistics, overall the data for avalanches in the dissipative limit appear  to be consistent with those obtained from the moving phase and with theoretical predictions.   In particular, if one includes corrections to scaling of the anticipated form, all the data is consistent.

\subsection{Avalanches with stress overshoots}

We now turn to avalanche properties for $M>0$.  Because the distribution of avalanche sizes will turn out to raise the most questions, we first study the duration versus area and moment versus area.  

\begin{figure}[h]
\begin{center}
\epsfxsize=8cm
\epsfysize=8cm
\epsfbox{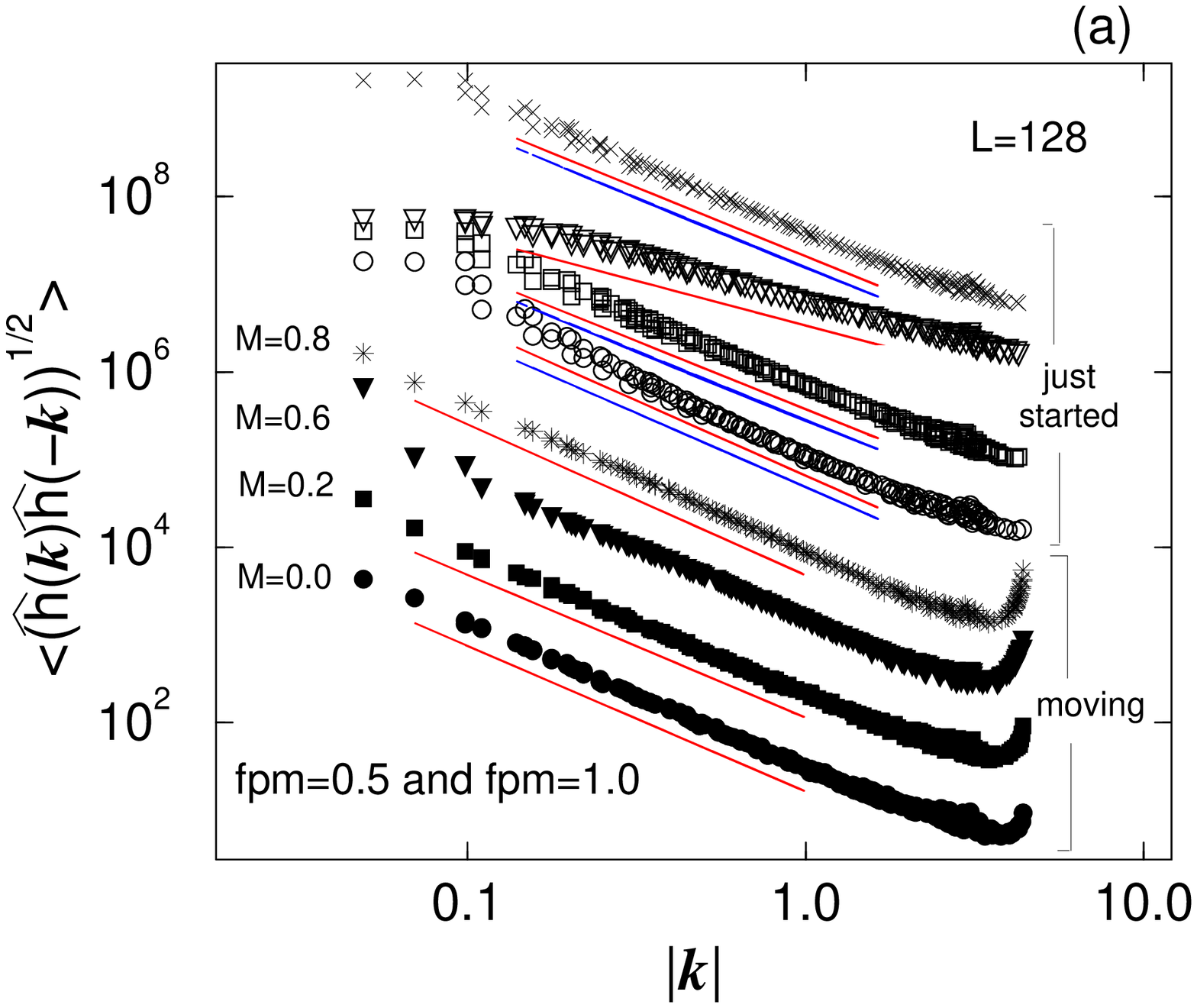}
\end{center}
\end{figure}
\vspace{-2cm}
\subsubsection{Durations and moments}

In Figs. \ref{ameandurplot}(a) and (b), the mean and median {\it durations} of avalanches whose area is within a factor of $\sqrt{2}$ of $a$ are plotted for various $M$.  Except possibly at the largest sizes, the data look remarkably similar to the data for $M=0$ and an exponent of $z\approx 1.4$ would be inferred from each. 
In Fig. \ref{ameanmomplot} the mean {\it moments} are plotted; again there seems to be remarkably little dependence on $M$.   For both the duration and the moment data, not only do the {\it exponents} look similar to those at $M=0$, but the {\it amplitudes} do also.  Given the relatively strong dependence of the velocity amplitude on $M$ (Sec. IV), this is particularly surprising.  

\begin{figure}[h]
\begin{center}
\epsfxsize=8cm
\epsfysize=8cm
\epsfbox{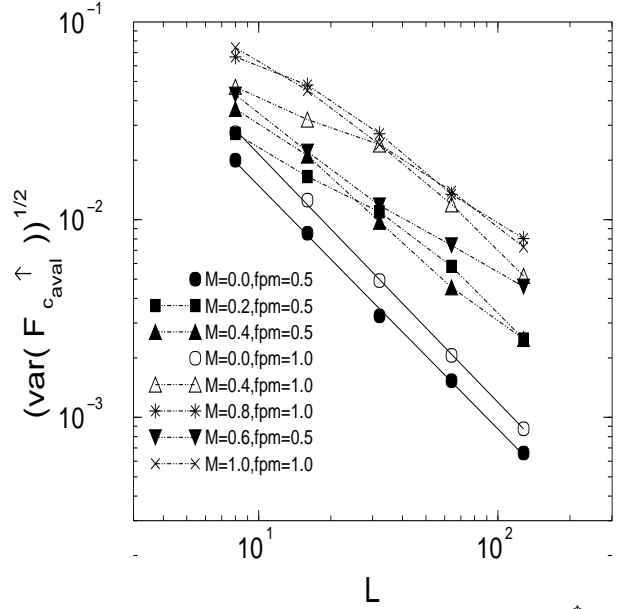}
\caption{\label {avarfcupplot}Log-log plot of  the rms variations of $F_{c_{aval}}
^{\uparrow}(M)$
as a function of the system length $L$. Filled symbols denote $\fpm=0.5$ and
 open symbols $\fpm=1.0$.  The stars represent $M=0.8$, 
$\fpm=1.0$.  Other symbols  are as in previous figures. The slopes of the two $M=0.0$ curves are $-1.22 \pm 0.04$ (filled) 
and $-1.28 \pm 0.02$ (open), shown by solid lines. For $M>0$, crossover is evident and he dotted lines are merely guides for eye.}
\end{center}
\end{figure}
An amplitude that would be expected to be proportional to the time-length scaling factor, $A_t$,  can be extracted from the mean duration of avalanches as a function of their area.  Although these would be expected to  differ by an unknown numerical factor from those extracted from  measurements in the moving phase, the ratios might have been expected to be universal.  The amplitude 
$A_t^{\rm aval}\approx 1.33$ for $M<0.4$ and $A_t^{\rm aval}\approx 1.44$ 
for $M=0.6, \fpm=0.5$. It is apparent that these amplitudes exhibit substantially less dependence on parameters
than do the  $A_t$ inferred from the moving phase.  Although at first this seems troubling, it should, perhaps, not be:  Even in the absence of overshoots, the properties of avalanches can depend on how the system is prepared.  Nevertheless, the differences here are probably related to some of the more puzzling differences that occur in  distributions of avalanche areas and distributions of critical forces; these we discuss further below.

\begin{figure}[h]
\begin{center}
\epsfxsize=8cm
\epsfysize=8cm
\epsfbox{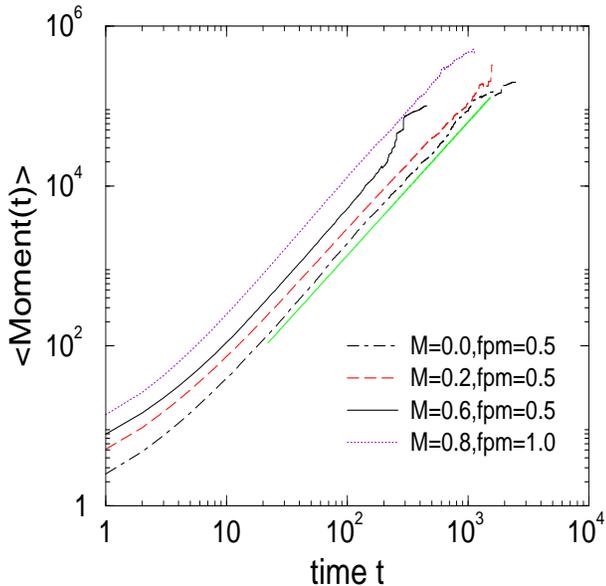}
\caption{\label {ameanmomplot} Log-log plot of the time evolution of the 
average moment of avalanches initiated between 
$F_c^{\downarrow}(M)$ and $\fa(M)$ (within a small 
force interval).  The 
normalization factor is the number of avalanches that still 
moving at the time $t$.  The light solid line displaced slightly 
below the $M=0.0$ curve 
is the result of a least squares fit to a power law whose exponent 
yielding an exponent of $1.63\pm0.01$ in the scaling regime.  In the 
quasistatic scaling 
regime, the exponent should be $(\zeta+d)/z$.  For all data, $L=128$. 
For clarity, $<{\rm Moment}(t)>$ for $M=0.2$ has been multiplied by $2$; 
for $M=0.6$, the multiplicative factor is $4$, and for the $M=0.8$ 
data, it is $6$. } 
\end{center}
\end{figure}

\subsubsection{Distributions of avalanche area}

In contrast to the properties of avalanches of a given area, the {\it distributions} of avalanche sizes depend strongly on $M$. In Figs. \ref{pcum0.5plot} (a) 
and \ref{pcum1.0plot} (a)  
the cumulative distributions of all avalanches are shown for $\fpm=0.5$ and  $\fpm=1.0$ respectively: the data are straight on a log-log plot over a substantial range of areas but the slope appears to vary continuously with $M$, for $\fpm=0.5$ from $0.89\pm0.02$ for $M=0$ to $1.53\pm0.02$ for $M=0.6$ and for $\fpm=1.0$,   from $0.90\pm0.01$ for $M=0$ to  $1.14\pm0.02$ for $M=1.0$.  At large areas, of order 10\% or so of the system area, the distributions of avalanches falls off increasingly more rapidly as $M$  increases.  Whether this is an intrinsic effect or a finite size effect is an important issue to understand.   

Data taken for stronger pinning, $\fpm=1.0$, are shown in Fig. \ref{pcum1.0plot}(a).  Surprisingly, these cumulative data depend far less on $M$ over the range shown.  Although the apparent exponent varies slightly, all are probably consistent with $K_{\rm cum}=1$ with corrections to scaling.  More strikingly, there is no dropoff seen for large sizes
even with $M=1.0$.

The dependence of the avalanche distributions on stress overshoot are more marked if we restrict consideration to avalanches that occur for forces {\it above} $\fd$, Fig. \ref{p0.5plot} and \ref{p1.0plot}. As mentioned earlier, for $M=0$ these yield an exponent, expected to be equal to $\kappa/d=\zeta/2$, of $0.38\pm.03$. But 
as $M$ increases the distributions appear to follow roughly the dissipative behavior for small avalanches only to fall substantially below it for large ones. This difference is  most pronounced for large $M$: the largest avalanches are about an order of magnitude rarer for $M=0.4$ than for $M=0$ for the weaker pinning and similarly for $M=1.0$ versus $M=0$ for the stronger pinning.  In contrast to the data for the cumulative numbers of avalanches, these data are suggestive of typical crossover behavior: from dissipative behavior for small sizes to something else for large sizes, with a crossover length that is long for small $M$ and shrinks as $M$ grows.  Note that this form of $M$ dependence of a crossover length is what would be expected if $M$ were a {\it relevant} perturbation about the dissipative depinning fixed point. 
But we must exercise caution before drawing such a conclusion.

\begin{figure}[h]
\begin{center}
\epsfxsize=8cm
\epsfysize=8cm
\epsfbox{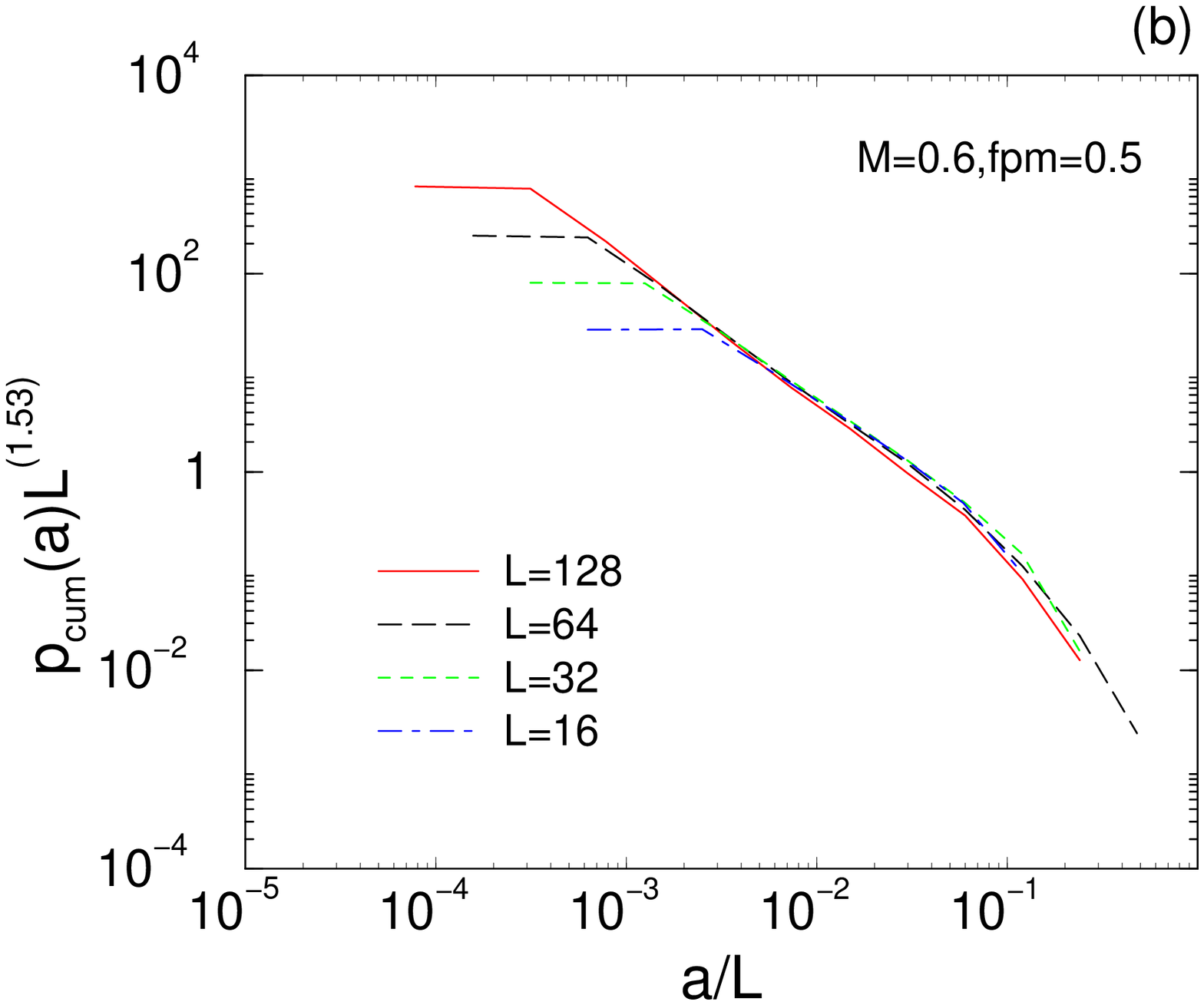}
\caption{\label {afinitesize1plot} \label {afinitesize2plot}(a) Finite-size scaling plot for cumulative avalanche area distribution during initial depinning. (b) Scaling plot of the same data with 
$\Upsilon=1/2$. }
\end{center}
\end{figure}

From the data of Fig. \ref {pcum0.5plot} for $\fpm=0.5,\ M=0.6$ it appears that there may be some interesting size dependence to the crossover: these data for $L=64$ fall somewhat below those for $L=128$ for large avalanches.  One can test whether the crossover is a finite size effect (in contrast to  one arising from putative relevance of $M$) by plotting the data versus the scaled area, $a/L^2$: this is done in Fig. \ref {afinitesize1plot}(a) 
for a wide range of system sizes with $M=0.6$.  It can be seen that the rollover at large areas does {\it not} scale simply with system size.  But neither is it consistent with a system-size independent crossover: there is much less than the expected factor of 64 difference in the crossover value of $a/L^2$ between $L=16$ and $L=64$.  A plausible intermediate conjecture  would be that the crossover scales as a power $1-\Upsilon$ of system area: this could arise from {\it dangerous irrelevancy} of $M$.  The plot in Fig. \ref {afinitesize2plot}(b)  versus $a/L$ appears 
roughly consistent with such a conjecture with $\Upsilon\approx 0.5$. But this requires the introduction of an extra scale and exponent into the interpretation of the data. Another interpretation, one that does not require such additional hypotheses. is that  systems with $\fpm=0.5,\ M=0,6$ are in the midst of some type of crossover for the range of system sizes investigated. As discussed later, we believe that this is probably the case.

\subsubsection{Roughness from avalanches}
The absence of large avalanches in the presence of substantial stress overshoots for $\fpm=0.5$ suggests that, right up to $\fa$, the manifold should be less rough on long length scales than in the dissipative limit.  This is indeed found to be the case: In Figures \ref{aspatialpowspecplot}(a) and (b), 
the spatial power spectrum of the roughness is shown as a function of wavevector just below $\fa$.  In the dissipative limit, a power law that is consistent with a roughness exponent of $\zeta=0.65\pm 0.02$ for $\fpm=0,5$ and $\zeta=0.71\pm0.05$ for $\fpm=1.0$  obtains down to the smallest wavevectors.  In contrast, for  $\fpm=0.5$, the spectra with stress overshoots are found to become flat at long length scales; this is particularly pronounced at the larger $M$.  
Although we have not studied the system size dependence of these spectra in detail, for $M=0.6$ the rough magnitude of the wavevector at which the crossover from power-law rough to flat occurs --- as seen in the blowups of the small $|{\bf k}|$ regime in Fig. \ref {aspatialpowspecplot}(b) --- is similar to the wavevector inferred from the inverse square-root of the system area at which the drop-off in the avalanche area distributions occurs.  But to again confuse the interpretation, different behavior is once again seen for the stronger pinning, $\fpm=1,\  M=0.8$, samples: these show apparent roughness exponents that are only slightly smaller than in the absence of overshoots and the roughness has less of a tendency to saturate at small wavevectors than the weaker pinning data.  

\begin{figure}[h]
\begin{center}
\epsfxsize=8cm
\epsfysize=8cm
\epsfbox{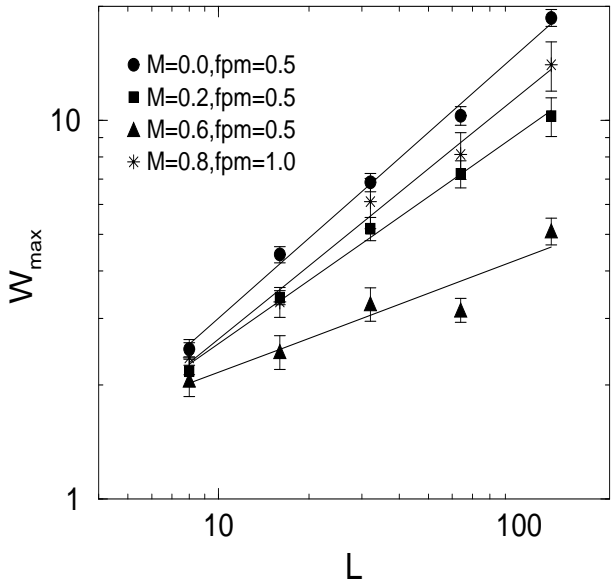}
\caption{\label {amaxwidplot}Log-log plot of the maximum width of the manifold just below the initial depinning at $\fa$. Data is averaged over 10 samples.  
For the $M=0.0$ curve, the slope is $0.70\pm0.03$; for the $M=0.2$ curve, 
it is $0.56=\pm0.02$.  For the $M=0.6$ curve, the result of the least 
squares fit is a slope of $0.29\pm0.06$.  And finally, for the $M=0.8$ 
curve, the slope is $0.64\pm0.06$.  }
\end{center}
\end{figure}

 Information can also be garnered from  the system size dependence of the average maximum width, $W_{\rm max}=<\max(|h(\bx)-\overline{h(\bx)})|>$, of the manifold 
at $\fa$. In the dissipative case, $W_{\rm max}\sim L^\zeta$ as expected for simple scaling. The stronger pinning samples with $\fpm=1.0,\ M=0.8$ show behavior for $W_{\rm max}$ that is relatively similar to the dissipative case, consistent with the spatial power spectra $\langle |\hat{h}(k)^2|\rangle$. For the samples with $\fpm=0.5,\ M=0.6$, however, much weaker size dependence is observed: Fig. \ref {amaxwidplot}.  We consider several possible reasons that this peculiar behavior for these parameter values might occur.

 If, as appears to be the case for samples with $\fpm=0.5,\ M=0.6$, the avalanche production were sharply cutoff  at a diameter of order $L^{1-\Upsilon}$ from its distribution in the dissipative case, the largest avalanches that occur would change $h$ locally by of order $L^{(1-\Upsilon)\zeta}$, yielding a $W_{\rm max}$ of this order.  If, instead,  the crossover were to a distribution decaying as a larger power-law of the area for large avalanches, the apparent roughness exponent extracted from $W_{\rm max}$  would be somewhat larger.  But if the largest avalanches were qualitatively different from dissipative ones ---- perhaps like failed nucleation bubbles with $W\sim L$  --- this could lead to an {\it increased} apparent roughness exponent for $W_{\rm max}$.   Unfortunately, at this point, it is hard to conclude much from this set of data except that something peculiar seems to be going on for $\fpm=0.5,\ M=0.6$.  But peculiarity in this range of parameters appears also in other quantities.

\begin{figure}[h]
\begin{center}
\epsfxsize=8cm
\epsfysize=8cm
\epsfbox{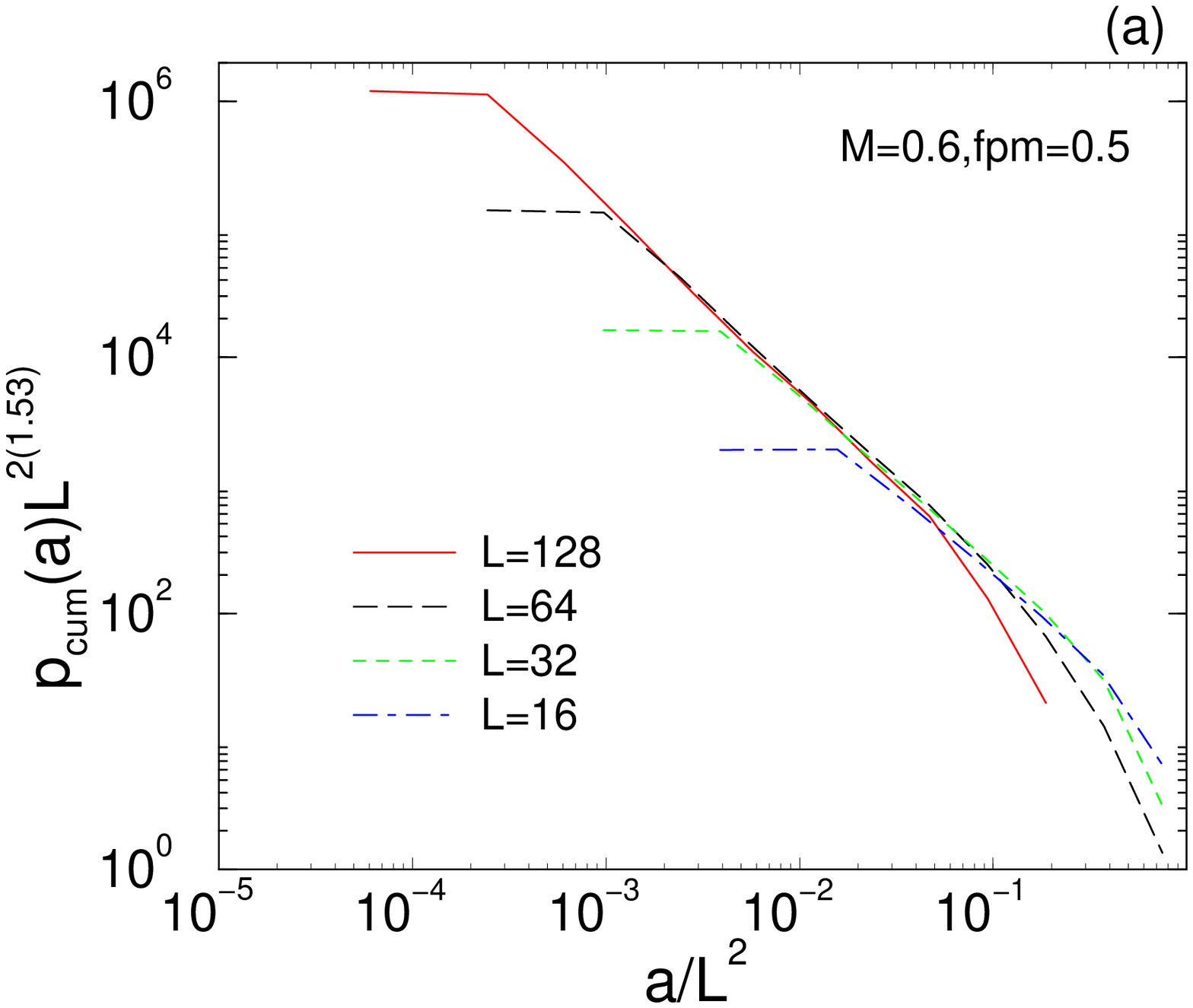}
\end{center}
\end{figure}

\subsubsection{History dependence of roughness}

The data at $\fa$ for both the spatial power spectrum of the roughness and for $W_{\rm max}$ should be contrasted with those in the steady state moving phase , Fig. \ref {spatialpowspecplot}(a), and with those in the ``just started" moving state that has not reached steady state:  Fig. \ref {aspatialpowspecplot}(a).   The latter two  both show power-law roughness with exponents that are consistent with being universal {\it independent} of $M$ and independent of whether the manifold has been slowly stopped from the moving phase or has barely started moving on initial depinning --- albeit with a velocity dependent cutoff at long scales in the latter case.  In contrast, the data just below $F_{c_{aval}}
^{\uparrow}$  exhibit roughness that depends substantially on $M$.  

The trends with changing $M$ are also different for the stopped samples (at $\fd$) and the samples at initial depinning ($\fa$): When the motion has been stopped from the moving phase, the roughness is slightly larger for larger $M$ at 
least for the weaker pinning.  
In contrast, the long length scale roughness in systems that started flat and had $F$ increased to $\fa$ tends to {\it decrease} with increasing $M$ because of the rarity of large system-roughening avalanches  with this history. 

\subsubsection{Variations of $\fa$}

We have seen that there appear to be interesting differences between the physics of avalanches in the dissipative limit and those with stress overshoots.  In the dissipative case, the critical force, $\fa$, defined from the first system-sized avalanche, and that, $\fd$, defined by when steady-state motion ceases, are essentially the same; the latter being slightly larger  and with a somewhat smaller variance because of the way it is defined (the system stops in a somewhat anomalously strongly pinned region).  See Table III.  With substantial stress overshoots, in contrast, while the lowest force at which a finite size system can have a system-sized avalanche is roughly $\fd(M)$, the {\it distribution} of $\fa$ extends far further up, into the region in which, once the motion starts, it is extremely unlikely to stop again. From Fig. \ref {avarfcupplot}. it is seen that 
for $\fpm=0.5$ and $L=128$,  the rms variations of $\fa$ are about a factor of four larger  for $M=0.4$  than those for $M=0$, and for $\fpm=1.0$, they are a factor of ten larger for $M=0.8$ than for $M=0$,  even though the corresponding rms variations of $\fd$ hardly differ.  This factor of ten difference in $F=\fd$ corresponds crudely to a factor of $10^\nu\approx 6$ difference in length scale suggesting that some scale of order a sixth of the system size might appear for the initial depinning history with $\fpm=1.0,\ M=0.8$.  Recall that such a scale was also apparent in the distribution of the force, $\fh$, at which the system restarts after being stopped from the moving phase.

\begin{table}
\caption{Comparison of $F^{\downarrow}_c(M)$ and $F^{\uparrow}_{c_{aval}}(M)$
for $L=128$.}
\begin{tabular}{cccccc}   
$M$ &$f_p^{max}$ &$\langle F^{\downarrow}_c\rangle$  
&$\langle F^{\uparrow}_{c_{aval}}\rangle$ 
&$({\rm var}(F^{\downarrow}_c(M)))^{1/2}$ 
&$({\rm var}(F^{\uparrow}_{c_{aval}}(M)))^{1/2}$ \\ \hline

$0.0$  &$0.5$ &$-0.0779$    &$-0.0785$   &$0.0004$ &$0.0006$  \\
$0.1$  &$0.5$ &$-0.1087$    &$-0.106$    &$0.0004$ &$0.003$  \\
$0.2$  &$0.5$ &$-0.1404$    &$-0.137$    &$0.0004$ &$0.003$  \\
$0.4$  &$0.5$ &$-0.2078$    &$-0.199$    &$0.0005$ &$0.002$  \\
$0.6$  &$0.5$ &$-0.2866$    &$-0.267$    &$0.0004$ &$0.005$   \\
$0.8$  &$1.0$ &$-0.0665$    &$-0.060$    &$0.0006$ &$0.009$   \\
$1.0$  &$1.0$ &$-0.1513$    &$-0.141$    &$0.0006$ &$0.007$  \\
\end{tabular}
\end{table}

\begin{figure}[h]
\begin{center}
\epsfxsize=8cm
\epsfysize=8cm
\epsfbox{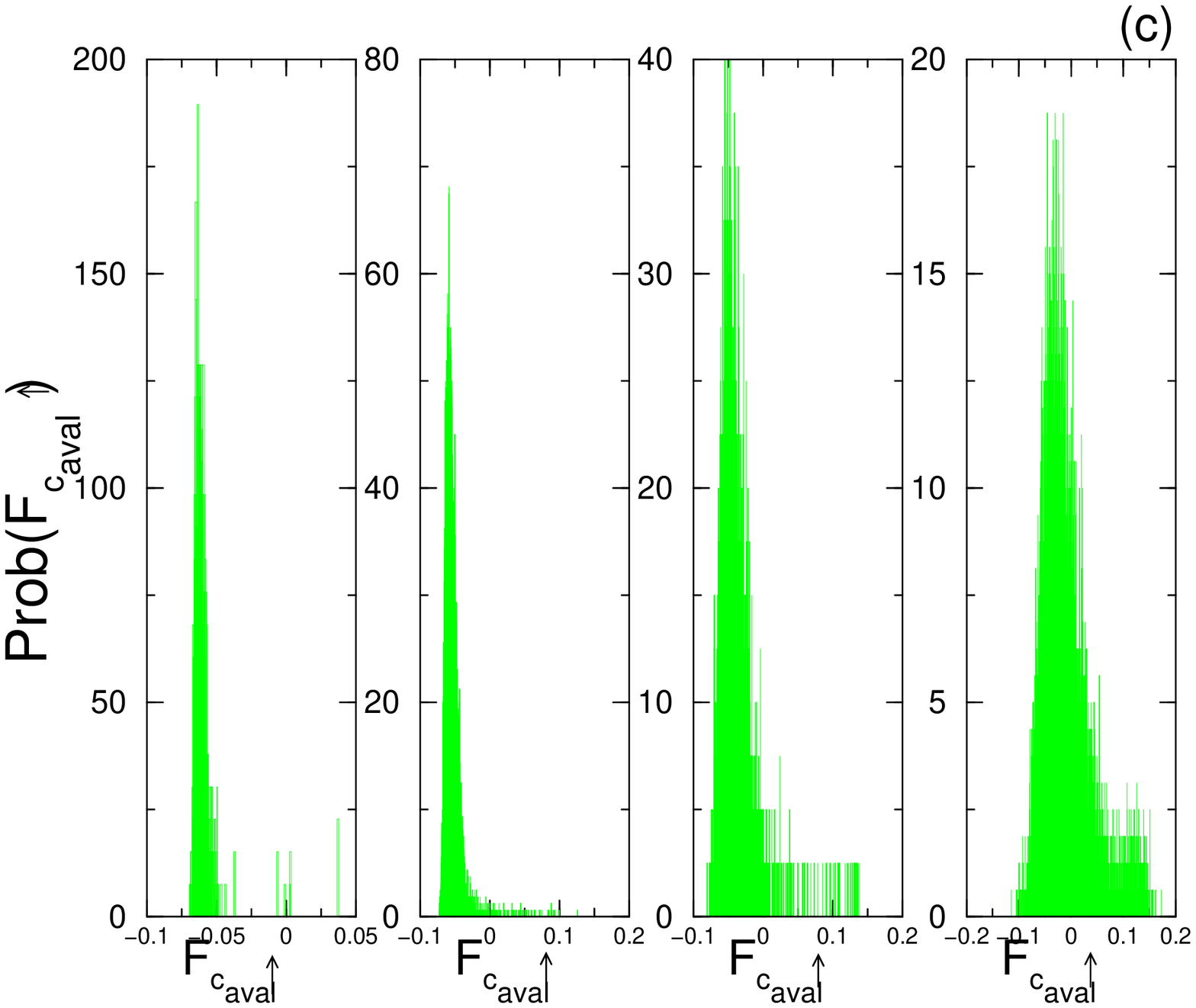}
\caption{\label {afcupplot}  Distributions of $\fa$ with, from left to right in each plot, $L=128,64,32,16$. (a)  $M=0.2,\fpm=0.5$. (b) 
$M=0.6,\fpm=0.5$.(c)$M=0.8,\fpm=1.0$.}
\end{center}
\end{figure}

The data for the rms variations of $\fa$ are shown in Fig. \ref {avarfcupplot} and histograms of the distributions in Figs. \ref {afcupplot}(a),(b),(c). 
Note the strong deviations from simple power law behavior of the {\it intermediate} $M$ data for the variances.  Surprisingly, the large $M$ samples with $\fpm=1.0$ have size dependence of the variance of $\fa$ that is, except for the overall amplitude, quite like that of the dissipative limit.  This again recalls the puzzling  behavior that was seen for the distribution of avalanche areas: the strong pinning large $M$ systems appear more similar to the dissipative limit than do intermediate $M$ ones with weak pinning.
\vspace{-1cm}
\begin{figure}[h]
\begin{center}
\epsfxsize=8cm
\epsfysize=8cm
\epsfbox{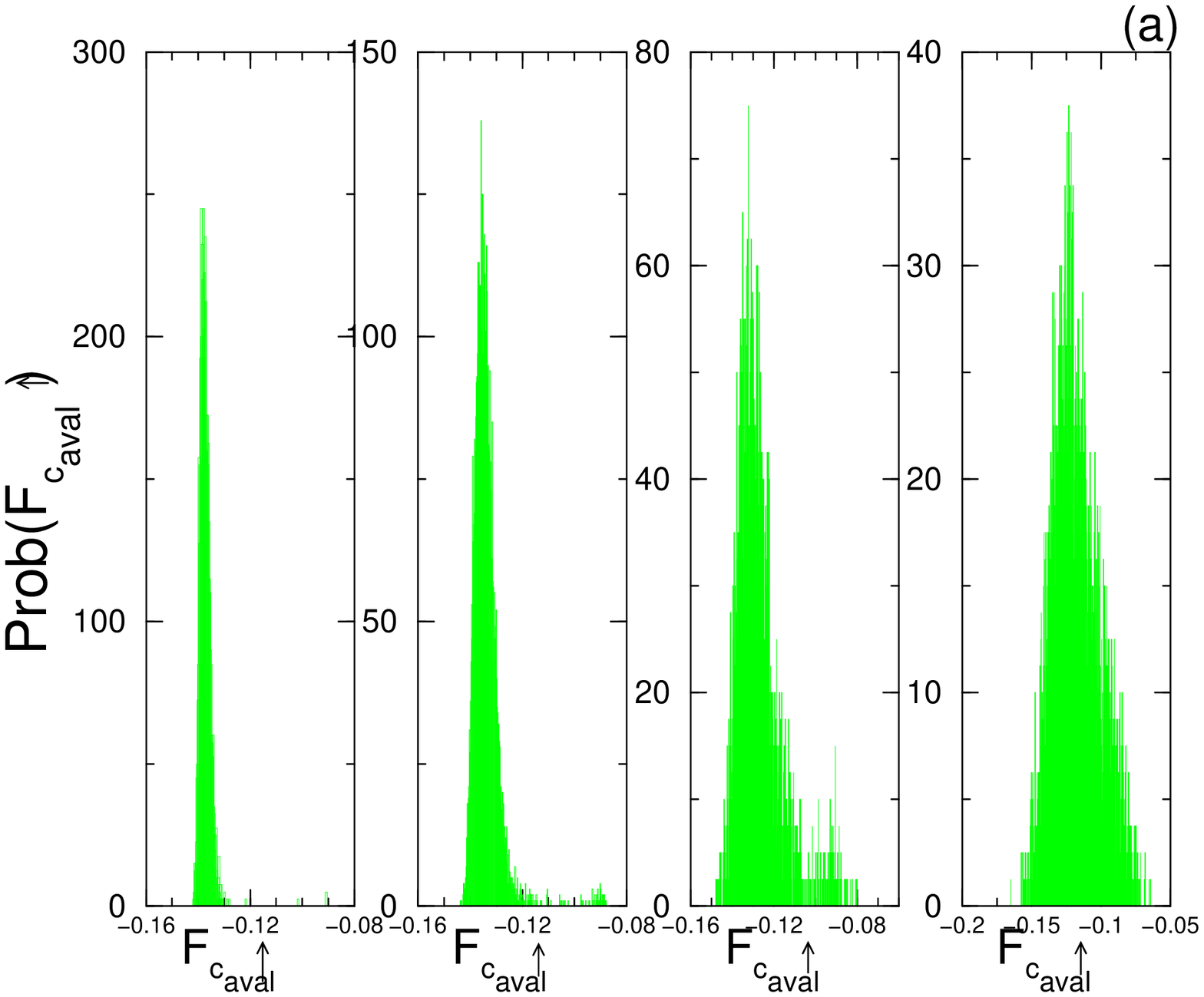}

\end{center}
\end{figure}
\vspace{-2cm}
\begin{figure}[h]
\begin{center}
\epsfxsize=8cm
\epsfysize=8cm
\epsfbox{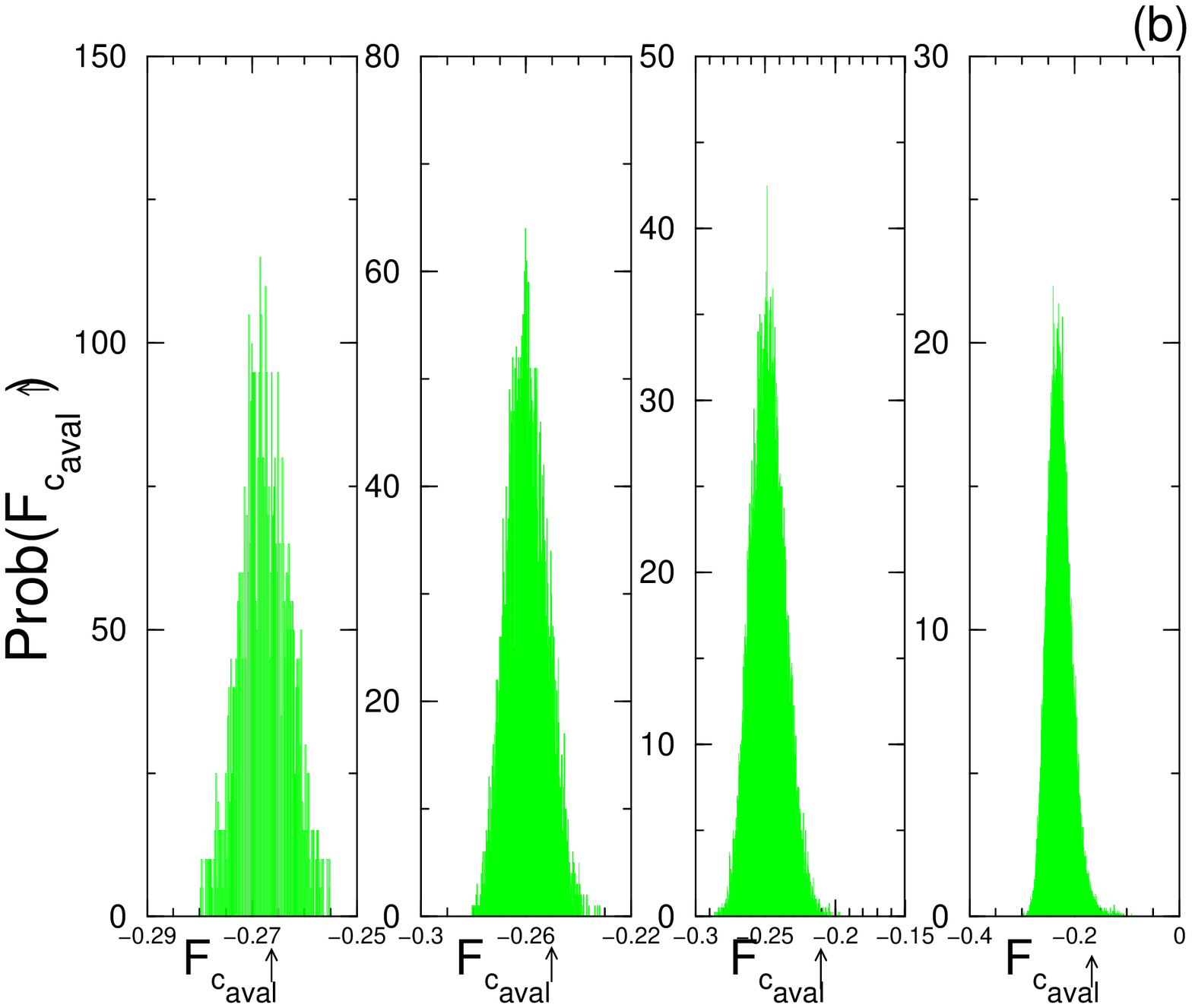}
\end{center}
\end{figure}
\vspace{-2cm}

\subsubsection{Crossover behavior}

The data we have presented on the statistics of avalanches and the roughness they induce show complicated dependence on the system size, on the magnitude  of the  overshoots, and on the strength of the random pinning.  But collectively they do suggest a plausible scenario involving a subtle crossover as a function of length scale:  For small $M$ one would expect dissipative-like behavior for small samples as the effects of small overshoots will need a substantial range of length scales to build up.  At the other extreme, the data for large $M$ and strong pinning suggest that very large samples show behavior characterized by the {\it same} exponents as the dissipative limit, but with very different amplitudes.  In between, at an $M$ dependent scale,  there must then be a crossover from one to the other amplitude.  On, for example,  a log-log plot of the rms variations of $\fa$ versus system size this would show up as a regime of slope $\frac{1}{\nu}$, followed by a crossover regime of lower slope to an asymptotic regime of slope again $\frac{1}{\nu}$.  The intermediate $M$ samples could then appear to have substantially different scaling by virtue of being in the crossover regime over the range of length scales studied,
In the next subsection we see how such crossover behavior might arise from the production and development of avalanches and the effects each has on subsequent
ones.

\subsubsection{Avalanche production and development}

We have seen that with stress overshoots, static configurations exist in finite size samples at forces that are substantially above (on the scale of variations of $\fd(L)$) the force, $\fd$,  at which the whole system can move. This suggests that in this regime there may be a change in character of the avalanches as they become large.  In contrast to below $\fd$ where all avalanches will stop, if an avalanche above $\fd$ becomes larger than some characteristic size, naively, perhaps the velocity correlation length, $\xi_v(F)$, in the moving phase,  it will runaway and the whole system will move.  This will strongly affect the distribution of avalanche sizes in this regime, yet it might not have much effect on the durations or moments of those that do {\it not} runaway: the evidence discussed above is that the durations and moments of these are indeed rather similar to the large ones that occur with $M=0$.   

The rates, $r(F)$, of avalanche production as $F$ is increased adiabatically are shown in Figs. \ref {afcupplot}(a),(b) and (c) for a variety of values of $M$ and $\fpm$.  These are normalized by the number of samples that are still pinned, i.e. still below their $\fa$ at the given $F$.
In contrast to the dissipative limit for which $r(F)$ increases sharply with $F$ as $F\to \fa\approx\fd$, the data for non-zero $M$ show a peak in the avalanche production rate somewhat below $\fd$ and then a marked decrease as $F$ increases into the region in which some samples have self sustaining avalanches and become depinned.  For the strongly pinned systems this suppression is strong; the behavior in the region in which $r(F)$ is small is shown blownup in the figure.  Note that in this regime the avalanche production rate is two orders of magnitude lower for $M=0.8$ and $M=1.0$ than for  $M=0$. 

Although we have not studied it in detail, this decrease in avalanche production rate is presumably related to the depletion of close-to-unstable sites in regions in which  moderately sized avalanches  have already occurred; this is  analogous to the depletion that occurs after the whole system is stopped from the moving phase.   Such depletion will suppress both the density of ``seeds" of avalanches and the probability of them becoming large.  Most new avalanches would be expected to occur in regions that have not yet had substantial sized avalanches.  The exception to this are avalanches that get up enough steam that they can run through the depleted regions. Once an avalanche does this, it will  continue growing until it sweeps through the whole system and becomes self sustaining.  As the chances of this happening will depend on the properties of various  regions of the system, how large it has to become to runaway is likely to be subtle.   

\clearpage
\begin{figure}[h]
\begin{center}
\epsfxsize=8cm
\epsfysize=8cm
\epsfbox{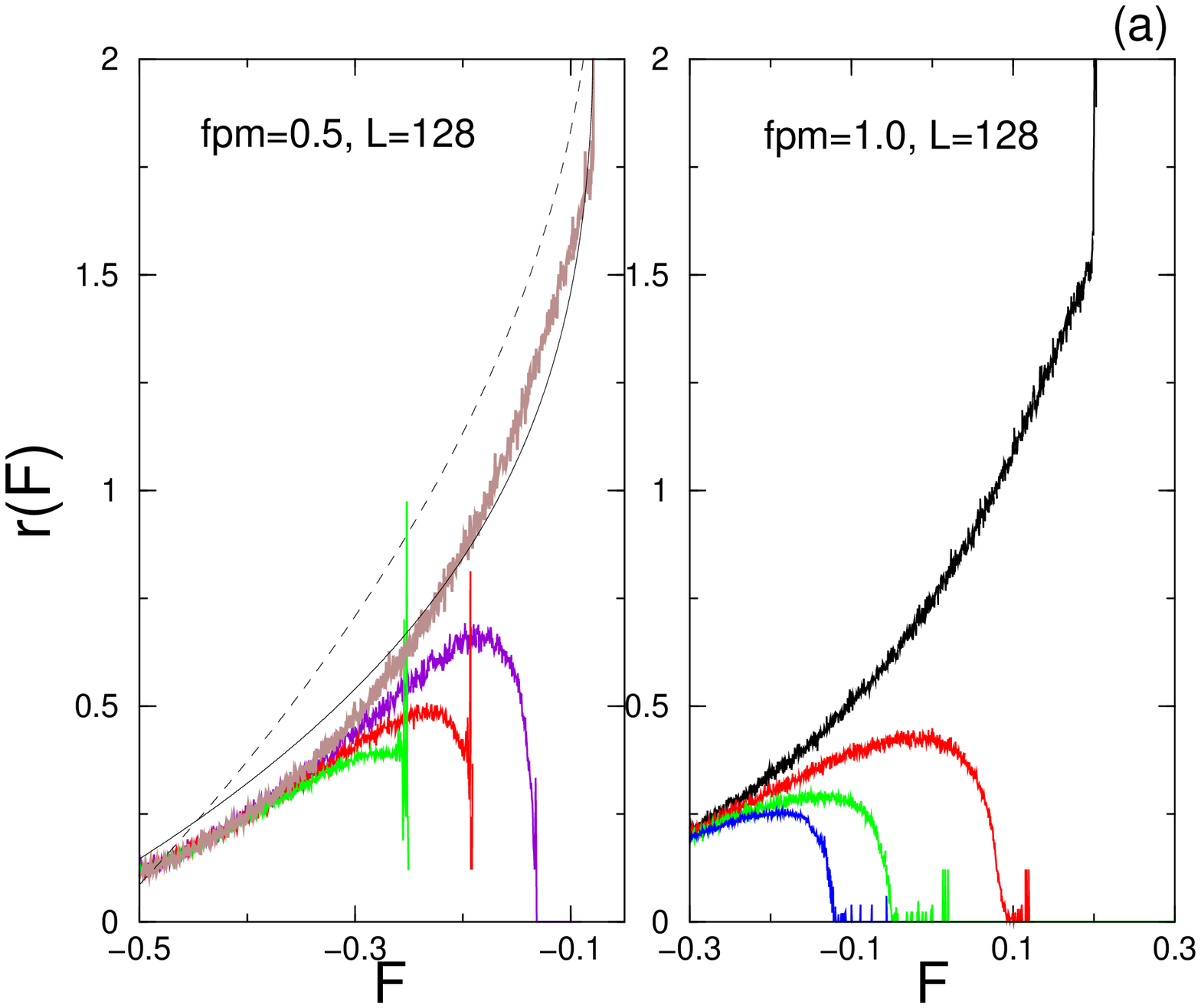}
\end{center}
\end{figure}

\begin{figure}[h]
\begin{center}
\epsfxsize=8cm
\epsfysize=8cm
\epsfbox{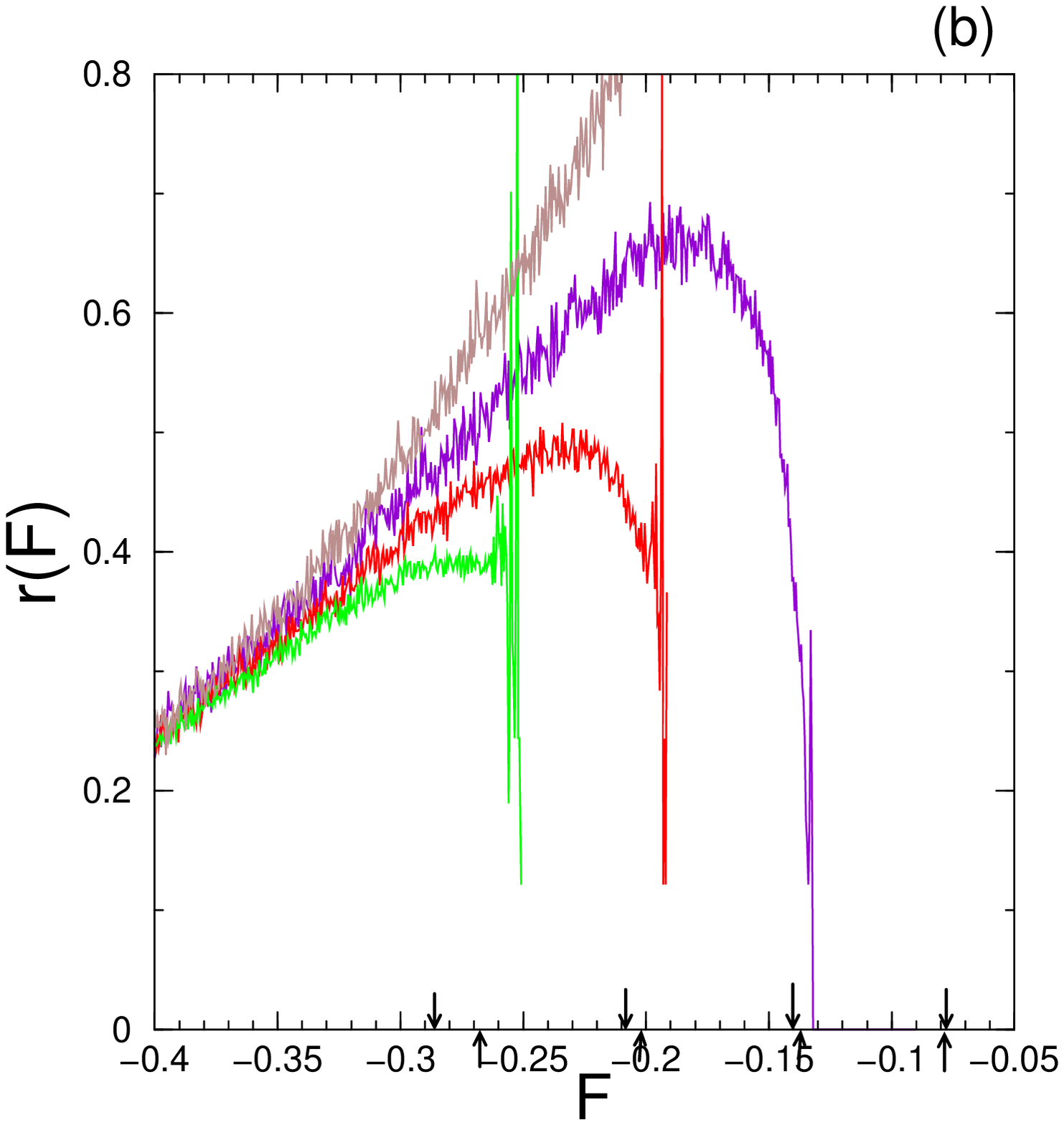}
\end{center}
\end{figure}

\begin{figure}[h]
\begin{center}
\epsfxsize=8cm
\epsfysize=8cm
\epsfbox{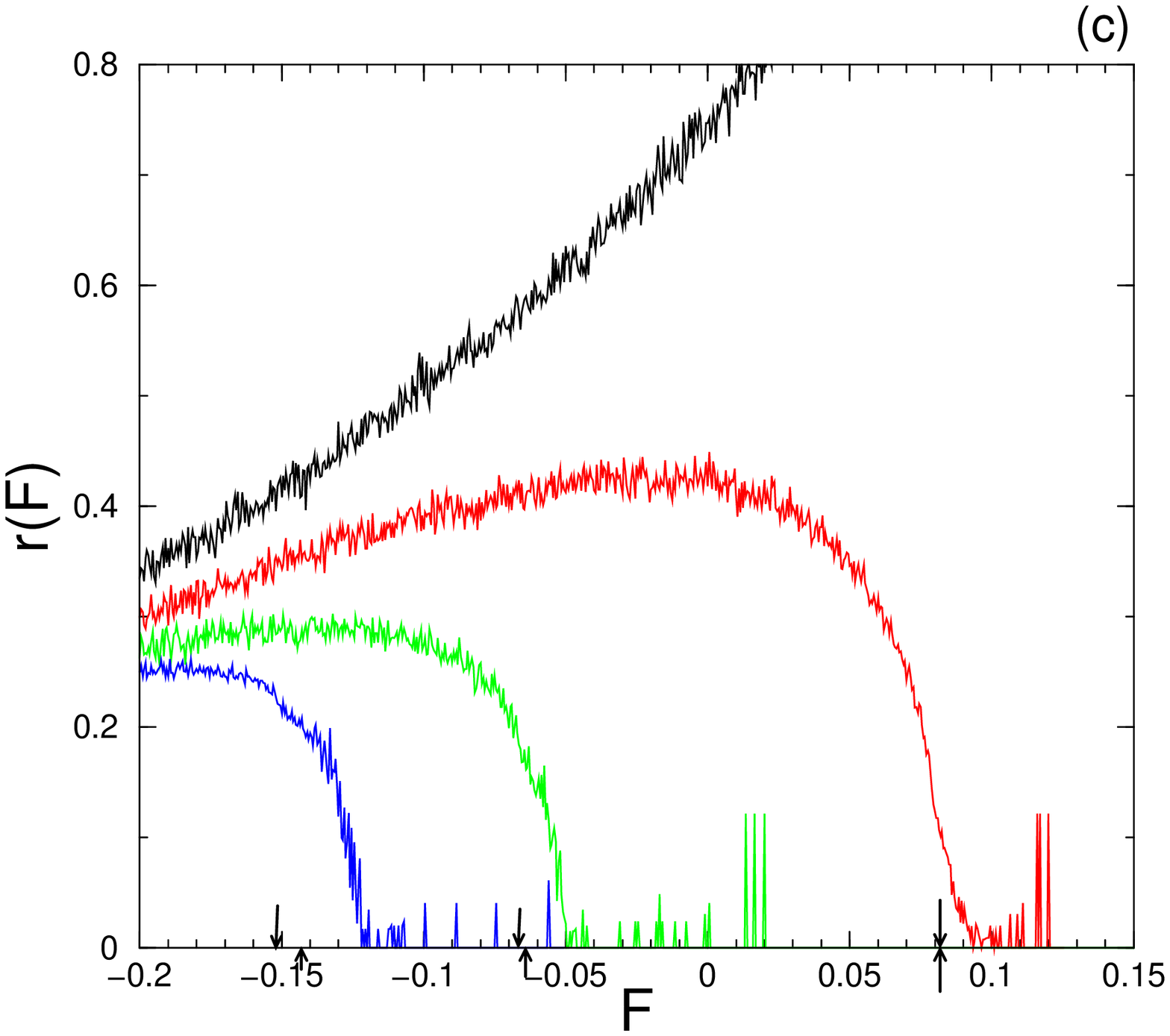}
\end{center}
\end{figure}
\begin{figure}[h]
\begin{center}
\epsfxsize=8cm
\epsfysize=8cm
\epsfbox{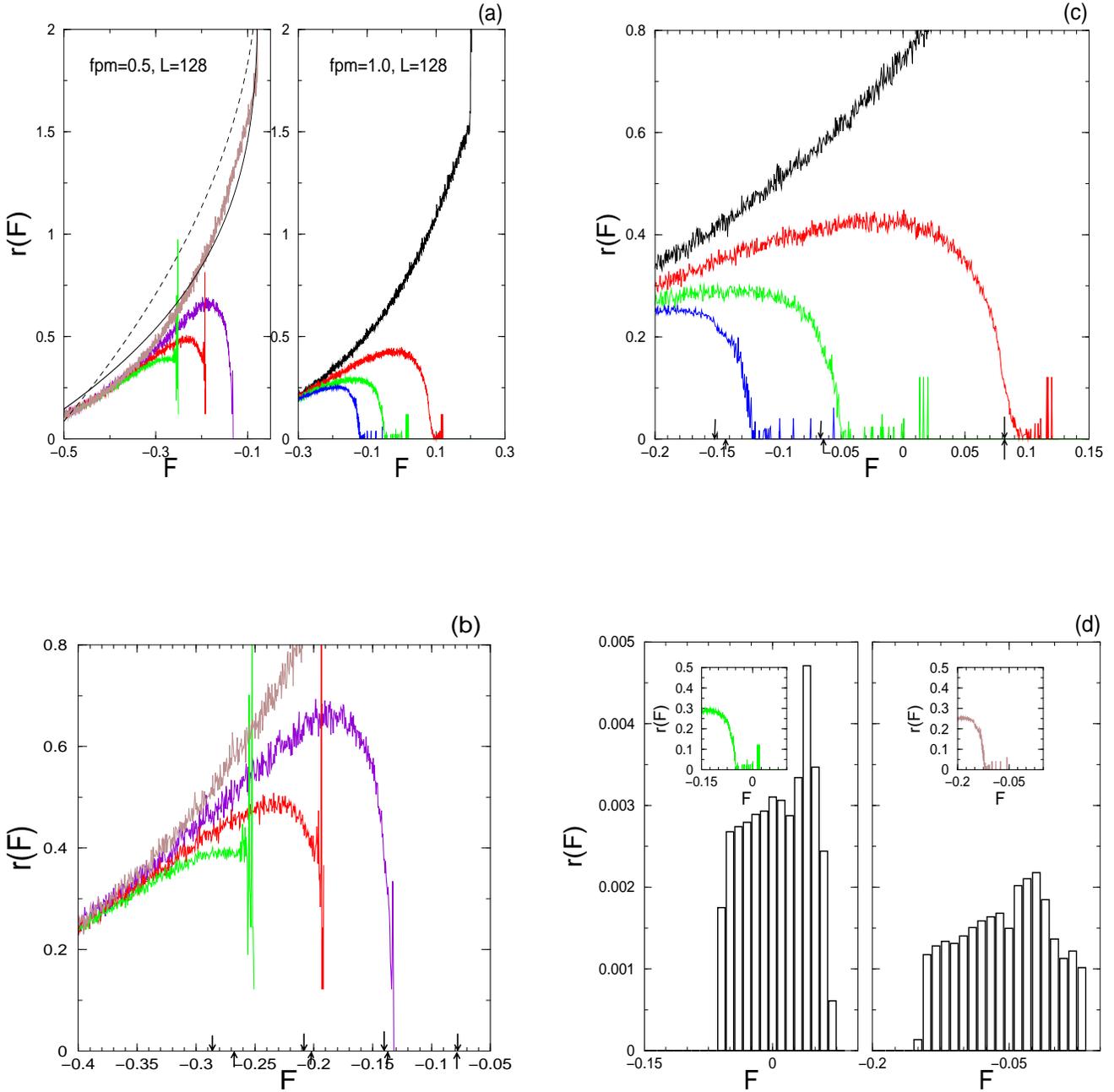}
\caption{\label {avalprodrateplot} (a) Avalanche production rate during initial depinning. From top to bottom of the left plot,  $\fpm=0.5$, are $M=0.0$, $M=0.2$, $M=0.4$ and $M=0.6$.   The solid line fit to the $M=0$ data is
  $2.35-3.3(-0.0785-F)^{0.30}$.   The dotted 
line is a fit in which  only $C_r$ (from Eq. (66)) 
is allowed to vary but $\alpha$ is 
fixed at $0.5$; note there is only one fitting parameter 
for the latter.  
 In the right plot, $\fpm=1.0$, top to bottom: $M=0.0$, $M=0.4$, $M=0.8$, $M=1.0$.  
  (b) Blow-up of (a) near the critical region for 
$\fpm=0.5$ data. Up arrows along the 
horizontal axis  indicate the median $\fa$, while down arrows indicate  $\langle\fd\rangle$. (c) Blow-up of (a) for $\fpm=1.0$ data.  (d) Comparison of $r(F)$ for 
initial depinning (insets) 
with that in 
the hysteresis loop. Note that for samples with  $L=128$ (shown) the hysteresis loops are considerably narrower than their naive width $M/Z$.    }
\end{center}
\end{figure}

There are three possible behaviors for the runaway process of large avalanches in the regime above $\fd$: (i) that there is a finite sample-size-independent length scale, $\xi_R(M)$  above which avalanches  typically run away, (ii) that the scale at which runaway occurs grows with system size, but as a decreasing {\it fraction} of system size, or (iii) that the runaway scale is proportional to system size in large samples.  In the dissipative limit, the last of these obtains with a proportionality constant close to unity.  In contrast, in the artificial model discussed earlier with overshoot stresses felt for the duration of an avalanche, once all segments have moved at least once, an increase of the force by $M/Z$ is needed to cause another avalanche and when one does occur it is likely to run away after its diameter  becomes of  order the velocity correlation length $\xi_v$ of the moving phase at that force.  Why this is the controlling length scale is explained in the next section.

 As discussed in the previous subsection, the peculiar data for avalanche distributions etc, for $\fpm=0.5,\ M=0.6$ appear to indicate  that in the regime $F>\fd$ the cutoff for the distribution of avalanche areas grows as a power of the area.  This would suggest the second scenario as such a  crossover in the distribution is presumably associated with a tendency of larger avalanches to runaway.    In contrast, the avalanche distributions observed in this regime  with strong pinning suggest that there may be a crossover from dissipative-{\it critical}-like behavior for smaller sizes to dissipative-{\it cumulative}-like behavior for larger sizes, with the crossover occurring for $\sqrt{a}$ of order $0.1-0.3L$; if this {\it fixed fraction} of $L$ persists for larger sizes, it would support the third scenario.
     
In the next section we develop these ideas further and test them in the context of the restarting of the manifold after it has come to a stop from the moving phase. 

\section{Dynamics of Nucleation}

In the previous section we studied some aspects of how the system becomes depinned when it has not previously undergone macroscopic motion.  We now return to the related question of how  the manifold restarts after it has been stopped by a slow decrease of the force from the moving phase.

As discussed earlier, the stopping process leaves behind an at-least-somewhat depleted region of small local forces. This means that as the force is increased back up, the rate of avalanche production, $r(F)$,  (which  in an infinite system is just the probability density of the local forces at zero) will be small.  In Fig. \ref {avalprodrateplot}(d),
$r(F)$ is plotted as a function of $F$ for manifolds that have previously been moving. Even with relatively strong pinning, $\fpm=1.0$, it is seen that for $M=0.8$ and $M=1.0$ the avalanche production rate is almost three orders of magnitude lower than its typical magnitude in a never-moved dissipative system.  Yet $r(F)$ is relatively flat over the regime in which almost all systems will restart: from $\fd$ to $\fd+M/Z$.   Note, however, that in the upper end of this range, at least for the data we have taken for $L=128$, 
there are no samples still pinned.  
For smaller samples, the range over which the distribution of $\fh$ extends 
is broader, i.e. from $\fd$ to $\fd + M/Z$.  
Note that for all these strong pinning samples, the range over which the restarting occurs is  
two orders of magnitude larger than the width of the distribution of $\fd$.

\subsection{Distribution of restarting avalanches}

Although some fraction of the samples of size $L=128$ will restart the first time a site is triggered, by no means all will and in general there will be a distribution of avalanche sizes before restarting.  In Fig. \ref {p1.0plot}(b) the distribution of the areas of these avalanches are shown for the same parameters as above.  The data are somewhat sparse due to the small number of avalanches that typically occur in a given sample, yet some are observed out to about a tenth of  the system area. Over the observed range, these avalanche distributions are, perhaps surprisingly, quite close to power laws, with exponents close to $0.9$ This should be compared with the much smaller avalanche area distribution exponent  $\kappa/2  \approx 0.39$  of the critical region avalanches in the dissipative limit.   But the measured exponent it is close to that of  the {\it cumulative} distribution of dissipative avalanches.

\subsection{Bubble nucleation}

We conjecture that in the stopped states the length scale above which an avalanche is likely to runaway and restart the overall motion is of order the velocity correlation length, $\xi_v(F)$, of the moving phase at the same force.  This can be rationalized from the behavior of the moving phase.

At a force that is only a small amount $f$ above $\fd$,
\be
f\equiv F-\fd \ ,
\ee 
the moving phase is strongly fluctuating on scales smaller than its correlation length, $\xi_v\sim \frac{1}{f^\nu}$.  On scales smaller than this, there is substantial starting and stopping and the motion is fractal. Given the $M$ independence of many of the qualitative properties of the moving phase, the motion on scales smaller than $\xi_v$ is probably qualitatively indistinguishable from a finite dissipative avalanche. 
Thus a reasonable guess is that an avalanche from the stopped state at $f$ will be fractal on small scales and is not likely to runaway unless it becomes of order $\xi_v(f)$. But if it does become bigger than this, it will ``think" that it is part of the moving phase and its local velocity will be unlikely to fluctuate to zero.  As its local motion continues, it will impart higher stresses to neighboring regions  and its size will grow.  With its interior moving at approximately the steady state velocity $\bar v (F)$, we would expect the diameter of the resulting ``bubble"  to grow linearly in time. 

\begin{figure}[h]
\begin{center}
\epsfxsize=8cm
\epsfysize=8cm
\epsfbox{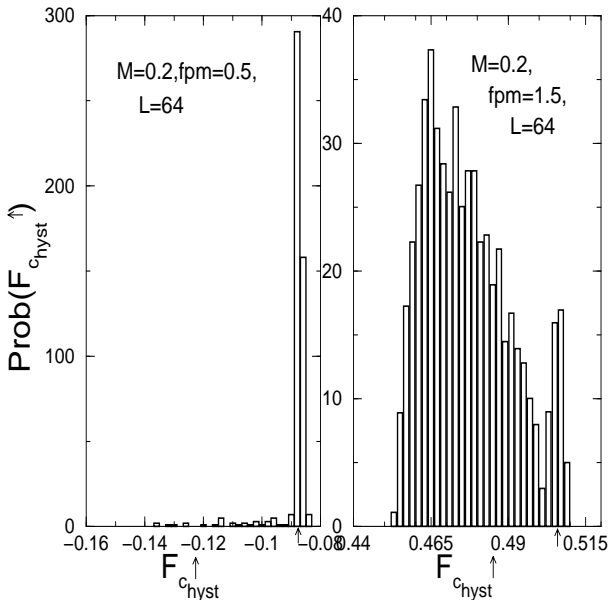}
\caption{\label {loopwidth3plot} Distribution of $\fh$'s  with $M=0.2$ and $L=64$ 
comparing $\fpm=1.5$ 
and $\fpm=0.5$.  The rms of 
$\fd$ for the these parameters are $0.4537\pm0.0009$ 
and $-0.139\pm0.001$ respectively.  The up arrows on the 
horizontal axis indicate   $<\fd>+M/Z$, the maximum $\fpm$ .}
\end{center} 
\end{figure}

\begin{figure}[h]
\begin{center}
\epsfxsize=6cm
\epsfysize=6cm 
\epsfbox{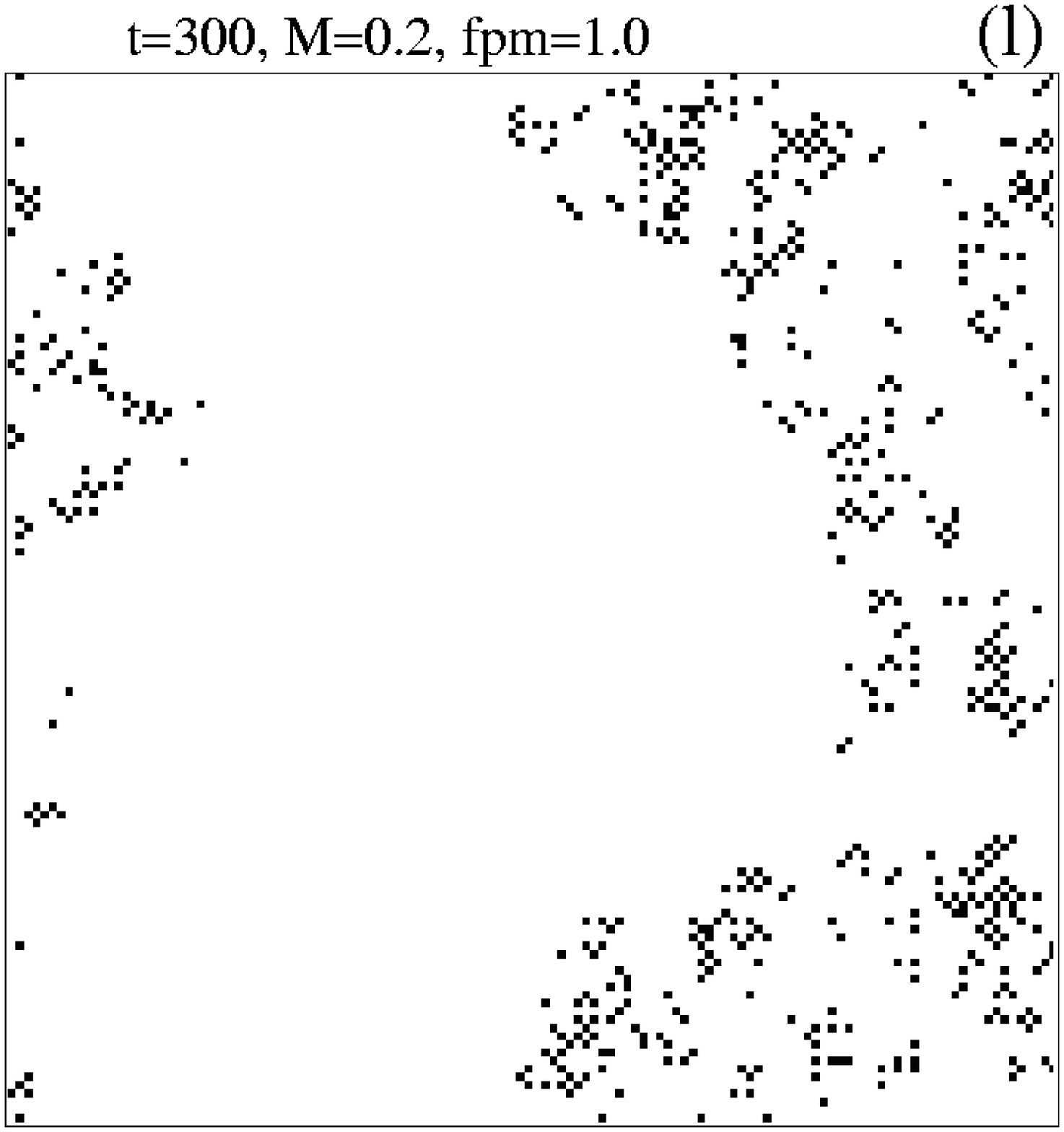}
\caption{\label {snapshots} Dynamics of nucleation: segments that are moving at  each of the indicated times $t$ after nucleation are shown.}
\end{center}
\end{figure}

 In order to test this hypothesis, we investigate in our simulations {\it how}  the
manifold begins to move when the applied force is increased after it has been stopped.  For $f_p^{max}=0.5$ and $M=0.2$, it can be seen from Fig. \ref {loopwidth3plot} that restarting will typically occur at $\fh=\fd+M/Z$ by which point the steady state velocity of the moving phase is already substantial and its correlation length short.  In Figs. \ref{snapshots} (a-l) the temporal evolution of the restarting process is shown; we see
 that it is \emph{not} fractal, but looks more like an expanding nucleation bubble as occurs after supercooling through an 
equilibrium first-order phase transition.  After a short initial transient. there is a front that propagates outward from an initial seed in a roughly deterministic manner.    
We observe that the radius of the nucleation bubble $R(t)$ grows
approximately linearly with time so that $R(t)=ct$, $c$ thus being an  expansion rate. The interior of the bubble is, as expected,  moving at roughly the steady state velocity, $\bar{v}(F)$ of the moving phase.  [Note that $c$ is a speed in the {\it spatial} rather than that in the displacement $(h)$ direction which is $v$.] 

\begin{figure}[h]
\begin{center}
\epsfxsize=8cm
\epsfysize=8cm
\epsfbox{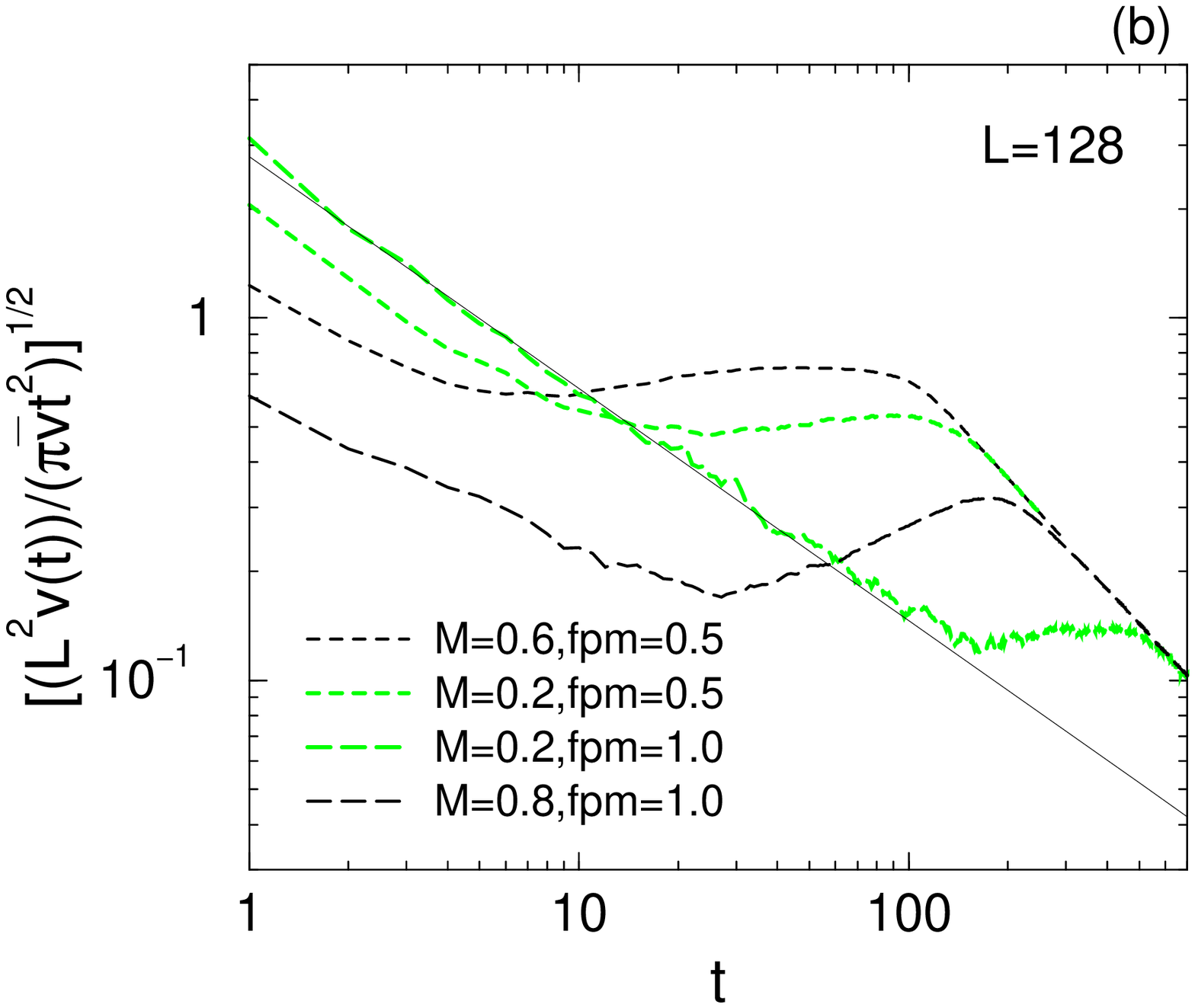}
\caption{\label {nucleateplot} \label {loglognucleateplot}(a) Effective bubble growth rate, $c(t)\equiv(\frac{L^2 v(t)}{\pi \bar{v} t^2})^{1/2}$ 
as a function of time averaged over $8$ $L=128$
restarting samples.  
The solid line is $(\frac{L^2}{\pi t^2})^{1/2}$, the long-time steady state behavior. (b) Log-log plot at early times.  The solid line is
the theoretical expectation in the fractal regime with a slope of $\frac{d+\zeta}{2z}-\frac{3}{2} \approx -0.64$.  }
\end{center}
\end{figure}

The expansion rate can be estimated from the time dependence of the spatially averaged velocity, $v(t)$ by assuming a circular bubble of radius $r(t)=ct$ so that
\be 
v(t)\approx  \frac{\pi (ct)^2}{L^2} \bar{v} \ .
\ee
  In Fig. \ref{nucleateplot} we thus plot 
\be
c(t)\equiv\bigg(\frac{v(t)L^2}{\bar{v}\pi t^2}\bigg)^{\frac{1}{2}}
\ee
as a function of the time $t$ after the restarting was triggered, for several different $M$'s and 
$f_p^{max}$'s.      
In general, at short times after nucleation, fractal growth that is like that in a dissipative avalanche occurs within an erratically  growing region of typical diameter $\ell(t)\sim t^\frac{1}{z}$ with resulting cumulative moment of the avalanche: $m(t) \sim \ell^{d+\zeta}$.  This results in an apparent $c(t)\sim \left(\frac{1}{t^2 \bar v}\frac{dm}{dt}\right)^\frac{1}{2}\sim t^{\frac{d+\zeta}{2z}-\frac{3}{2}}\sim \frac{1}{t^{0.64}}$. See Fig. \ref {loglognucleateplot} (b).  When the size of the avalanche has reached $\xi_v(F)$, $c(t)$ stops decreasing and the flattish parts of the curves at intermediate times are indicative of approximately uniform expansion;  the inferred $c(t)$ in this regime is roughly the expansion rate, $c$, of the 
bubble.  If the expansion rate is small, as will occur if the steady state velocity is low, then we expect $c\sim \frac{\xi_v}{\tau_v}\sim \xi_v^{1-z}$ with $\tau_v$ the characteristic relaxation time in the moving phase; this is also of order the time at which  the crossover from fractal to bubble growth occurs.  

At long times, the whole system is moving and $c(t) \approx \frac{L}{\sqrt{\pi}t}$ as shown. This saturation occurs at $t\sim L/c$.  Because the region near the origin of the bubble has already been moving for some time when the regions far away start moving,  at the time when the furthest regions start moving, the roughness of the manifold will be {\it greater} than it is in steady state  The elasticity will diffusively smooth out this roughness at longer times leaving behind only the logarithmic roughness characteristic of the moving phase.  This excess roughness near the crossover time from bubble nucleation to steady state motion is the cause of the peak in the width, $w^2(t)$, of the manifold shown in Fig. \ref {wtplot}.  Naively, we expect the peak 
$w^2(t)$ to scale as  system diameter, $L$, but subtleties associated with elastic slowing down of the velocity inside the bubble might need to be taken into account to understand the process more fully. 

For $\fpm=0.5$, the forces at which the restarting occurs are substantially above $\fd$  and bubble growth appears to occur with an expansion rate for $M=0.6$ of $c\approx 0.7$, close to the maximum possible rate of unity, while  for $M=0.2$, it is somewhat slower: $c\approx 0.5$.  In the former case, the motion inside of the bubble is essentially alternating sublattices, behavior characteristic of the plateau in the steady state velocity at $\bar v\approx \frac{1}{2}$ that occurs with substantial overshoots; see Fig. \ref {vfplot}(a).  For the
stronger randomness, $\fpm=1.0$, the hysteresis loops are narrower and the nucleation occurs to a lower velocity state. For $M=0.2$, the expansion rate is quite small, $c\approx 0.13$ but the flat constant $c$ regime in the figure is observable over a factor of five in time. For these parameters, at early stages the evolution appears fractal, consistent with expectations.   
For $M=0.8,\ \fpm=1.0$, there is a much narrower flat region and the evolution is not obviously bubble-like.  This can be seen in the snapshots of the local motion as the bubble expands shown in Figs. \ref {snapshots}.  
At long times, all of the samples show, as they must, approach of the spatially averaged velocity to its steady state value: this asymptotic behavior is indicated in Fig. \ref {nucleateplot}(a) by the solid line.  We see that the time at which $v(t)$ becomes close to $\bar{v}$ is roughly that that would be expected for a bubble to encompass the whole system and run into its periodic ``images".

 More investigation of the dynamics of nucleation, in particular of the transition from fractal on small scales to bubble-like on larger, and the dependence of the crossover scale on system size, on $F-\fd$, and on the other parameters, is clearly needed.
 
\begin{figure}[h]
\begin{center}
\epsfxsize=8cm
\epsfysize=8cm
\epsfbox{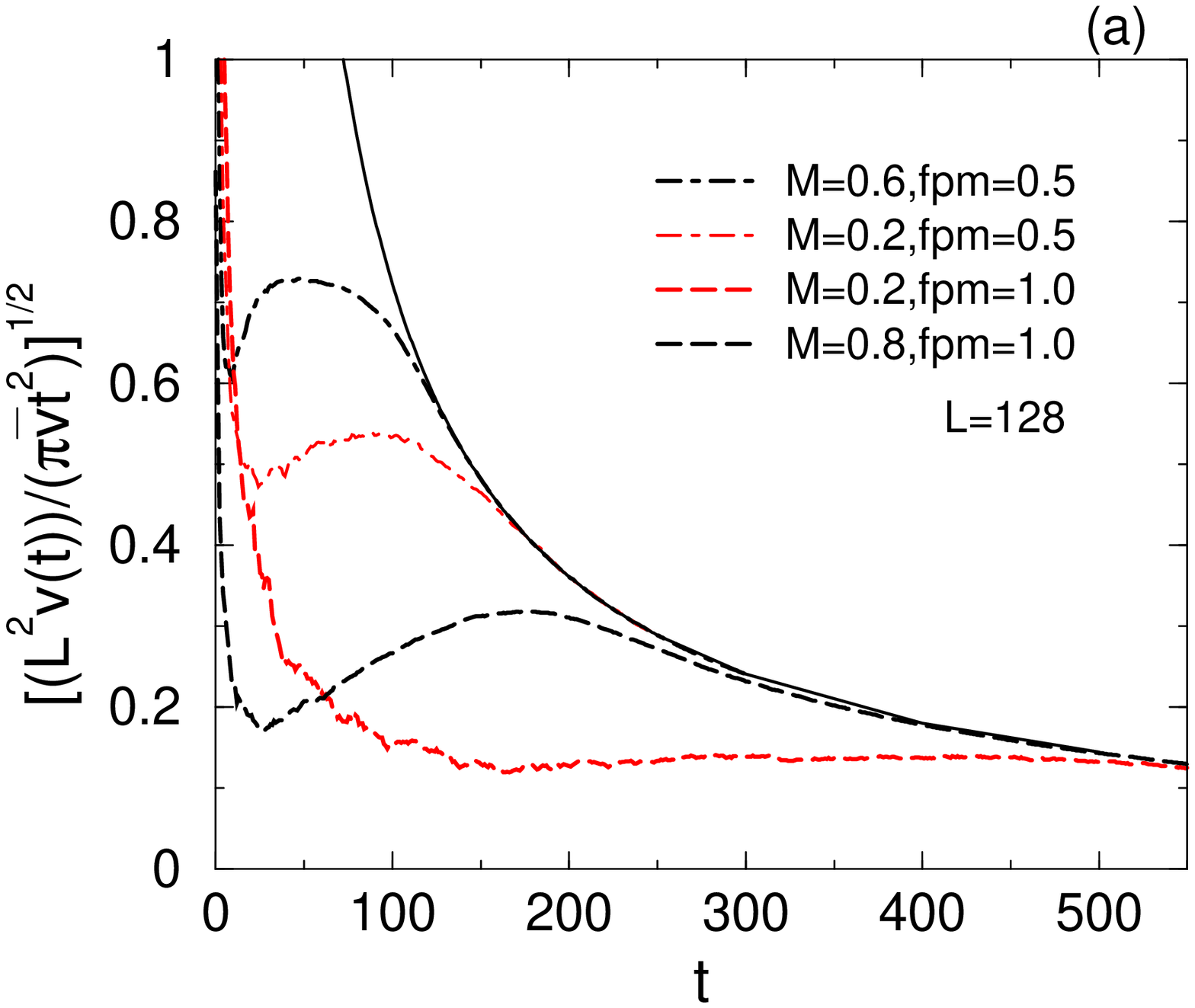}
\end{center}
\end{figure}
\vspace{-2cm}
\begin{figure}[h]
\begin{center}
\epsfxsize=6.0cm
\epsfysize=6.0cm
\epsfbox{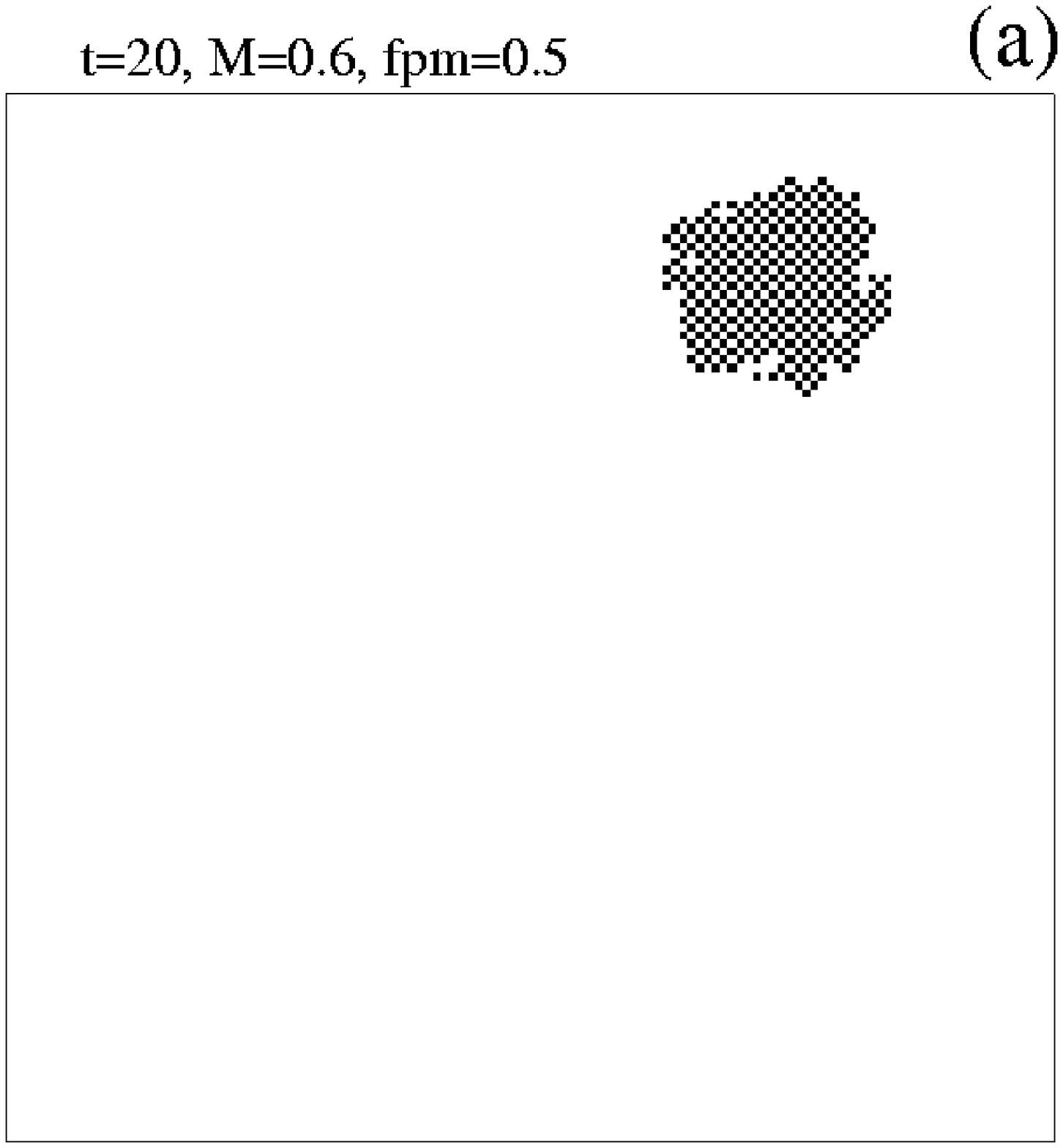}
\end{center}
\end{figure}
\begin{figure}[h]
\begin{center}
\epsfxsize=6.0cm
\epsfysize=6.0cm
\epsfbox{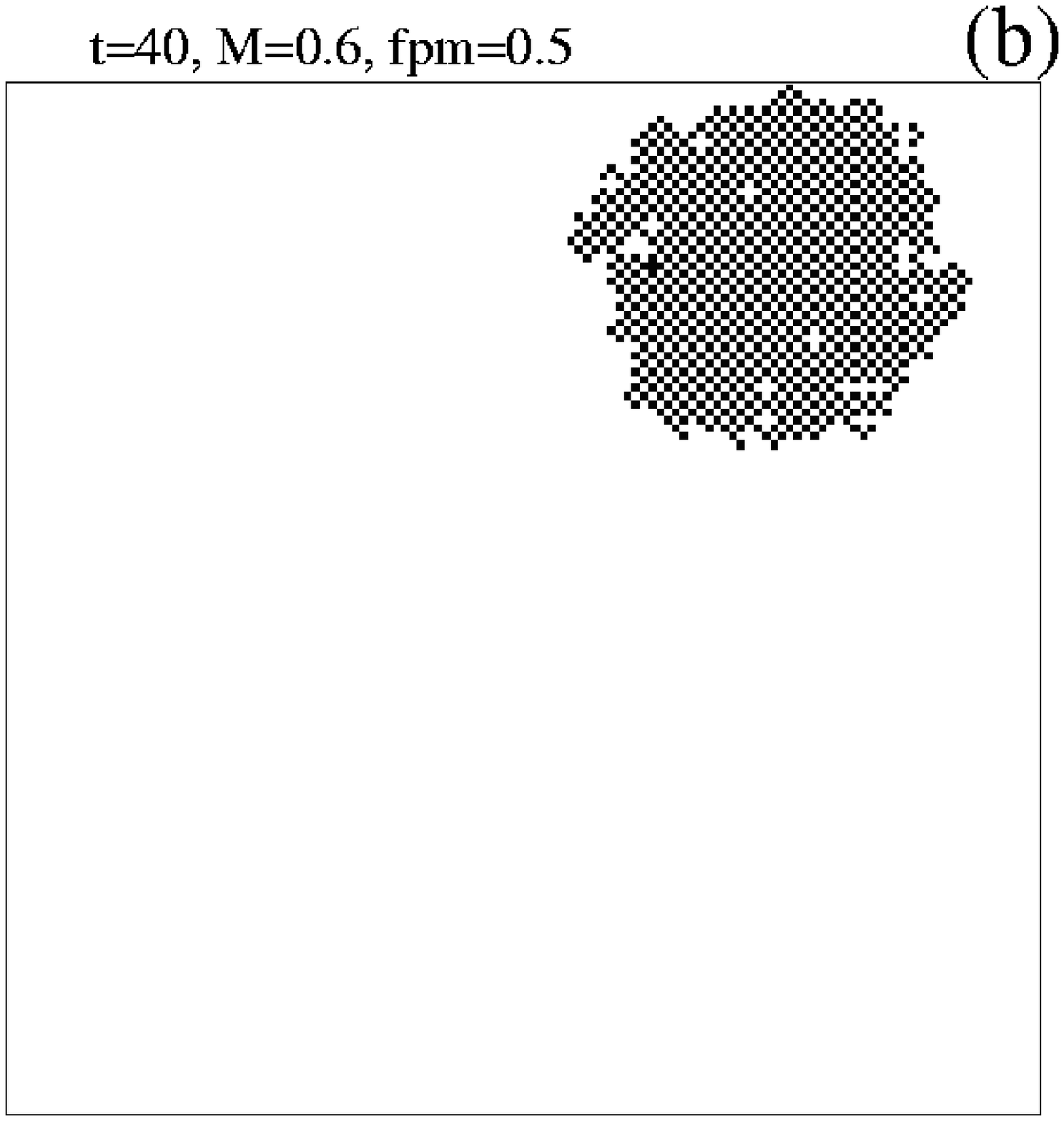}
\end{center}
\end{figure}
\begin{figure}[h]
\begin{center}
\epsfxsize=6.0cm
\epsfysize=6.0cm
\epsfbox{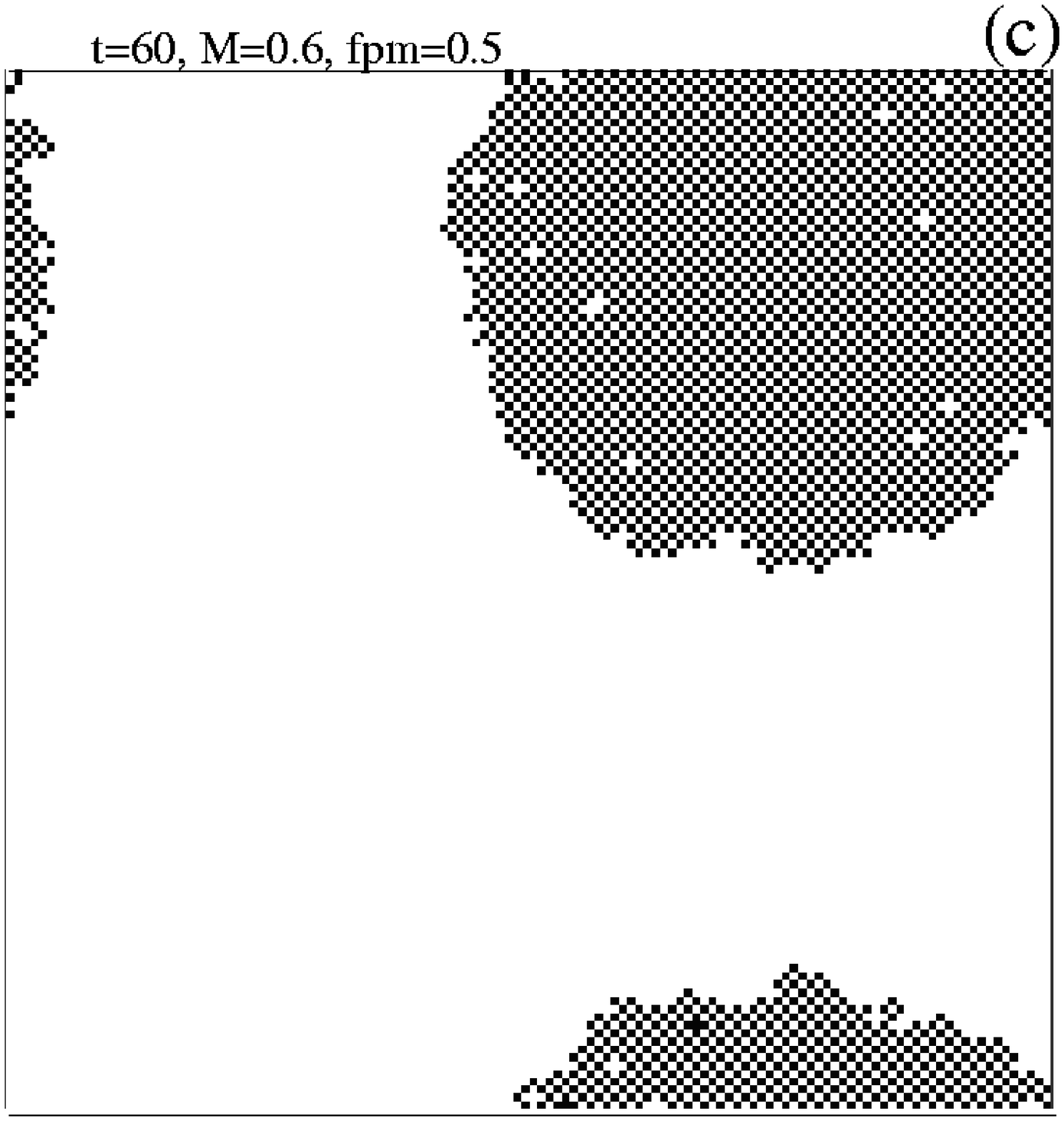}
\end{center}
\end{figure}
\clearpage
\begin{figure}[h]
\begin{center}
\epsfxsize=6cm
\epsfysize=6cm
\epsfbox{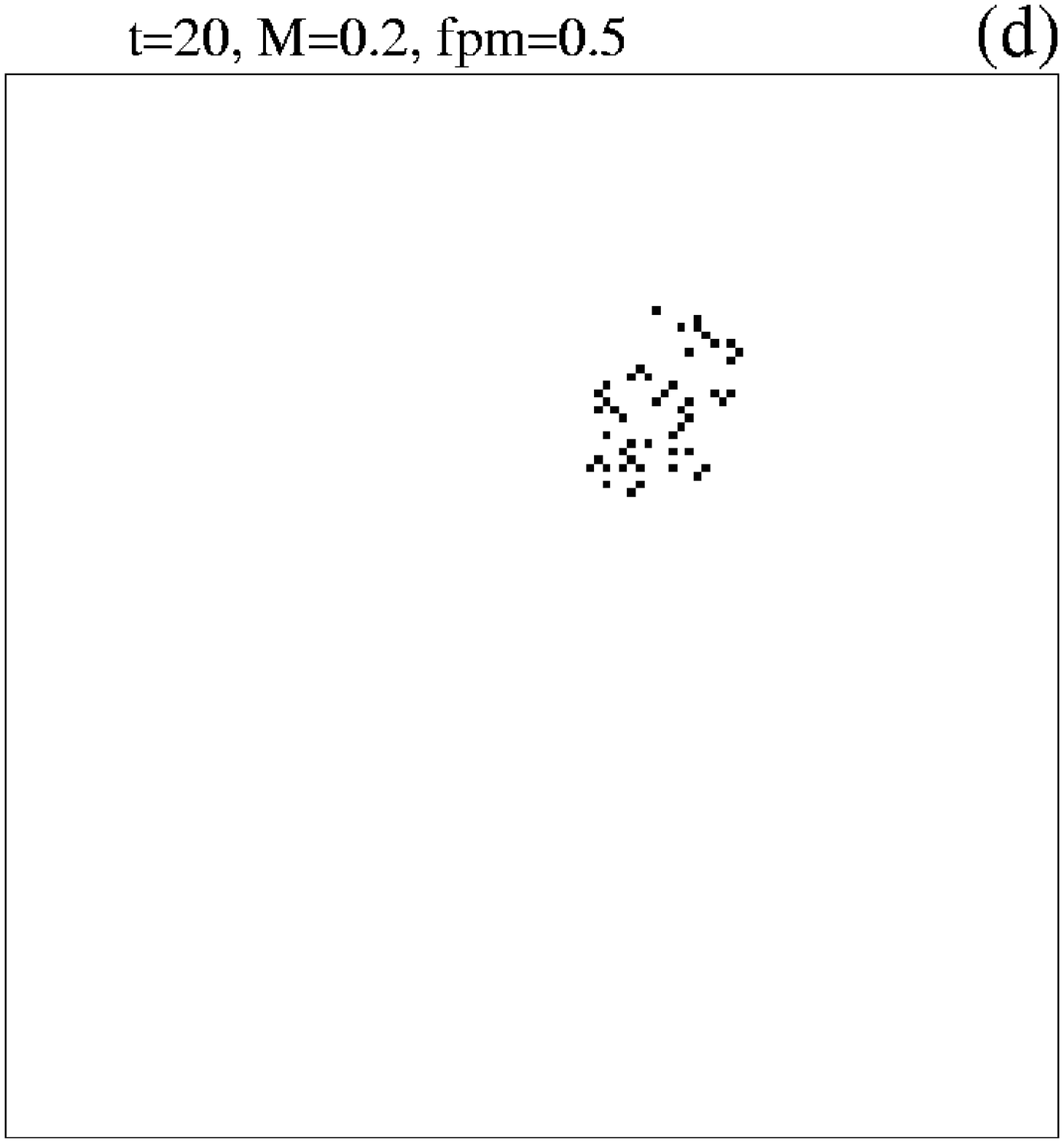}
\end{center}
\end{figure}
\begin{figure}[h]
\begin{center}
\epsfxsize=6cm
\epsfysize=6cm
\epsfbox{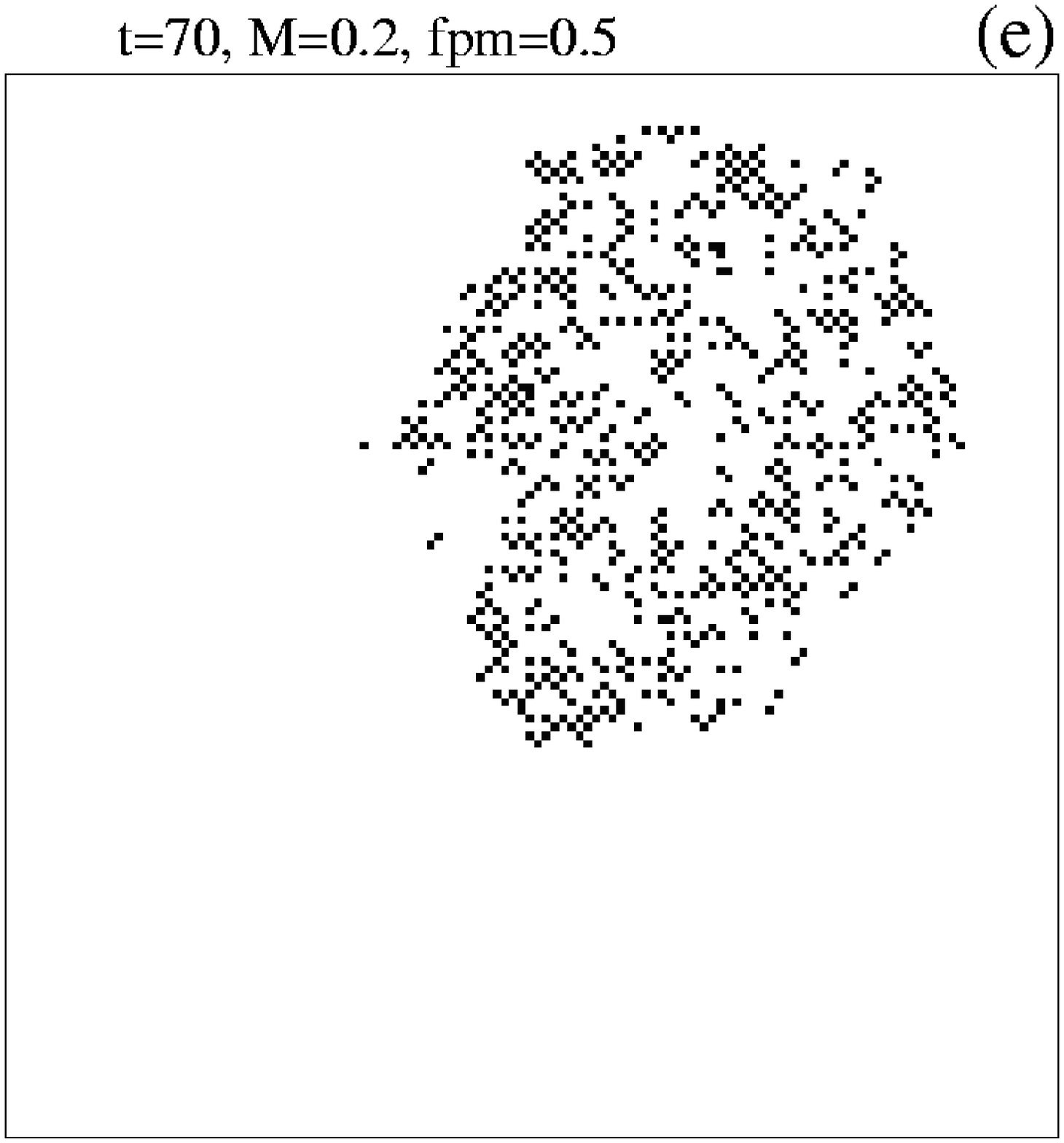}
\end{center}
\end{figure}
\begin{figure}[h]
\begin{center}
\epsfxsize=6cm
\epsfysize=6cm
\epsfbox{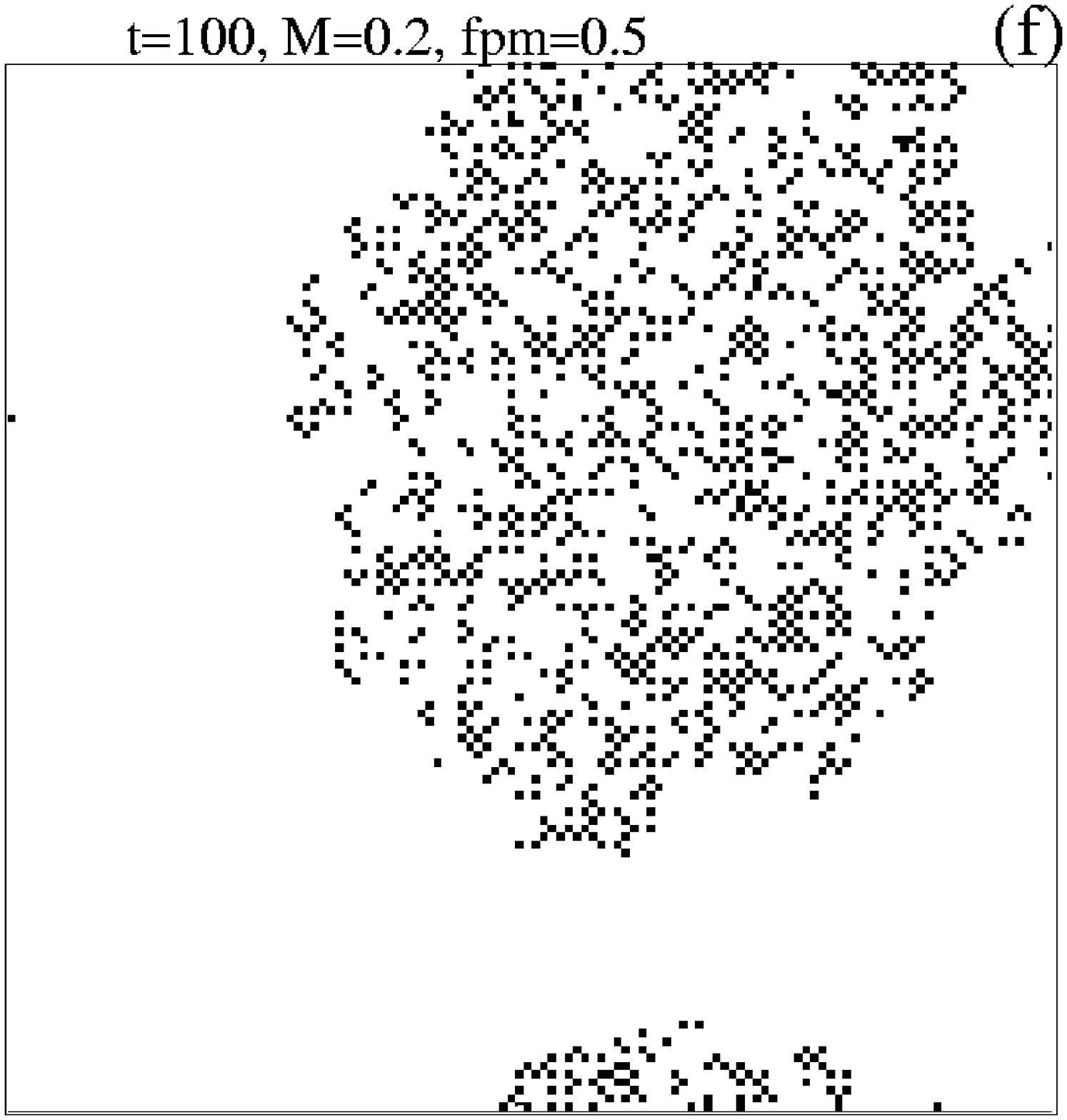}
\end{center}
\end{figure}
\begin{figure}[h]
\begin{center}
\epsfxsize=6cm
\epsfysize=6.5cm
\epsfbox{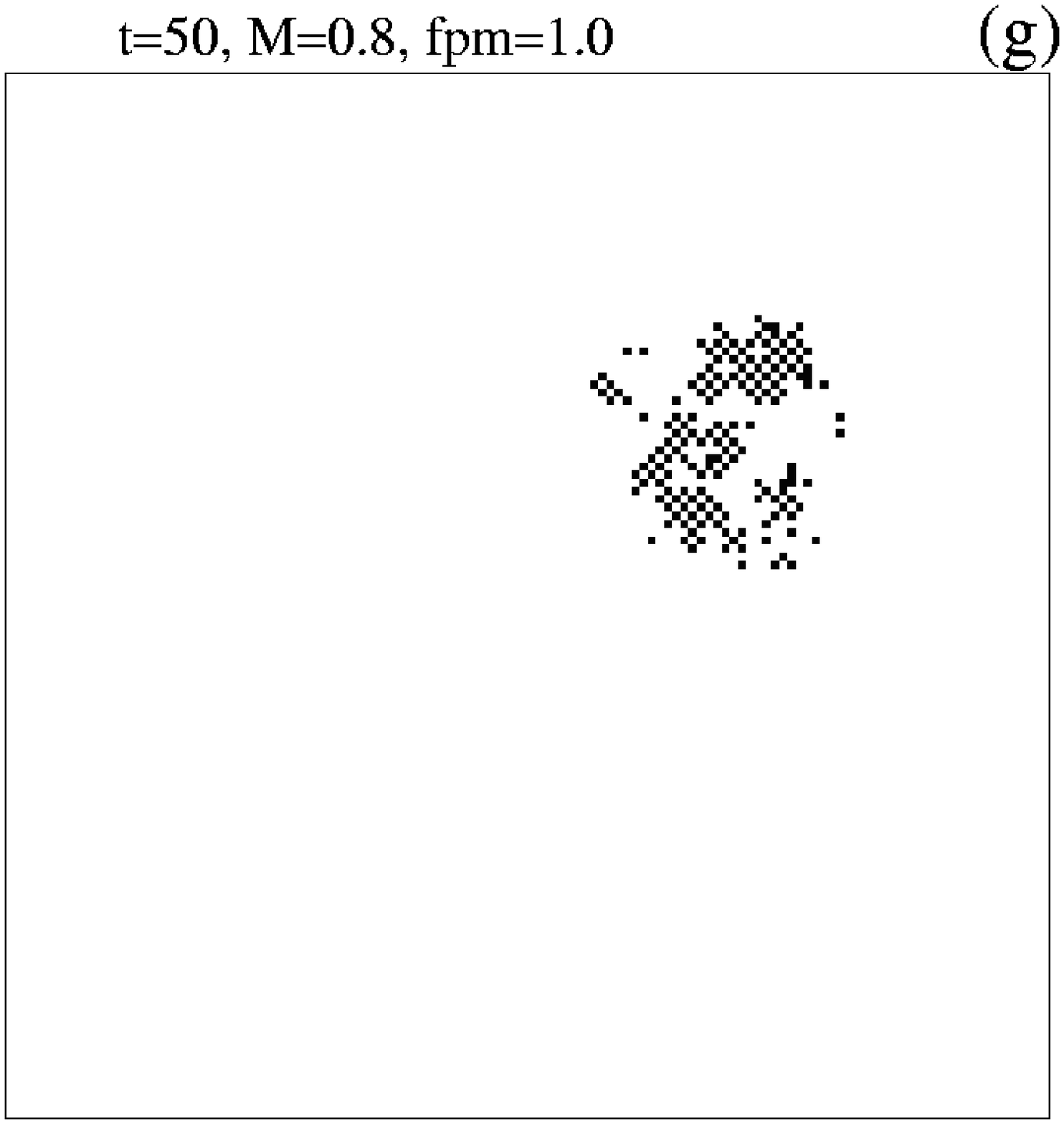}
\end{center}
\end{figure}

\begin{figure}[h]
\begin{center}
\epsfxsize=6cm
\epsfysize=6.5cm
\epsfbox{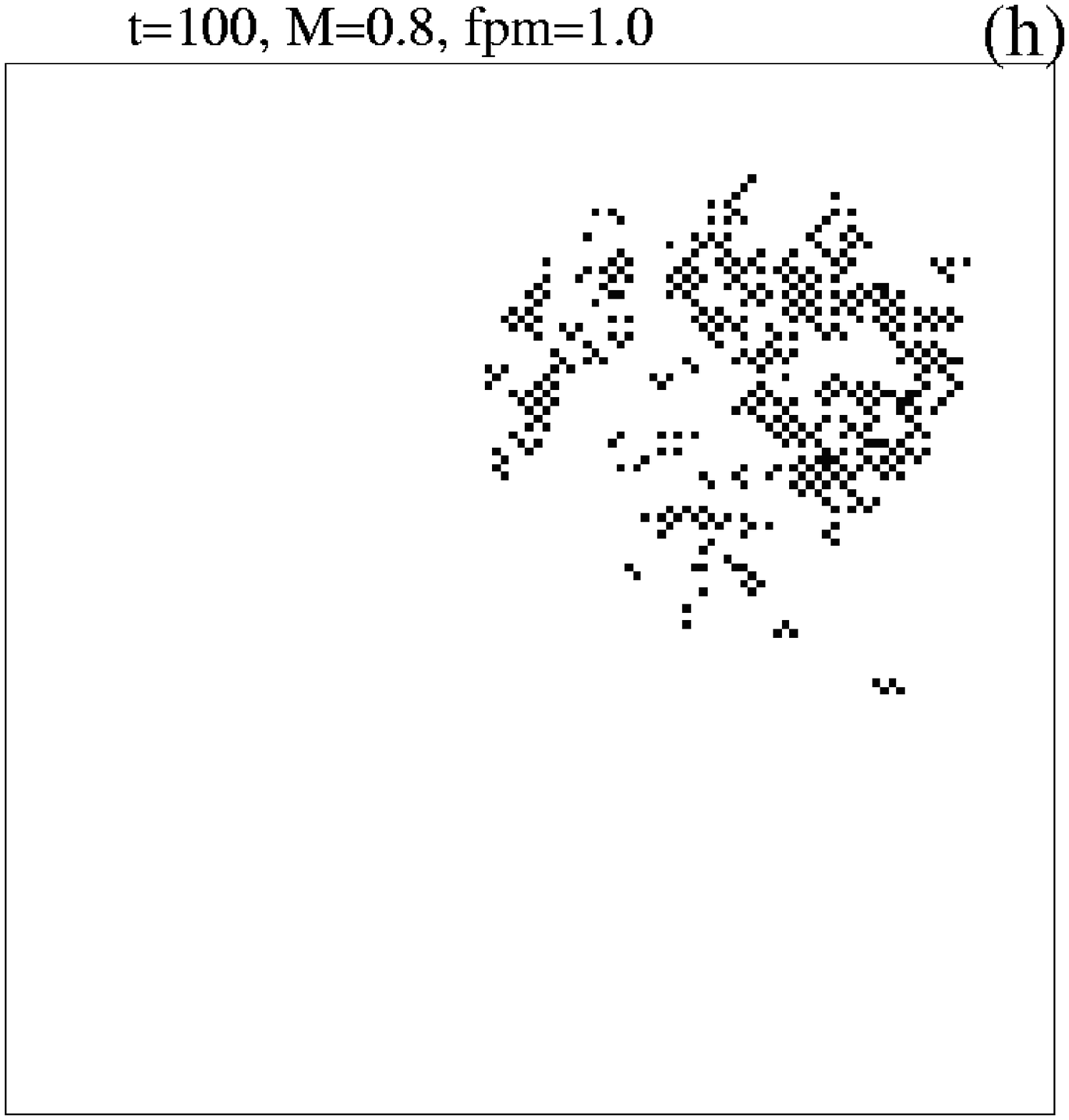}
\end{center}
\end{figure}

\begin{figure}[h]
\begin{center}
\epsfxsize=6cm
\epsfysize=6cm
\epsfbox{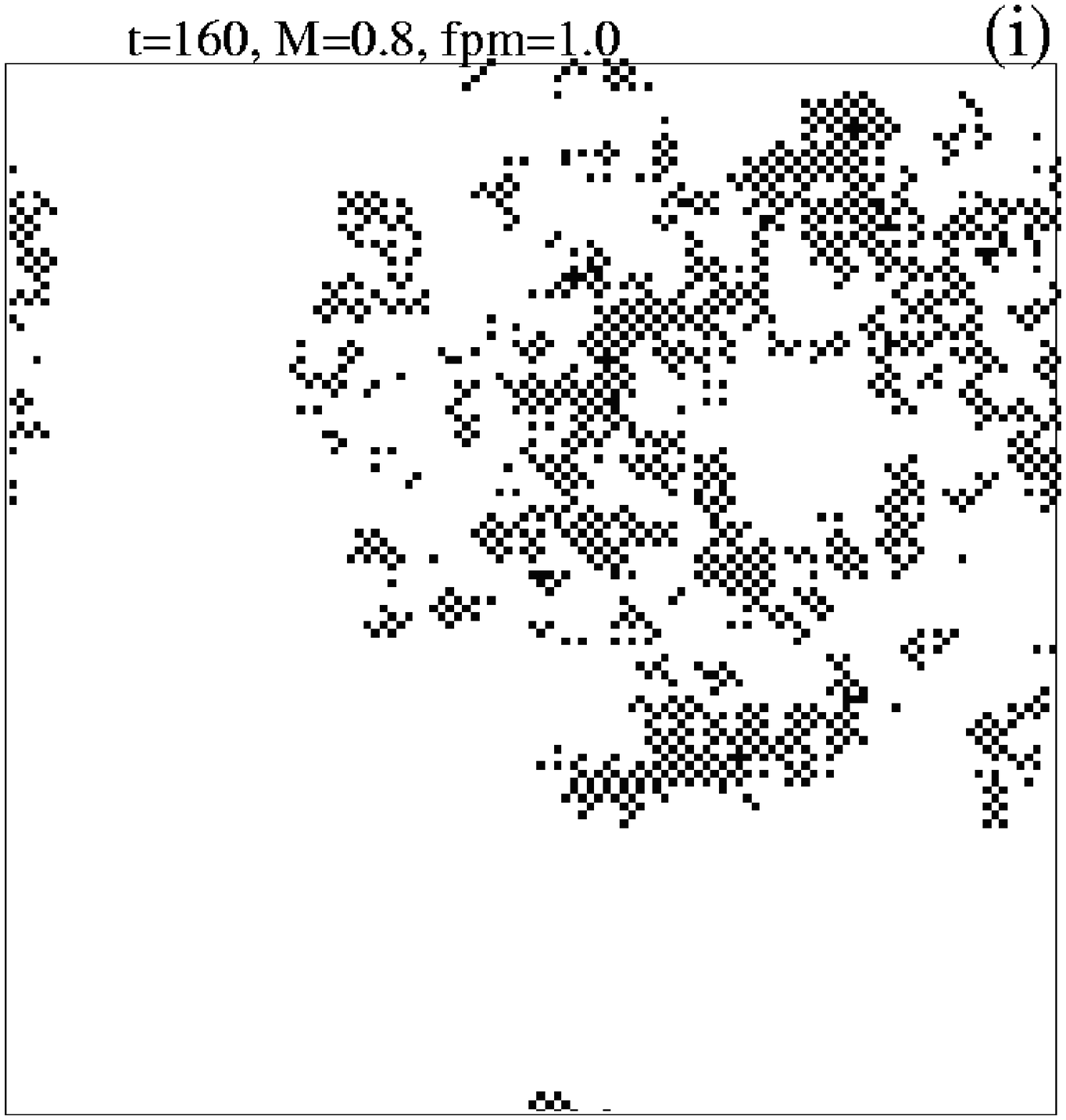}
\end{center}
\end{figure}
\begin{figure}[h]
\begin{center}
\epsfxsize=6cm
\epsfysize=6.5cm
\epsfbox{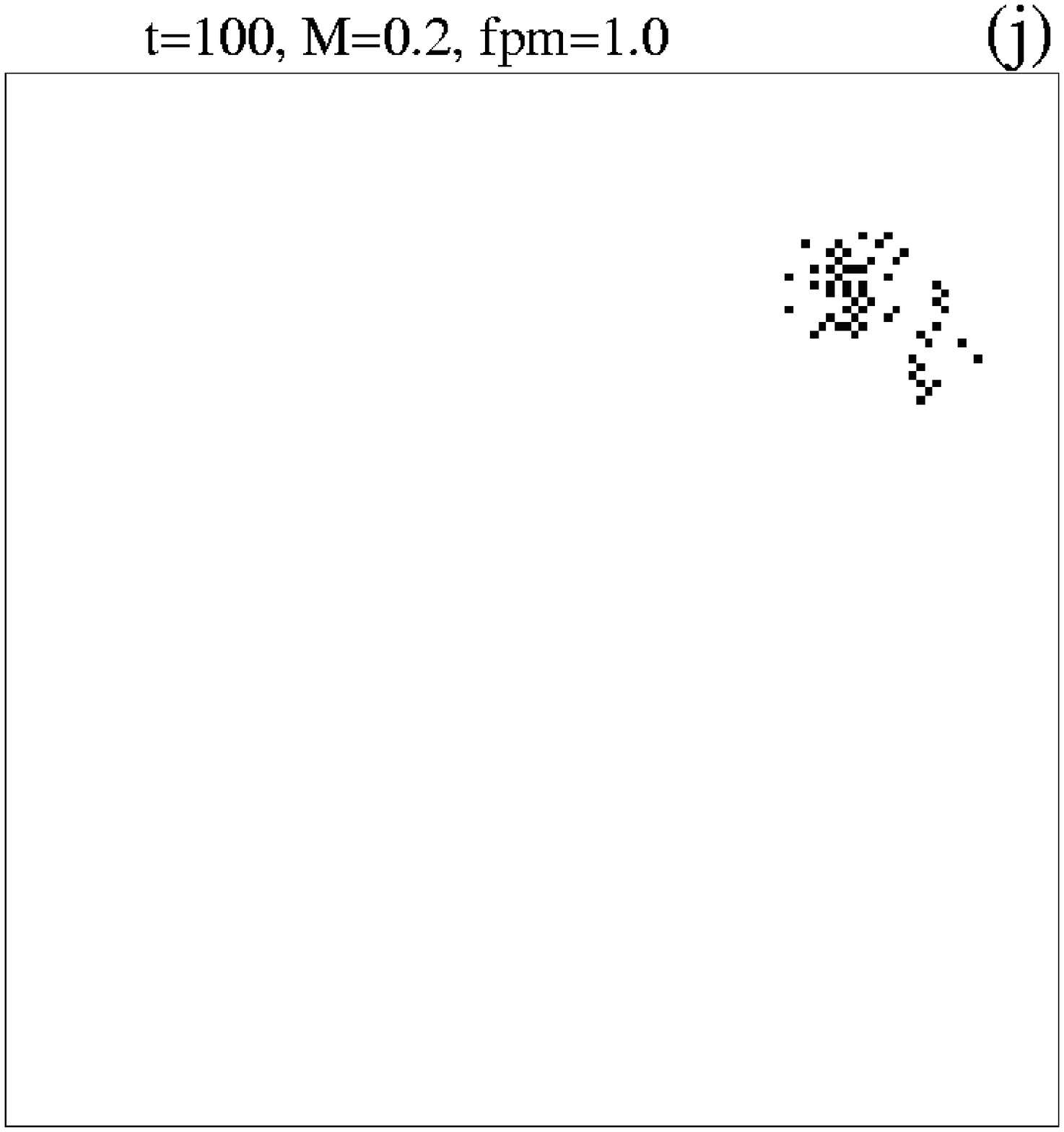}
\end{center}
\end{figure}

\begin{figure}[h]
\begin{center}
\epsfxsize=6cm
\epsfysize=6.5cm
\epsfbox{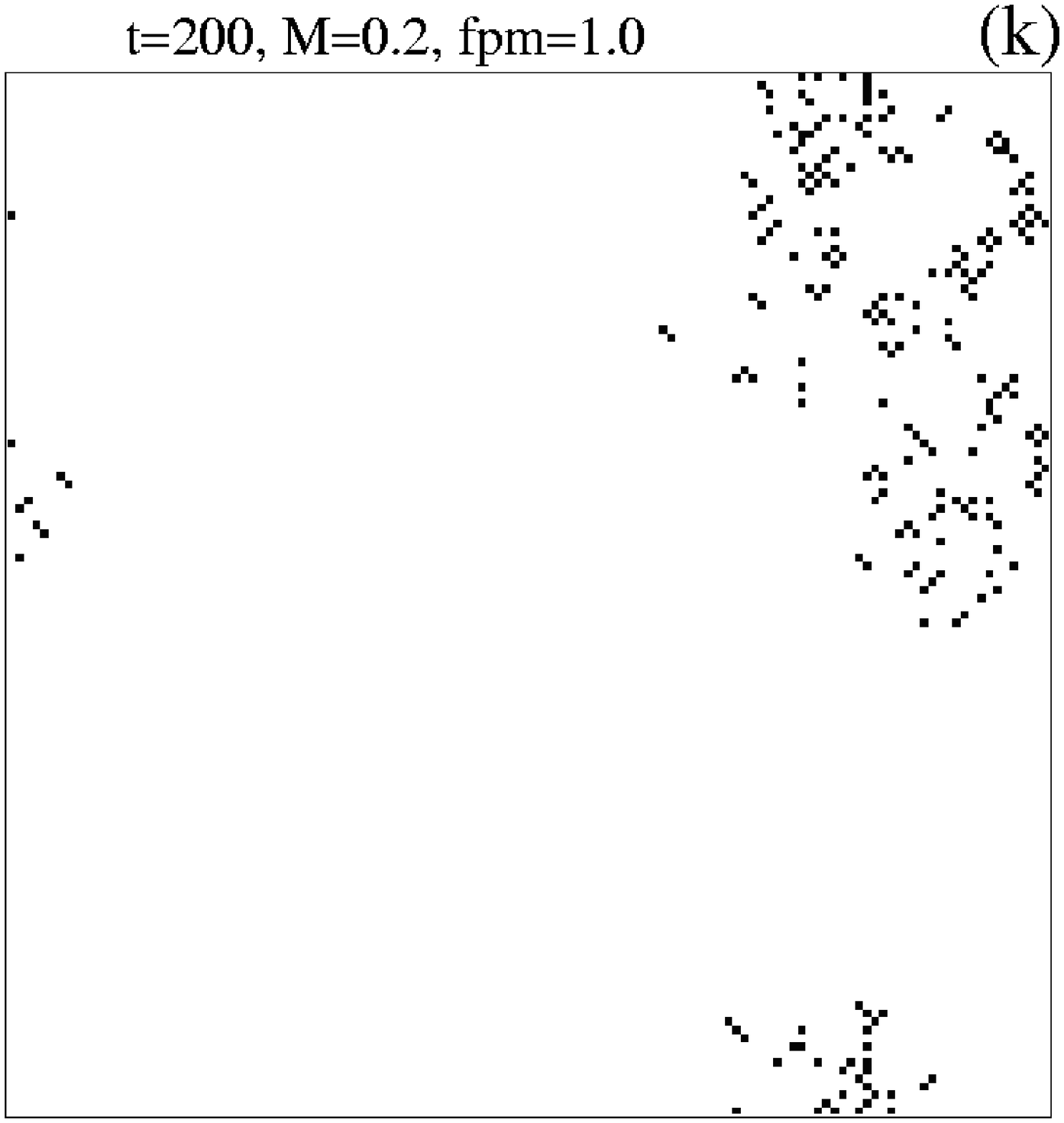}
\end{center}
\end{figure}

\subsection{Nucleation avalanche statistics}

Armed with the picture of avalanches in the hysteresis loops as fractal on scales out to of order $\xi_v(F)$ and bubble-like on larger scales, we can understand the key aspects of the avalanche statistics in this regime.  Avalanches that are triggered very close to $\fd$ --- at forces within the distribution of  $\fd$ for that system size --- will be fractal on all scales and very unlikely to run away.  We expect their statistics to be similar to those that occur in the regime near $\fd$ when the force is increased from an initially flat configuration: the probability of their area being larger than $a$ will decay only as $a^{-\frac{\kappa}{d}}$ out to the largest sizes.  But as the force is increased, the small rate of avalanche production means that there is still a substantial probability that no system sized avalanche will have occurred until a force well outside the distribution of $\fd$.  In this regime, avalanches that do occur will still have a probability of reaching area $a$ that decays as $a^{-\frac{\kappa}{d}}$ out to scales of order $\xi_v(F)$, but this is now considerably smaller than the system size.  Those that do reach areas  of order $\xi_v^d$ will tend to runaway and so the distribution of areas, $p(a,F)da$, will be cutoff at this $F$-dependent crossover scale. The probability of runaway of an avalanche triggered at $F$ is roughly the probability that its area become larger than $a_v(F)\sim\xi_v(F)^d$:
\be
P_{\rm run}(F)\approx 1- \int_0^{a_v(F)} p(a,F)da \sim \frac{1}{\xi_v(F)^\kappa}
\ee    
where $p(a,F)da$ is the fraction of avalanches that occur near $F$ which have areas within $da$ of $a$. In order to obtain the cumulative distribution of avalanches in the hysteresis loop we must multiply the distribution $p(a,F)$ by the probability that runaway has not yet occurred, 
\be
P_{\rm pinned}(F,L)= \exp\big(-\int_{\fd}^F r(F')L^d P_{\rm run}(F')dF'\big)
\ee
 and then integrate this over $F$. If the rate of avalanche production per unit area were of order unity, then $P_{\rm pinned}$ would be small for $F-\fd>L^{-\frac{1}{\nu}}$ and the cumulative avalanche distribution would be dominated by avalanches in the regime very near $\fd$ for which the distribution is characteristic of the critical point; the cumulative distribution of avalanche areas in the hysteresis loop would then decay with the exponent  $\kappa/d$ consistent with what we observe for the dissipative case.    But with $r(F)$ very small for strong pinning and $M=0.8$ or $M=1.0$ --- almost three orders of magnitude smaller than for the dissipative case, (see Fig. \ref {avalprodrateplot}(d)) $P_{\rm pinned}$ will only become small when $\xi_v^d < rL^d$ corresponding to  $F-\fd> r^\frac{-1}{d\nu} L^{-\frac{1}{\nu}}$ a factor of fifty or so further from $\fd$ than in the dissipative case, roughly consistent with what is observed.

\section{Crossovers and Asymptopia}

\subsection{Hysteresis loops}

The form of the avalanche distribution in hysteresis loops with a very low density of nucleation segments should be apparent from the discussion at the end of the previous section: Most avalanches will take place when $F$ is close to the $\fh$ at which the system restarts, thus it is the distribution $p(a,\langle\fh\rangle)da$ in this regime that will dominate the cumulative statistics. This results in  a distribution of avalanche areas that for small avalanches is critical with exponent $\frac{\kappa}{d}$, but for large avalanches decays with the larger cumulative exponent $K_{\rm cum}=1$. With small $r$, as occurs in the strong pinning systems with $M=0.8$ and $M=1.0$,  the crossover in a system of linear size $L$ occurs at a length scale of order $Lr^\frac{1}{d}$ or corresponding crossover area of order $rL^2$.  For the size ranges investigated here --- up to $L=128$ --- this means that in practice almost the full range of avalanche areas will be in the large avalanche  tail region with exponent $K_{\rm cum}=1$.  But extrapolating to very large systems from these data would be highly misleading: in the limit of large systems, the  distribution should scale like the critical dissipative limit over most of the range of avalanche sizes with the rapid decay observed in our data obtaining only for the largest few orders of magnitude of avalanche areas. 

The dynamics of a typical runaway event will reflect this crossover.  Initially, it will start as a fractal avalanche similar to those in the dissipative limit.  But when --- and if --- an avalanche reaches the crossover length of order $Lr^\frac{1}{d}$, it will crossover to bubble like expansion with its interior behaving like the moving phase and its radius expanding linearly in time at a rate, $c$, that scales as $(Lr^\frac{1}{d})^{1-z}$.  As the distribution of $\fh-\fd$ is broad, however, there will be substantial run-to-run variations in both the crossover size  and the rate of growth of the bubble.

The behavior of the size dependence of the  distribution of widths of the hysteresis loops, as shown in Figs. \ref {loopwidth1.0plot}(a) (b),  can also be better understood in this framework.   In small systems, the force will typically have to be increased by $M/Z$ from $\fd$ in order to trigger restarting.  In contrast, in very large samples, we expect that the 
restarting will typically occur at  forces $F=\langle\fd\rangle +g L^{-\frac{1}{\nu}}$ which are  above $\langle\fd\rangle$ by of order  the width of the distribution of $\fd$.  Yet because of the strong suppression of the density, $r(F)$,  of seeds for nucleation of avalanches in the stopped system, in practice the numerical factors, $g$  can be very large.  The distribution of $g$ in the small $r$ limit is broad, and has a shape close to a Weibull distribution with characteristic scale $1/r^\frac{1}{d\nu}$ and  shape parameter  $\frac{1}{d\nu}$ 
but we do not expect it to be exactly of this form.  The reason is due to the breakdown of the argument given in Section V: In sufficiently small subregions of size $\big(\frac{L}{b}\big)^d$ of a system of size $L^d$, the possible nucleation processes are {\it not} independent from one subregion to the next.  This is because the diameter of a critical nucleation avalanche when it crosses over to approximately deterministic bubble-like growth is of order $\xi_v(F)$ which is of order a {\it fixed} --- albeit $r$ dependent ---  fraction of $L$ at the typical forces at which the first runaway avalanche occurs in a large system.   Thus the division into approximately independent subregions will not work for $b>1/r^\frac{1}{d}$.

Because the hysteresis loops vanish in the limit of large systems, the total number of avalanches per unit area that will occur in a hysteresis loop decreases as a power of system size.  This will also occur in the dissipative limit, however in that case the ``runaway" avalanches that occur for $F\approx \fd$ will not have all that much larger moment than ones that only involve, say,  half the system area.  But due to the strong suppression of avalanche production after a large event, in the presence of substantial stress overshoots the total number of avalanches will be further reduced in spite of the hysteresis loop becoming wider.  The above analysis implies that the total number of avalanches per unit area that occur in the hysteresis loop will be of order:
\be
n_{aval} \sim r^{1-\frac{1}{d\nu}} \frac{1}{L^\frac{1}{\nu}}\ .
\ee
Furthermore, the runaway avalanche will typically run for a very long way --- probably exponentially far in $1/r$ --- before stopping at a displacement, $h$, for which the total pinning is anomalously strong.

At this point, it is not clear that the numerical data are consistent with this picture. In particular, it appears that the size dependence of the widths of the hysteresis loops is somewhat weaker than would be expected asymptotically.  Nevertheless, we believe that the deviation of the scaling of the width of the hysteresis loops from the $L^{-\frac{1}{\nu}}$ scaling expected asymptotically is most likely associated with the crossover from the small system to large system behavior.  In sufficiently large systems, we conjecture that 
\be 
\mu=\nu\ .
\ee 

\subsection{Initial depinning}

The complicated avalanche behavior observed when depinning is approached for the first time starting from almost flat initial conditions is probably due to a hybrid of effects.  For the strong pinning samples with substantial-sized overshoots, in systems of size $128^2$ enough avalanches typically occur before runaway that a strong suppression of the avalanche production rate occurs over the whole system.  This suppression enables the force to be increased well past the range in which runaway would occur in the dissipative limit, although not as far as it can for samples that have been stopped from the moving phase.   The statistics of the avalanches in this regime will have a similar form to that of  avalanches in the hysteresis loops discussed above, with a crossover from critical-like to cumulative-like at a length scale that depends on how strongly $r(F)$ has been suppressed.  

For the weaker pinning samples, for example with $\fpm=0.5$ and $M=0.6$, the rate of avalanche production is not suppressed nearly as strongly and it is most likely that the crossover regime is right in the middle of the range of sample sizes studied.  

Further systematic tests to investigate this overall scenario would be very useful.  In particular, one would like to understand on what aspects of the model, parameters, and history, crucial quantities such as the suppression of avalanche production rate after a large avalanche, depend. 
 
\section{Discussion and Conclusions}

In this paper we have introduced and studied numerically and by scaling analyses a simple model of depinning transitions in which there are  dynamic stress overshoots caused by rapid motion of segments of the manifold.

\subsection{Macroscopic behavior}

We have produced a substantial amount of evidence that the macroscopic behavior of this model with various kinds of generic initial conditions is {\it not} hysteretic for a wide range of stress overshoots in  
spite of the coexistence over a range of  the driving force of static configurations and steady state moving configurations.   In particular, the critical force, $F_c(M)$ is {\it uniquely defined and history independent} in the limit of large systems; the distributions of the critical force, $\fd$, at which steady state motion ceases, that at which it restarts when the force is increased from such a stopped state, $\fh$, and that at which macroscopic motion commences when the force is increased from a flat initial condition for the first time, $\fa$, are all narrow and converge to $F_c(M)$ in the limit of large systems. Furthermore, the steady state velocity, $\bar v(F)$, is a unique function of the driving force, $F$ that vanishes as a universal power of $F-F_c$.
Nevertheless, in principle there are static configurations that could be reached by careful enough control of the dynamics of all the system  for all forces up to $F_{c0}=F_c(M=0)>F_c(M)$.  But if the system were to be started in such a configuration and the force increased by an amount that is arbitrarily small in the limit of a large system, an avalanche would be triggered that would runaway and cause the whole system to start moving.

The universality class of the depinning transition for all $M$ less than a multicritical value $M_c$ appears to be that of the purely dissipative limit without stress overshoots that has been analyzed by renormalization group methods previously.  This universality class is thus much broader than had been conjectured.  

The crucial feature of the dynamics on which this macroscopic uniqueness and universality relies is the existence of a non-zero density of nucleation sites for new avalanches after an avalanche has run through a region.  These must exist for arbitrarily small increases of the applied force even if their density, $r$, is much lower than it would have been in the absence of stress overshoots.   If these did not exist --- more precisely if there were a depletion layer of local forces after an avalanche--- then there would be macroscopic hysteresis with the force needing to be increased by the width of the depletion layer to restart the system once it has stopped after  having moved previously.

\subsection{Finite-size effects}
 
In the remainder of this last section we summarize the behavior that occurs in  our model for a range of $M$ in which there is a strongly suppressed density of nucleation seeds. We also consider briefly under what circumstances similar behavior will obtain; and discuss some of the consequences for a system in which understanding the finite size effects is crucial: earthquakes on a geological fault.

In the dissipative limit of our model, all finite size effects in the vicinity of the depinning transition occur within a range of forces around $F_c$ of width of order $1/L^\frac{1}{\nu}$.   In contrast, in the presence of stress overshoots finite-size effects can be important over a much wider range of forces.  These effects will be particularly pronounced if the suppression of nucleation seeds for further avalanches is very strong, i.e. if after a system spanning avalanche the density of nucleation segments, $r$, is very small.  

With $r$ small, unless samples are sufficiently large, $L^d> Z/(Mr)$ there will usually be no nucleation until the force has been increased by the magnitude, $M/Z$, of the overshoots.  Unless $M$ is very small, the nucleation will then be bubble-like with rapid motion in a linearly expanding bubble  occurring as soon as one segment is triggered.    For larger systems, the width of hysteresis loops will start to decrease and eventually, if the scenario presented above is correct, decrease as $1/L^\frac{1}{\nu}$ but with a large coefficient that is a randomly varying multiple of  $1/r^\frac{1}{d\nu}$.  Sites that are triggered as $F$ is increased after the whole system has been stopped may result in finite avalanches with a distribution that crosses over from one power law to another as a function of their size.  Eventually one of these avalanches will runaway: initially it will grow in a fractal manner, but when its diameter reaches of order  $L/r^\frac{1}{d}$, it will began to expand in a more deterministic manner at an approximately constant rate until it covers the whole system and the macroscopic motion restarts.   In the limit of small $r$ this regime of bubble growth will obtain over a wide range of length scales.    

\subsection{Model earthquake dynamics and statistics}

The picture presented here for effects of stress overshoots on the dynamics of finite size systems that occur as the driving force is varied have particularly interesting consequences for models  of earthquakes on disordered faults.    The appropriate driving to model a geological fault is not a constant force, by rather driving by a weak spring.  This is roughly equivalent for our case with short range elastic interactions to replacing the applied force by a ``pulling spring" with  
\be
F(t)=\frac{G}{L^2}\big(v_s t-\bar h(t)\big)
\ee
where $G$ is an effective elastic constant of order unity, the $1/L^2$ factor arises from the $\nabla^2$ elasticity, and 
\be
\bar h(t)=\frac{1}{L^d}\sum_{\bf x} h({\bf x},t)
\ee
is the spatially averaged displacement.  

As has been discussed previously, \cite{narayan2} in the absence of stress overshoots a system driven with an infinitesimal ``shearing" velocity, $v_s$, ``self-organizes" into 
a statistically steady state with a power law distribution of avalanche diameters that falls off as $\frac{1}{\ell^\kappa}\frac{d\ell}{\ell}$ out to length scales of order $L$, (the cutoff being affected by the magnitude of $G$). There is no particular qualitative distinction between the character of ``earthquakes" that have substantial $\ell\ll L$ and those with $\ell\sim L$.  

With stress overshoots that give rise to a substantial reduction  of the density of seeds for nucleating new events, the behavior is quite different.   As the pulling spring advances there will be a string of avalanches until one runs away and the whole system moves.  This motion will only stop when the force $F(t)$ has decreased as a result of the motion to $\fd$, a force  that will have variations of order $1/L^\frac{1}{\nu}$.  As the spring slowly restretches, the pulling force will increase again and a series of quakes of various sizes will occur at a rate given by $rL^d/v_s$.  These events will not relieve much of the accumulating stress and so the force will continue to increase until a force $\fh$ at which a runaway event occurs. The typical magnitude of $\fh-\fd$ and the width of its distribution are both of order $\frac{1}{r^\frac{1}{d\nu}}\frac{1}{L^\frac{1}{\nu}}$, The evolution of the runaway event will at first be fractal, but then a crossover to bubble-like growth will occur and it will grow with constant expansion rate until the whole system is moving.  

The initial slip velocity of the large earthquake will be of order $\bar v(\fh)$ which will vary substantially, but have typical magnitude 
\be
v_R\sim \frac{1}{(r^\frac{1}{d}L)^{z-\zeta}}
\ee
As it runs, the driving force will gradually decrease until it gets down to another $\fd$ at which point it will stop.  The total displacement in such an event will be of order 
$\Delta h\sim L^2\big(\fh-\fd\big)$ and its moment hence of order 
\be
m_R\sim L^d\Delta h\sim \frac{L^{d+\zeta}}{r^\frac{1}{d\nu}}
\ee
bigger by the $r$ dependent factor than  the typical largest events in the dissipative limit.  [In obtaining this result, the scaling law $\zeta+\frac{1}{\nu}=2$ was used.]

The distribution of earthquake moments in this model will be a composite of characteristic earthquake and Gutenberg-Richter-like statistics \cite{gutenberg},
\cite{wesnousky}.  Small 
events will have a power law distribution with exponent $\kappa/(d+\zeta)$ that will obtain out to moments of order 
\be
m_X\sim \left(Lr^\frac{1}{d}\right)^{d+\zeta}\  ,
\ee
much smaller than $m_R$ for small $r$, The distribution of larger events will fall off more rapidly with exponent $d/(d+\zeta)$ out to events with moments of order $m_0\sim L^{d+\zeta}$.  Events larger than this will not occur unless they are runaway events.  As these run for a considerably larger $\Delta h$ before stopping,  there will be a strong suppression of the distribution of moments from $m_0$ out to of order $m_R$, but the runaway events will have a non-trivial distribution.

This picture implies interesting anticorrelations between the occurrences of  the largest of the intermediate size and the runaway events: the former are more likely to occur relatively soon after a runaway event rather than just before one because at later stages in the ``earthquake cycle", once an event gets to a small fraction of the system size, it will almost certainly runaway.  Some information on the statistics of the largest events can also be inferred from the approximate Weibull form of the distribution of $\fh$ which will primarily determine the moments of the largest events.

In order to translate these results to models more appropriate for geological faults, the elasticity needs to be long-ranged, with static stress transfer falling off as
\be
J(\bf x,\bf y)\sim\frac{1}{|\bf x-\bf y|^3}
\ee
in the two dimensional case of interest for all but the biggest earthquakes. 
This modifies scaling laws to $\zeta+\frac{1}{\nu}=1$ and $\kappa=d-1+\zeta$.
In the absence of long range correlations in the random properties of the fault, in the dissipative limit the predicted value of $\zeta$ is zero, up to logarithms.  But with long-range correlated randomness it could be substantially larger. 

The consequences of the scenario suggested by our results for geological fault dynamics certainly merits further investigation along these and other lines. But to conclude this paper, we turn to another important issue: how much can be carried over from the stress overshoot model to other types of local dynamic effects?  We should remark that this scenario is valid within the context of 
$\mu=\nu$ for very large systems.  

\subsection{Other dynamic overshoot effects}

At the beginning of this paper we argued that some of the effects of local inertia of  a  driven, pinned elastic manifold would be qualitatively similar to the effects of stress overshoots.  A key question to raise at this point is whether the finite density of seeds for future avalanche events will be left behind in inertial systems.  We conjecture that this will be the case if the inertia is not too large, as we found here for stress overshoots, but this bears more careful thought.

In the context of earthquakes, a more important issue is  dynamic frictional weakening.  An extreme form of frictional weakening was considered in \cite{fisher3}: if once a segment has moved it is always easier for it to move again at any time in the same earthquake, this will certainly give rise to macroscopic hysteresis. As long as there is healing back up to a higher strength between events, a system spanning event will leave behind a configuration that cannot start moving again until the driving force has increased again by enough to overcome the difference between the static and dynamic friction.  In reality, the history dependence of frictional forces is complicated and some healing will occur already during an event on those segments that slow down or stop while  other parts are still moving.  But whether this, combined with dynamic stress overshoots, will leave behind a finite density of nucleation sites for easily triggerable events is a difficult but important question.  

Although this paper has perhaps raised as many questions as it has answered, the progress it represents in understanding of the dynamics of driven elastic manifolds in the presence of both randomness and non-dissipative dynamics should help frame and address some of the key questions that remain. 

\section{Acknowledgments}
The authors would like to thank Jim Sethna and Ron Maimon for very useful 
discussion.  
This work was supported in part by the National Science Foundation via grants DMR-9976621, 9809363.  J.M.S. was also supported by NSF via 
grants DMR-0109164, 9805818.

\end{document}